\documentclass[fleqn,usenatbib]{mnras}

\usepackage{subfig} % for having figures side by side
\usepackage[T1]{fontenc}

\DeclareRobustCommand{\VAN}[3]{#2}
\let\VANthebibliography\thebibliography
\def\thebibliography{\DeclareRobustCommand{\VAN}[3]{##3}\VANthebibliography}

\usepackage{graphicx}	% Including figure files
\usepackage{amsmath}	% Advanced maths commands
\usepackage{amssymb}	% Extra maths symbols

\newif\ifblackandwhite
\usepackage[svgnames]{xcolor}
\usepackage{siunitx}
\usepackage{tikz}

\usepackage{spverbatim}

\usepackage{etoolbox}
\usepackage{hyperref}
\usepackage{longtable}%
\makeatletter
\let\oldlt\longtable
\let\endoldlt\endlongtable
\def\longtable{\@ifnextchar[\longtable@i \longtable@ii}
\def\longtable@i[#1]{\begin{figure}[t]
\onecolumn
\begin{minipage}{0.5\textwidth}
\oldlt[#1]
}
\def\longtable@ii{\begin{figure}
\onecolumn
\begin{minipage}{0.5\textwidth}
\oldlt
}
\def\endlongtable{\endoldlt
\end{minipage}
\twocolumn
\end{figure}}
\makeatother

\usepackage{placeins}
\usepackage[normalem]{ulem}
\usepackage{pdflscape}
\usepackage{xcolor}
\usepackage{colortbl}%

\usepackage{booktabs}

\ifblackandwhite

\else

\fi

\usepackage{newtxtext,newtxmath}

\defcitealias{rodriguez2019optical}{RG19}

\mathchardef\Re="023C
\mathchardef\Im="023D
\usepackage[utf8]{inputenc}

\title[Generative modelling for galaxy images]{Realistic galaxy images and improved robustness in machine learning tasks from generative modelling} 

\author[B. Holzschuh et al.]{
Benjamin J. Holzschuh,$^{1,2}$\thanks{E-mail: benjamin.holzschuh@tum.de}
Conor M. O'Riordan,$^{2}$
Simona Vegetti,$^{2}$
\newauthor Vicente Rodriguez-Gomez,$^{3}$ and Nils Thuerey$^{1}$
\\
$^{1}$Technical University Munich, Department of Informatics, Boltzmannstraße 3, 85748 Garching bei München, Germany\\
$^{2}$Max Planck Institute for Astrophysics, Karl-Schwarzschild-Straße 1, 85748 Garching bei München, Germany\\
$^{3}$Instituto de Radioastronom\'ia y Astrof\'isica, Universidad Nacional Aut\'onoma de M\'exico, Apdo. Postal 72-3, 58089 Morelia, Mexico
}

\date{Accepted XXX. Received YYY; in original form ZZZ}

\pubyear{2022}

\begin{document}
\label{firstpage}
\pagerange{\pageref{firstpage}--\pageref{lastpage}}
\maketitle

% Abstract of the paper
\begin{abstract}
We examine the capability of generative models to produce realistic galaxy images. We show that mixing generated data with the original data improves the robustness in downstream machine learning tasks. We focus on three different data sets; analytical S\'ersic profiles, real galaxies from the COSMOS survey, and galaxy images produced with the SKIRT code, from the IllustrisTNG simulation. We quantify the performance of each generative model using the Wasserstein distance between the distributions of morphological properties (e.g. the Gini-coefficient, the asymmetry, and ellipticity), the surface brightness distribution on various scales (as encoded by the power-spectrum), the bulge statistic and the colour for the generated and source data sets. With an average Wasserstein distance (Fr\'echet Inception Distance) of $\num{7.19e-2}$ (0.55), $\num{5.98e-2}$ (1.45) and $\num{5.08e-2}$ (7.76) for the S\'ersic, COSMOS and SKIRT data set, respectively, our best models convincingly reproduce even the most complicated galaxy properties and create images that are visually indistinguishable from the source data. We demonstrate that by supplementing the training data set with generated data, it is possible to significantly improve the robustness against domain-shifts and out-of-distribution data. In particular, we train a convolutional neural network to denoise a data set of mock observations. By mixing generated images into the original training data, we obtain an improvement of $11$ and $45$ per cent in the model performance regarding domain-shifts in the physical pixel size and background noise level, respectively.
\end{abstract}

\begin{keywords}
galaxies: structure; methods: data analysis, statistical; techniques: image processing
\end{keywords}

%%%%%%%%%%%%%%%%%%%%%%%%%%%%%%%%%%%%%%%%%%%%%%%%%%

%%%%%%%%%%%%%%%%% BODY OF PAPER %%%%%%%%%%%%%%%%%%

\section{Introduction}

Astronomy and cosmology are entering a new era where the size of available data sets will dramatically increase. Surveys such as Euclid \citep{laureijs2011euclid}, the Vera Rubin Observatory \citep{brough2020vera}, and the Square Kilometer Array \citep[SKA,][]{dewdney2009square} will cover an unprecedented amount of the sky with comparable data quality to earlier, smaller surveys. For example, the Euclid survey will cover a $40\,\text{deg}^2$ contiguous area, compared to $2\,\text{deg}^2$ for the COSMOS field \citep{scoville2007cosmic} at a similar angular resolution.

The employment of machine-learning techniques has been proposed to deal with this influx of data in several different fields. For example, these techniques have already produced valuable results in e.g. star-galaxy classification \citep[][]{kim2016star},  morphological classification \citep[][]{huertas2015galaxy}, measuring the cosmic shear \citep[][]{tewes2019weak}, and identifying strong gravitational lens systems \citep[e.g.][]{lanusse2018cmu}. Nonetheless, machine learning remains a data-driven method, i.e., the performance of any machine-learning model critically depends on the data set used in training. This can be an issue in astronomy, where data is often scarce or suffers from biases and selection effects. 

For example, this fundamental problem arises when applying machine learning to the task of finding strong gravitational lens systems, where spatially highly resolved images of source galaxies are required for training.
However, observations of sufficient resolution only exist for low redshift galaxies, which are fundamentally unlike the high redshift source galaxies in strong lens systems. Using low redshift galaxies in training, because their observations are readily available, and analysing high redshift galaxies in testing constitutes a domain shift. Methods to deal with the domain shift problem are the subject of ongoing research in machine learning \citep[][]{wilson2020survey, wang2018deep}.

We propose to use generative modelling to address these kinds of problems. The focus of this paper is twofold. First, we train several recent generative models on different data sets of galaxy images. In assessing the models' performance we focus on physically motivated metrics and compare these metrics to those traditionally used in computer vision. Second, we show that the generated data can improve the robustness of other models which make use of the generated data during training.

We make use of three different data sets in this paper: synthetic observations using S\'ersic profiles, real galaxy observations from the COSMOS field \citep{Mandelbaum.2012}, and a data set of synthetic high-resolution galaxy images at redshift $z=0$ from the IllustrisTNG simulation \citep[][]{nelson2019illustristng, springel2017, nelson2017, naiman2018, marinacci2018, pillepich2017}, created using the SKIRT radiative transfer code \citep[][]{baes2011skirt, baes2015skirt} as described in \citet[][]{rodriguez2019optical}, which we refer to as `the SKIRT data set'. The spatial resolution of galaxy images from each data set increases in this order while the data set size decreases. While we can create analytic S\'ersic profiles  with randomly drawn parameters indefinitely, there are $\sim$20,000 images from COSMOS and $\sim$9,000 synthetic high-resolution images in the SKIRT data set. This resembles the trade-off between data quality and data availability often seen in practice.

We focus on variational and adversarial-based methods for generative modelling, using three model architectures specifically: a generative adversarial neural network \citep[StyleGAN, ][]{karras2020analyzing}, adversarial latent autoencoders \citep[ALAE, ][]{pidhorskyi2020adversarial}, and variational autoencoders \citep[VAE, ][]{kingma2013auto, kingma2019introduction}. Recently, other methods like score-based models \citep[][]{song2020score} and normalising flows \citep[][]{grcic2021densely} have shown competitive results compared to traditional approaches, while also learning to approximate the probability of data samples.

In machine learning and computer vision, there are several metrics for comparing and ranking different generative models, but an approach that scores well according to these metrics may not necessarily produce physically realistic images. We discuss more physically motivated metrics that are based on the $\mathcal{W}_1$-Wasserstein distance between 1D distributions of morphological measurements and transformations of the 2D power spectra on which we base our evaluation.

As a further test for our trained models, and generative modelling in general, we evaluate how successfully generated data can replace or expand the original data in downstream machine learning tasks. As an example task, we consider the problem of denoising images. Real observations are affected by different sources of noise and blurred by the point spread function (PSF) of the instrument. We train a convolution neural network (CNN) in order to remove the noise and PSF in a supervised manner. For this purpose, we consider the SKIRT data set and data generated from the StyleGAN to which we artificially add a PSF and background noise. We evaluate the performance of denoising networks trained with different mixing factors for the SKIRT data and the generated data and analyse the robustness of the trained models by measuring how domain shifts in the test data affect the performance.

In Section \ref{sec:gen_modelling}, we introduce the generative models considered in this paper. In Section \ref{sec:datasets}, each data set is described in more detail. In Section \ref{sec:metrics}, we describe our metrics for evaluating the generative models. This is followed by an evaluation of the different generative models on the three training data sets in Section \ref{sec:evaluation}. 
In Section \ref{sec:generalization}, we analyse how mixing the SKIRT data set with data generated from StyleGAN affects the performance of the trained denoising models on test data with domain shifts.

\section{Generative Modelling}
\label{sec:gen_modelling}

Generative modelling with deep learning has been used successfully in areas such as: face generation and reconstruction \citep[e.g.][]{karras2020analyzing}, precipitation nowcasting \citep[e.g.][]{Ravuri.2021}, and solving inverse problems in medical imaging \citep[e.g.][]{song2021solving}.
Generative models have been considered also for astronomical data. For example, \citet{arcelin2021} employ VAEs to learn probabilistic models of galaxies for deblending. \citet{bretonniere2021} propose to train VAEs for generating synthetic data for the upcoming Euclid survey, which can help with the preparation and calibration of algorithms. \citet{smith2022ddpm} train denoising diffusion probabilistic models on galaxy images and demonstrate applications for in-painting of occluded data and domain transfers. 

GANs represent another popular method for generative models and were successfully adopted in a variety of applications \citep[e.g.]{radford2016dcgan,xie2018tempoGan,karras2020analyzing}.
In the context of astronomy, \citet{fussell2019forging} have, for example, used chained 
GANs to demonstrate that the distributions of morphological parameters from generated galaxies are close to those of the GAN training data set. \citet{yip2020} have used synthetic data generated by GANs to train neural networks for exoplanet discovery.   

In general, a generative modelling approach aims to learn the probability distribution $P(X)$ of some data $X$. The distribution $P(X)$ is approximated by a model $P_\phi(X)$ with parameters $\phi$, i.e. for some observed data $x$, we want to have $x \sim p_\phi(x)$. Drawing samples from the learned distribution $p_\phi(x)$ then produces new, synthetic data with a similar distribution to the original data. As a means to break down this task and to provide an efficient way to sample from $P_\phi(X)$, we introduce latent variables $z$ \citep{kingma2019introduction} and write 
\begin{equation}
	p_\phi(x) = \int_z p_\phi(x,z) dz = \int_z p_\phi(x|z) p_\phi(z) dz\,.
\end{equation}
The collection of latent variables can be thought of as a compact, meaningful representation of the data. 

In the rest of this section, we introduce the three generative modelling techniques which we use in this paper; variational autoencoders, generative adversarial neural networks, and adversarial latent autoencoders. The reader already familiar with these concepts can skip-over this section, and move to Section \ref{sec:datasets} for a description of the data considered in this paper.

\subsection{Variational autoencoders} \label{subsection:vae}
Variational autoencoders split learning the data distribution into two parts. A decoder model for $p_\phi(x,z) = p_\phi(x|z)p(z)$, where $p(z)$ is a fixed prior distribution, and an encoder model $p_\phi(z|x)$, which is intractable to compute but is approximated by  $q_\phi(z|x)$.  We define the Kullback-Leibler divergence between two probability distributions with densities $q$ and $p$ as 
\begin{equation} 
    \text{KL}(q||p) = \int q(x) \log\left[ \frac{q(x)}{p(x)} \right] dx\,.
\end{equation} 
We can then write the data probability as 
\begin{equation}
	\log p_\phi(x) = \mathrm{E}_{z \sim q_\psi(z)}
	\left\{ \log \left[ \frac{p_\phi(x,z)}{q_\psi(z)}\right] \right\}
	+ \mathrm{KL}(q_\psi(z|x)||p_\phi(z|x))\,,
\end{equation}
where the expectation $\mathrm{E}$ on the right-hand side is called the evidence lower bound \citep[ELBO, see][for a derivation]{kingma2019introduction}. Maximizing the ELBO for $\phi$ and $\psi$  has two expected effects: the decoder model is improved by maximizing $p_\phi(x|z)$, and, the encoder model is improved by minimizing $\text{KL}(q_\psi(z|x)||p_\phi(z|x))$.
In cases where both the encoder and decoder are neural networks, the ELBO maximization can be written as maximizing
\begin{equation} \label{eq:vae_objective}
	\log p_\phi(x|z) + \log p_\phi(z) - \log q_\psi(z|x)\,,
\end{equation}
where the first term $\log p_\phi(x|z)$ is the negative reconstruction loss of an autoencoder, and the two other terms act as a regularization for the network \citep{goodfellow2016deep}. See Appendix \ref{subsec:vae_detailed} for a detailed description of the encoder and decoder architecture.

A natural reconstruction loss for physical data is the squared pixel-wise distance, or L2 distance, between the original data and the reconstruction. However, the L2 distance leads to overly-smooth images.
%}the decoder network learning to produce smeared outputs in practice. 
This is because the fine textures and details cannot reasonably be encoded in the latent representation. Therefore, the reconstruction loss is often lower when the decoder generates smeared pictures than when it produces fine details that do not align perfectly with the original. 
\begin{figure}
	\centering
	\includegraphics[width=.5\textwidth,clip=True, trim= 2 2 2 2]{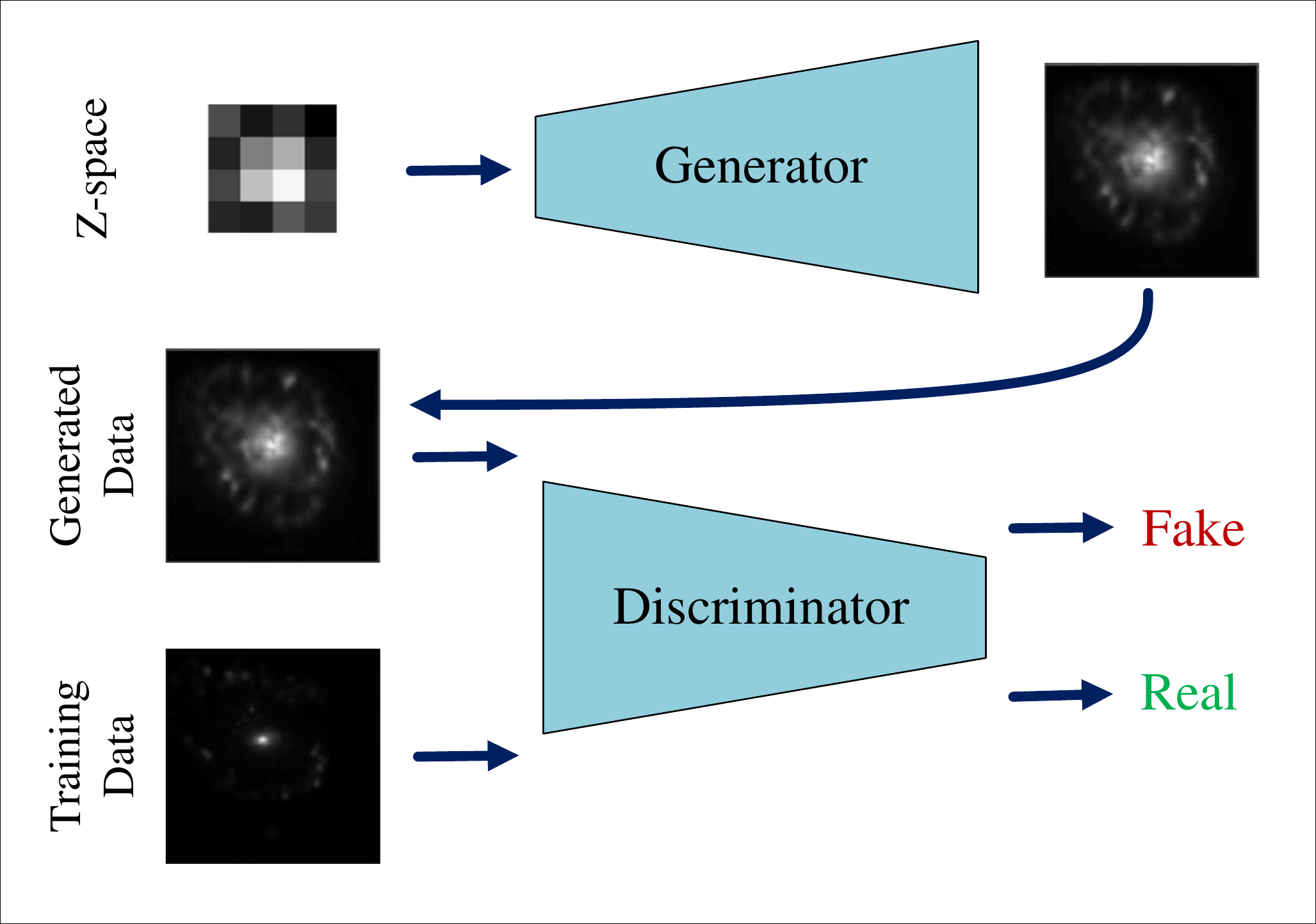}
	\caption{GAN architecture overview. $Z$-space refers to a set of vectors $z$, which are randomly drawn from a Normal distribution. Generated data is obtained by feeding $z$ to the generator network. The discriminator classifies whether its input is drawn from the training distribution that we try to learn (Real, output 0), or was generated by the generator (Fake, output 1).}
	\label{fig:gan_training}
\end{figure}
A remedy for this problem is provided by an adversarial loss $\mathcal{L}_\mathrm{adv}$, which is used in generative adversarial networks (GANs). We discuss the latter in more detail in the following section.

\subsection{Generative adversarial networks} \label{subsection:gan}

GANs are comprised of a discriminator network $D$ and a generator network $G$  parameterized by $\Phi_D$ and $\Phi_G$.  New samples are generated by drawing a random vector $z$ with entries from a standard Normal distribution and feeding it to the generator network. The optimization objective of a GAN is twofold.  The discriminator network is trained to minimize the classification loss $\mathcal{L}_\mathrm{adv}$ on whether an input image is sampled from the real data distribution or generated by the decoder network. The generator network, on the other hand, is optimized to maximize this classification loss. The training of the network happens in two steps: (1) the weights of the generator are frozen and the discriminator is trained, (2) the weights of the discriminator are frozen, and the generator is trained. 
This process is repeated until the generator and discriminator converge to a Nash equilibrium, i.e. neither the generator can increase $\mathcal{L}_\mathrm{adv}$ nor the discriminator can decrease $\mathcal{L}_\mathrm{adv}$. See Fig. \ref{fig:gan_training} for an illustration. The training objective can be written as
\begin{equation}
	\min_{\Phi_G} \max_{\Phi_D} \mathbf{E}_{x\sim p(x)} [\log(D_{\Phi_D})] + \mathbf{E}_{z \sim p(z)}[\log[1 - D_{\Phi_D}(G_{\Phi_G}(z))]]\,.
\end{equation}
In this work, we employ the StyleGAN architecture \citep{karras2020analyzing}, which has shown impressive results in synthesizing high-resolution images. There are several differences between the standard GAN architectures and StyleGAN, although the main training loop and network components remain the same. For example, the generator network in StyleGAN is stochastic, i.e. there is an additional noise input to the generator that causes variations in the synthesized data on a small scale.
As discussed by \citet{goodfellow2014generative}, GANs attain visually better results and resolve finer details than VAEs. On the other hand, it is much more difficult to train GANs until convergence. Moreover, GANs often suffer from mode collapse, i.e., the GAN focuses on learning some small regions of the training dataset distribution very well while not learning about other parts. Therefore, GANs might not cover the training distribution as well as VAEs do. To address this issue, efforts have been made to combine GANs and VAEs in a new architecture with all the advantages of both methods.

\subsection{Adversarial latent autoencoder} \label{subsection:alae}
The adversarial latent autoencoder (ALAE) is one such attempt at a combined architecture, and extends the GAN architecture with a reconstruction loss \citep[][]{pidhorskyi2020adversarial}. ALAEs introduce an additional abstraction by distinguishing between the $Z$-space, where data is sampled, and a space for representing data in a compact but meaningful way, called $W$-space. This is coupled with a reconstruction loss $\mathcal{L}_\mathrm{rec}$ that is optimized together with the adversarial loss $\mathcal{L}_\mathrm{adv}$. The generator is effectively split into two parts, a dense neural network $F$ that maps from the $Z$-space to the $W$-space, and the decoder $G$ that maps from the $W$-space to the data space. 
The discriminator is also split into two parts, an encoder $E$ that maps from the data space to the $W$-space, and a smaller discriminator $D$ that maps from the $W$-space to a single output.  
The probability of the data can be written as follows
\begin{equation}
	p_\phi(x)  = \int \int p_\phi(x|w) p_\phi(w|z) p(z) dw dz\,,
\end{equation}
where $p_\phi(x|w)$ corresponds to the decoder $G$ and $p_\phi(w|z)$ to $F$. The reconstruction loss is based on establishing the reciprocity of the encoder and decoder, i.e. $G^{-1} = E$
\begin{equation}
	\mathcal{L}_\text{rec} = \mathbb{E}_{z \sim p(z)} [ \left| \left| \text{E}(\text{D}(F(z))) - F(z) \right|\right|_1]\,, 
\end{equation}
where $||\cdot ||_1$ denotes the L1-norm. The adversarial latent autoencoder is trained by alternately optimizing two objectives 
\begin{equation}
    \max_{G,F} \min_{E,D}  \mathcal{L}_\text{adv}(G,F,E,D),
\end{equation}
and 
\begin{equation}
	\min_{G,E} \mathcal{L}_\text{rec}(G,F,E)\,.
\end{equation}
Some authors impose additional regularizations on the distribution of the latent representations similar to VAEs \citep{srivastava2017veegan, ulyanov2018takes}.
We refer to \citet{pidhorskyi2020adversarial} for more details on this topic. 

\begin{figure*}
	\centering
	\subfloat[SKIRT synthetic images]{ \begin{tabular}[b]{c}%
	\includegraphics[width=\textwidth]{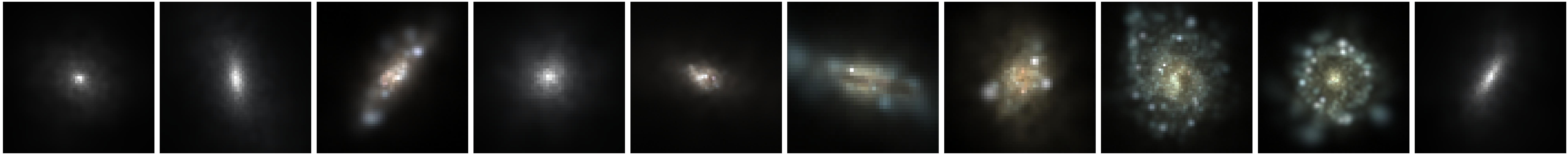}\\
	\includegraphics[width=\textwidth]{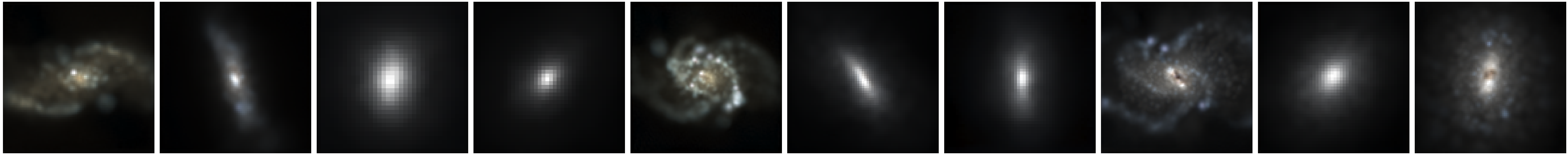}
	\end{tabular}} \\
	\subfloat[COSMOS field observations]{ \begin{tabular}[b]{c}%
	\includegraphics[width=\textwidth]{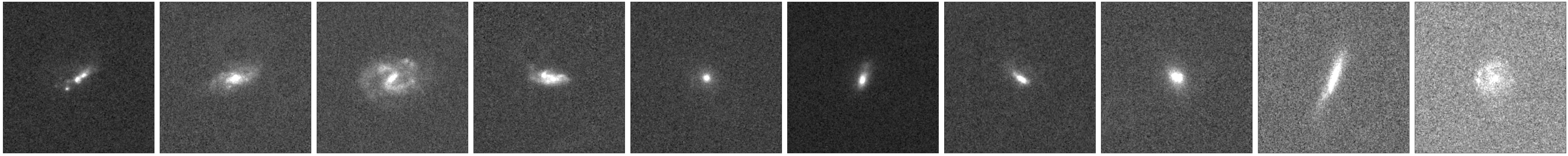} \\
	\includegraphics[width=\textwidth]{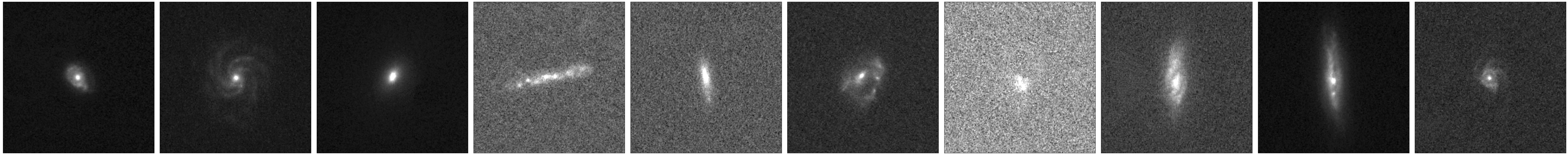}
	\end{tabular}} \\
	\subfloat[S\'ersic profiles]{ \begin{tabular}[b]{c}%
	\includegraphics[width=\textwidth]{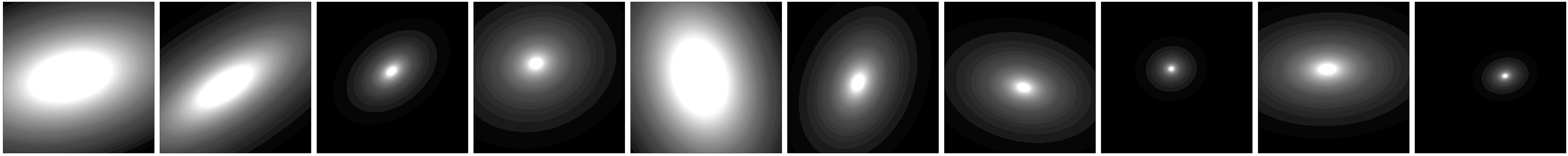} \\
	\includegraphics[width=\textwidth]{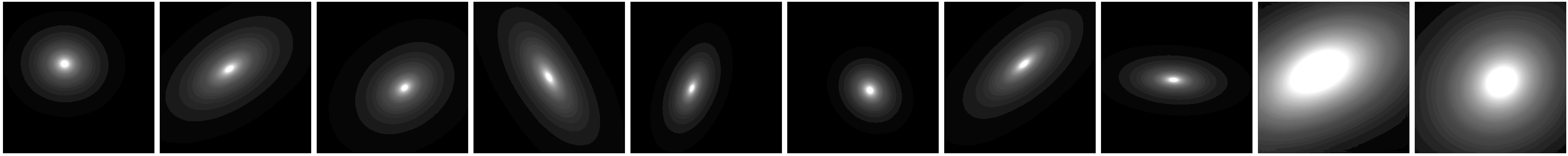}
	\end{tabular}}
    \caption{Original data (\textbf{top row}) and StyleGAN generated data (\textbf{bottom row}) for different galaxy image data sets.}
    \label{fig:dataset_examples}
\end{figure*}

\section{Data sets} 
\label{sec:datasets}

The first data set consists of purely analytic S\'ersic profiles with fixed physical pixel size and one filter. The COSMOS data set contains observations from the Hubble Space Telescope (HST). The galaxy images are obtained by cropping those from the catalogue by \citet{Mandelbaum.2012}. Finally, we consider synthetic high-resolution images from the IllustrisTNG simulation \citep[\citetalias{rodriguez2019optical} hereafter]{rodriguez2019optical} with four filters that match the SDSS survey.  In Fig. \ref{fig:dataset_examples}, we show an example of galaxies taken from each data set as well as a preview of the galaxies generated by the StyleGAN model, which we discuss in Section \ref{sec:evaluation}. Additionally, in Table \ref{tab:dataset_statistics}, we summarise essential quantities characterising each galaxy sample.

\begin{table}
                     \begin{tabular*}{.5\textwidth}{@{}l@{\hspace*{9pt}}l@{\hspace*{7pt}}l@{\hspace*{7pt}}l@{\hspace*{7pt}}l@{\hspace*{7pt}}l@{}}
                      \hline
                      Data set & \# objects & Filters & Pixels & Pixel scale & PSF/noise \\
                      \hline
                      S\'ersic profiles & 50,000 & F606W & $256 \times 256$ & \ang{;;0.04} & no \\  
                      COSMOS & 20,114 & F814W & $256 \times 256$ & \ang{;;0.03} & yes \\ 
                      SKIRT & 9,564 & \emph{g,r,i,z} & $N \times N$ & \ang{;;0.04} & no\textsuperscript{1} \\
                      \hline
                     \end{tabular*}
                     \caption{Data set summary statistics. The number of pixels for the SKIRT data set depends depends on the half-mass radius of each galaxy.}
                     \label{tab:dataset_statistics}
\end{table}

\footnotetext[1]{For the morphological measurements we simulate actual observations and include background noise and PSF. See Section \ref{sec:appendix_morphological measurements} in the appendix. }

\subsection{S\'ersic profiles} \label{subsec:sersic_profiles} The S\'ersic profile is frequently used to describe the surface brightness distributions of elliptical galaxies. The intensity $I$ at a distance $r$ from the centre of the galaxy is given by
\begin{equation}
         I(r) = I_0 \exp\left\{-n_\mathrm{s}\left[\left(\frac{r}{r_\mathrm{eff}}\right)^{\frac{1}{n}}-1\right]\right\},
\end{equation}
where $I_0$ is the intensity at $r=0$, $n_\mathrm{s}$ is a normalising constant \citep[][]{ciotti1999analytical} and $r_\mathrm{eff}$ is the effective radius of the galaxy. We can create an unlimited number of galaxies with Sérsic profiles, but limit the training data set size to $50,000$ images. All images have a size of 256 x 256 pixels.

To avoid numerical errors due to the discretisation, we use sub-pixelisation to better resolve the subgrid around the peak. During sampling, we place uniform priors on the parameters of the profile, which do not necessarily resemble the actual physical distribution from observations of real galaxies. We draw the central position coordinates from $\mathcal{U}(-1,1)$ arcsec, the effective radius $r_\mathrm{eff}$ from $\mathcal{U}(1, 4)$ arcsec, the S\'ersic index $n_\mathrm{s}$ from $\mathcal{U}(1, 4)$, the axis ratio $q$ from $\mathcal{U}(0.4, 1.0)$ and finally the position angle from $\mathcal{U}(0, \pi)$. The AB magnitude is fixed for all S\'ersic profiles at $22$.

\subsection{COSMOS} 
\label{subsec:COSMOS} 

This data set consists of individual galaxies from the contiguous COSMOS field \citep{Mandelbaum.2012, bretonniere2021}. The field was observed using the F814W filter with a drizzle pixel scale of $0.03$ $\mathrm{arcsec}$ $\mathrm{pixel}^{-1}$ and limiting point source depth at $5 \sigma$ of $27.2$ mag. 
As each galaxy image has a different number of pixels, which depends on the size of the galaxy itself, we crop them to a final size of $256 \times 256$ pixels (from the centre) and discard all galaxies with a smaller number of pixels. Otherwise, we apply no further pre-processing steps to the data.

\subsection{SKIRT} 
\label{subsec:skirt} 

This data set comprises synthetic galaxy images obtained by running the SKIRT radiative transfer code on galaxies from the IllustrisTNG (TNG100) simulation \citepalias{rodriguez2019optical} at snapshot $z=0$. The SKIRT code adds realistic radiation to dusty galaxies by simulating various physical processes. 
The output images have a physical pixel size of $0.276$ comoving kpc/h, units of $e^{-}s^{-1}\mathrm{pixel}^{-1}$. The size of each galaxy image is equal to 15 times its stellar half-mass radius. As part of our pre-processing, we crop out pixels corresponding to one fourth of either the width or height at each border. To account for the different number of pixels $N$, we augment the models discussed in Section \ref{sec:gen_modelling} with dynamic rescaling, which allows us to generate image data with a varying number of pixels $N$. This approach is detailed in Appendix \ref{sec:detailed_model}. We remove all images with size $N \leq 64$. There remain $\sim$9,000 galaxies of which we exclude 25 per cent as a test set for the denoising task described in Section \ref{sec:results_denoising} and train the generative models on the remaining 75 per cent. 

\section{Metrics} 
\label{sec:metrics}

In order to quantify and compare the performance of the different generative models, we first need to specify which image properties to consider and how to compare the distribution of these quantities in the original and generated data sets. While this problem has been successfully addressed
for natural images in computer vision \citep{heusel2017gans,sajjadi2018assessing,binkowski2018demystifying}, it is still unclear how meaningful these metrics are when applied to astronomical data.
In this paper, we first consider a set of physically-related galaxy properties, and compare how well these are reproduced by the different generative models in terms of the  $\mathcal{W}_1$-Wasserstein distance (see Section \ref{subsec:optimal_transport} for more details) between their distributions and those of the training data. Additionally, we quantify the relative performance of the generative models in terms of  traditional computer vision metrics and test how appropriate these are for galaxy images. 

\subsection{Physical properties} 

\subsubsection{Morphology}
\label{sec:morph_measurements}

We employ the \verb statmorph $\,$ package \citepalias{rodriguez2019optical} to compute optical morphological measurements of the galaxy images from the training and the generated data sets. In particular, we focus on the asymmetry, the smoothness, the concentration, the Gini-coefficient, the $M_{20}$ statistic, and the half-light radius. We also consider parameters obtained by fitting a S\'ersic profile to the light distribution, specifically the S\'ersic index, the orientation, ellipticity, and elongation calculated with respect to the centroid. See Section \ref{sec:appendix_morphological measurements} for an overview as well as for some further pre-processing steps that we apply only before computing the morphological measurements. 

\begin{figure*}
  \includegraphics[width=\textwidth, clip=True, trim=5 5 5 5]{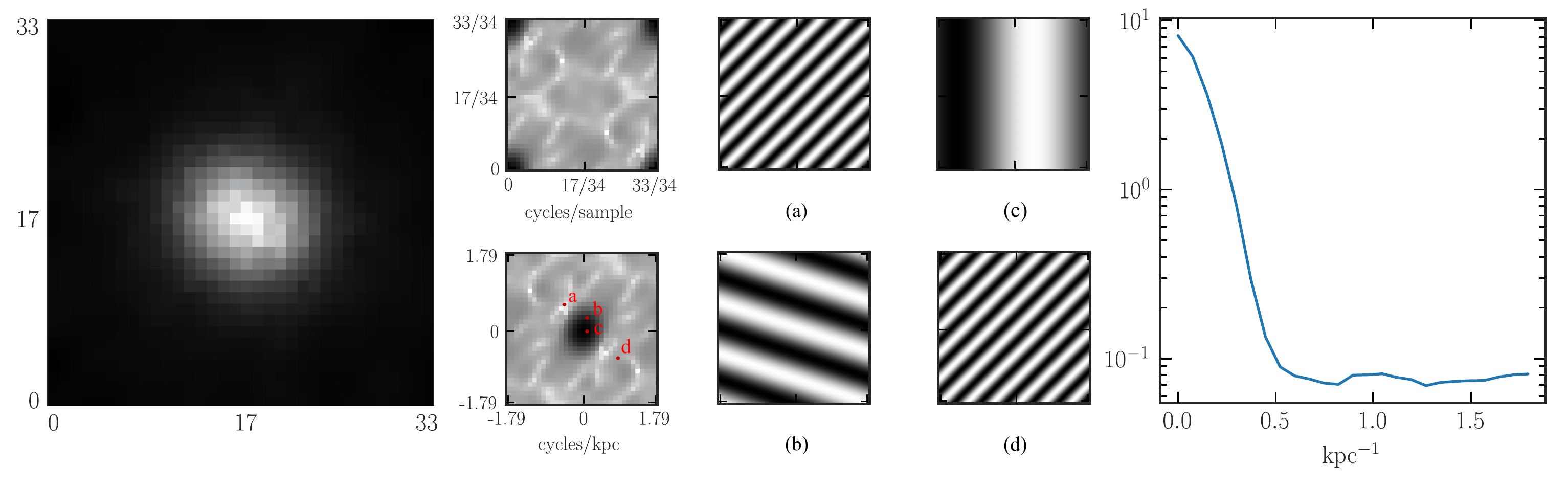}
  \caption{The 2D power spectrum and its connections to wavelengths corresponding to the physical pixel size. The leftmost panel shows the $r$-band of a galaxy from the SKIRT data set with $34 \times 34$ pixels. The two panels right to it show its 2D power spectrum (top) and the shifted 2D power spectrum (bottom). The value at each pixel combined with the coordinates gives the magnitude of a $2$-dimensional wave and its frequency (number of cycles per sample), see panels (a) to (d) and the corresponding points in the shifted 2D power spectrum. Since the pixel size corresponds to a physical length, we can write the average magnitude as a function of the wave frequency by eq. \ref{eq:radial_power_spectrum}, which is shown in the rightmost panel. }
  \label{fig:powerspectra_example}
\end{figure*}

\subsubsection{2D power spectrum} \label{subsec:2d_power_spectrum}

We also consider transformations of the 2D power spectrum of the surface brightness distribution of the images.
Our process involves computing the 2D discrete Fourier transform (DFT) $F(x,y)$ given an $N \times N$ pixel grid and we require that the physical size of each pixel is the same for all images. See Fig. \ref{fig:powerspectra_example} for an overview. 

The frequencies of the 2D power spectrum correspond to wavelengths depending on either the physical pixel size or angular pixel scale $L_\mathrm{pixel}$, which is the distance between two grid points in the pixel grid. The maximum frequency that corresponds to the smallest wavelength is given by 
\begin{equation}
    f_\mathrm{max} = \frac{1}{\sqrt{2}L_\mathrm{pixel}}\,.
\end{equation}
We define the average power spectrum given a radial distance (or frequency $f$) from the centre in the shifted 2D power spectrum as
\begin{equation} 
   AP(f) = \frac{1}{2\pi r^2}\int_{(x^2 + y^2)^{1/2} = f} |F(x,y)| dxdy\,. \label{eq:radial_power_spectrum}
\end{equation}
We partition the frequency range $[0, f_\mathrm{max}]$ into $15$ intervals, which we call modes. The size of each interval is proportional to its frequency, and mode $0$ corresponds to frequency $0$, i.e. $|F(0,0)|$. See Table \ref{tab:quantitative_metrics} for the exact partitioning. 

\subsubsection{Colours}

As a final physically-motivated quantity we consider the $g-i$ galaxy colour. In this paper, the SKIRT data set is the the only one with multiple bands. For objects with single-band-only information, a comparison of the distribution of total flux between the generated data and source data is already covered by the comparison of the 2D power spectra. 
We define the $g-i$ colour index according to the SDSS conventions \citep[][]{york2000sloan} as the ratio between the total flux in the $g$-band and the $i$-band. The total flux in a specific band is obtained by measuring the Petrosian radius of the galaxy in that band and integrating the flux within a circular aperture twice the Petrosian radius and pixel coordinates that minimize the galaxy asymmetry $A$ as the centre. See Fig. \ref{fig:colors_plot} for the $g-i$ colour distribution in the SKIRT data set. 

\subsection{Wasserstein distance} \label{subsec:optimal_transport}

Often, two distributions are compared in terms of their mean and standard deviation. However, in many cases, these distributions cannot be approximated by simple Gaussians and a comparison that relies on the mean and standard deviation may be misleading. For this reason, we use the Wasserstein distance instead. 

Let $\mu$ denote a probability measure for the training data on the space of all possible images $M$, i.e. we may assume that the training data is randomly sampled from $\mu$. Similarly, let $\nu$ be the probability measure corresponding to the generator. The $p$-th Wasserstein distance between $\mu$ and $\nu$ is defined as 
\begin{equation}
    \mathcal{W}_p(\mu, \nu) = \left[ \inf_{\gamma \in \Gamma(\mu, \nu)} \int_{M \times M} d(x,y)^p d\gamma(x,y) \right]^{\frac{1}{p}}\,,
\end{equation}
where $d : M \times M \to \mathbb{R}_0^+$ is a distance function for images, e.g. the average pixel-wise squared difference and $\Gamma(\mu, \nu)$ denotes the set of all probability measures on $M \times M$ with $\mu$ and $\nu$ as marginals.

There are two constraints on $\gamma(x,y)$:
 \begin{equation}
    \int \gamma(x,y) dy = \mu(x)
\end{equation}
and 
 \begin{equation}
    \int \gamma(x,y) dx = \nu(y)\,.
\end{equation}

Computing the $\mathcal{W}_1$-Wasserstein distance between two distributions quickly becomes computationally infeasible for high dimensions. To solve this issue, in this paper, we take the approach to not compute the Wasserstein distance on $\mu$ and $\nu$ directly, but to compute it on the so-called push-forward measure of $\mu$ and $\nu$. 

For a given function $g : M \to \mathbb{R}$ the push-forward measure $\mu^{(g)}$ is defined by $\mu^{(g)}(x) = \mu(g^{-1}(x))$. Since these measures are 1-dimensional, we can compute the distance efficiently. For this, we use the package \verb Python $\:$ \verb Optimal $\:$ \verb Transport $\:$ \citep[][]{flamary2021pot} with the Euclidean distance for the underlying metric. 
We define a pseudo-metric based on push-forward measures with functions from a set of critic functions $\mathcal{F}$ by 
\begin{equation} \label{eq:wasserstein_pseudo_metric} 
    d_\mathcal{F}(\mu, \nu) := \sup_{f \in \mathcal{F}} \mathcal{W}_1 \left(\mu^{(f)}, \nu^{(f)}\right)\,. 
\end{equation}
Since we can only consider a finite number of functions $f \in \mathcal{F}$, it makes sense to choose functions that have application-specific advantages. A natural choice for the push-forward measures are the optical morphological measurements discussed in Section \ref{sec:morph_measurements} and the frequency ranges consider in the 2D power spectrum of the galaxy images, discussed in Section \ref{subsec:2d_power_spectrum}. 

\subsection{Computer vision-based metrics} \label{subsec:computer_vision_metrics}
Metrics for evaluating generative models that are commonly used in computer vision are based on feature-extraction networks \citep[][]{kohl2020learning, zhang2018unreasonable}.
Feature-extraction networks in this context are convolutional neural networks pre-trained for classification tasks on large data sets. We can obtain compressed representations of the input images by extracting intermediate activation layers from these networks. These compressed representations can be computed for both the training data set and generated data. The number of activations in the compressed representation is typically between $10^3$ and $10^5$. It is not possible to perfectly reconstruct the input images from the compressed representation, but the activations recognize basic geometric shapes as well as more complex compositions of shapes and patterns. Therefore, the content of images is in general very well described by the compressed representation. Instead of defining the similarity between sets of images, the problem is simplified to finding a similarity metric for two sets of real-valued vectors. This is conceptually very similar to the push-forward measures, but pushing-forward to $\mathbb{R}^n$ instead of $\mathbb{R}$, where $n$ is the size of the compressed representation. We will discuss the Fr\'echet Inception Distance \citep[FID,][]{szegedy2016rethinking} and the Kernel Inception Distance \citep[KID,][]{binkowski2018demystifying}, which are based on two different approaches to compare the distributions of compressed representations. In general, there is strong evidence that the FID and KID correlate with the perceptual similarity of image-like data sets \citep[][]{zhang2018unreasonable}.

\subsubsection{Fr\'echet Inception Distance}

The compressed representations are obtained from different layers of the InceptionV3 model pre-trained on ImageNet \citep{szegedy2016rethinking} for the training data set and the generated data.
Then, the $\mathcal{W}_1$-distance is computed for the two sets of compressed representations. To make this process computationally feasible, a simplifying assumption that each set is drawn from a multivariate Gaussian distribution is made. In this case, the FID can be computed as 
\begin{equation} \mathrm{FID} := |\mu_\mathrm{train} - \mu_\mathrm{gen}|^2 + \mathrm{tr} \left[\Sigma_\mathrm{train} + \Sigma_\mathrm{gen} - 2(\Sigma_\mathrm{train} \Sigma_\mathrm{gen})^{1/2}\right]\,, \end{equation}
where $\mu_\mathrm{train}$ and $\mu_\mathrm{gen}$ are the mean of the compressed representations of the training data set and generated data. Analogously, for the covariances $\Sigma_\mathrm{train}$ and $\Sigma_\mathrm{gen}$. The FID correlates strongly with perceived image quality and diversity.
\subsubsection{Kernel Inception Distance}
The KID relies on compressed representations from the same InceptionV3 model as the FID. The KID is based on the maximum mean discrepancy, which is defined as
\begin{equation} \label{eq:kr_duality}
    \mathrm{MMD}(\mu, \nu) := \sup_{||f||_\mathcal{H} \leq 1} \left|\mathbb{E}_{x \sim \mu}[f(x)] - \mathbb{E}_{x \sim \nu}[f(x)]\right|\,.
\end{equation}
where $\mathcal{H}$ is some kernel Hilbert space. In the case of the KID, the chosen reproducing kernel Hilbert space is defined by the kernel 
\begin{equation}
    K(x,y) = \left[\frac{1}{d}\phi(x)^T\phi(y)+1\right]^3\,, 
\end{equation}
where $\phi$ represents the InceptionV3 \citep[][]{szegedy2016rethinking} network mapping to the compressed representations. The reproducing kernel property of $\mathcal{H}$ implies that for all $f \in \mathcal{H}: f(x) = \langle f, K(x, \cdot)\rangle$. This can then be used to show 
\begin{equation}
\begin{split}
    &\text{KID}(\mu, \nu) := \text{MMD}^2(\mu, \nu) = \\
  &\qquad \mathbb{E}_{x,\Tilde{x} \sim \mu}[K(x, \Tilde{x})] + \mathbb{E}_{y,\Tilde{y} \sim \nu}[K(y, \Tilde{y})] - \mathbb{E}_{x\sim \mu, y \sim \nu}[K(x,y)]\,,
\end{split}
\end{equation}
see \cite{gretton2012kernel}. The advantages of the KID over the FID are that there exists an unbiased estimator for the KID and it does not assume a parametric form for the distribution of compressed representations. 

\subsubsection{Feature-extraction for physical data with pre-trained CNNs}

A difficulty in overcoming when using the FID and KID metrics is that images need to be compatible with the expected input of the feature-extraction networks. In the case of the InceptionV3 network, the model expects inputs of the shape $256 \times 256$ with three channels and values between 0 and 1 corresponding to red, green and blue. We considered the following strategy to convert galaxy images to RGB ones. Each image is normalized by dividing by its maximum pixel intensity. With this approach disregards differences in the absolute distribution of luminosity and focuses on the relative distribution of luminosity for each galaxy.  Afterwards, we pick a single band from the data and apply a colourmap. In this paper, we have used the \textit{magma} colourmap from the \verb matplotlib $\:$ package in Python \citep[][]{Hunter:2007}. For the SKIRT data set with four filters, we always pick the $r$-band.  

\begin{table*}
 \begin{tabular*}{\textwidth}{@{}l@{\hspace*{16pt}}c@{\hspace*{10pt}}c@{\hspace*{10pt}}c@{\hspace*{10pt}}|@{\hspace*{10pt}}c@{\hspace*{10pt}}c@{\hspace*{10pt}}c@{\hspace*{10pt}}|@{\hspace*{10pt}}c@{\hspace*{10pt}}c@{\hspace*{10pt}}c@{\hspace*{10pt}}}
  \hline
   & \multicolumn{3}{c}{\hspace*{-28pt}SKIRT} & \multicolumn{3}{c}{\hspace*{-28pt}COSMOS} & \multicolumn{3}{c}{\hspace*{-20pt}S\'ersic Profiles} \\
  \hline
  & VAE & StyleGAN & ALAE & VAE & StyleGAN & ALAE & VAE & StyleGAN & ALAE \\[2pt]
  & $\left[\times\num{e-2}\right]$ & $\left[\times\num{e-2}\right]$ & $\left[\times\num{e-2}\right]$ & $\left[\times\num{e-2}\right]$ & $\left[\times\num{e-2}\right]$ & $\left[\times\num{e-2}\right]$ & $\left[\times\num{e-2}\right]$ & $\left[\times\num{e-2}\right]$ & $\left[\times\num{e-2}\right]$ \\[2pt]
  \hline
  \hline
  Morphological properties & & & & & & & & &\\[2pt]
  $\quad$ \makebox[0pt][l]{Asymmetry}\phantom{Orientation xxxxxxxx xx} & 95.12 & 17.09 & 15.09 & 54.78 & 11.41 & 33.17 & 97.87 & 57.93 & 48.94 \\[2pt]
  $\quad$ \makebox[0pt][l]{Smoothness}\phantom{Orientation xxxxxxxx xx} & 115.44 & 4.85 &  6.86 & 57.61 & 5.06 & 29.12 & 69.46 & 4.83 & 14.47 \\[2pt]
  $\quad$ \makebox[0pt][l]{Concentration}\phantom{Orientation xxxxxxxx xx} & 45.05 & 3.92 & 31.12 & 46.69 & 8.31 & 58.56 & 22.15 & 5.09 & 5.36 \\[2pt]
  $\quad$ \makebox[0pt][l]{Gini coefficient}\phantom{Orientation xxxxxxxx xx} & 51.15 & 14.76 & 36.30 & 48.60 & 3.26 & 50.63 & 21.63 & 3.05 & 6.22 \\[2pt]
  $\quad$ \makebox[0pt][l]{$\text{M}_{20}$}\phantom{Orientation xxxxxxxx xx} & 65.85 & 5.61 & 28.03 & 74.74 & 5.20 & 69.65 & 21.90 & 6.28 & 4.06 \\[2pt]
  
  $\quad$ \makebox[0pt][l]{Half-light radius}\phantom{Orientation xxxxxxxx xx} & 41.27 & 5.87 & 9.60 & 66.63 & 2.75 & 48.01 & 30.10 & 8.08 & 2.64 \\[2pt]
 
  $\quad$ \makebox[0pt][l]{S\'ersic index $n$}\phantom{Orientation xxxxxxxx xx} & 39.56 & 5.22 & 35.84 & 105.40 & 3.67 & 60.34 & 18.28 & 8.01 & 5.90 \\[2pt]
 
  $\quad$ \makebox[0pt][l]{Orientation}\phantom{Orientation xxxxxxxx xx} & 2.94 & 1.19 & 1.61 & 4.82 & 2.33 & 2.47 & 1.55 & 3.24 & 2.17 \\[2pt]
  $\quad$ \makebox[0pt][l]{Ellipticity}\phantom{Orientation xxxxxxxx xx} & 85.79 & 13.95 & 7.08 & 105.75 & 13.83 & 34.92 & 21.99 & 10.8 & 10.96 \\[2pt]
  $\quad$ \makebox[0pt][l]{Elongation}\phantom{Orientation xxxxxxxx xx} & 69.24 & 17.97 & 7.00 & 74.64 & 11.38 & 36.30 & 20.05 & 11.52 & 9.94 \\[2pt]
  \hline
  $\quad$ \makebox[0pt][l]{Mean}\phantom{Orientation xxxxxxxx xx} & 61.14 & 9.04 & 17.85 & 63.97 & 6.72 & 42.32 & 32.50 & 11.89 & 11.07 \\[2pt]
  \hline
  \hline
  Power spectra - wavelength ranges & &  &  &  &  &  &  &  & \\[2pt]
  $\quad$ 0: $\,\:$ \makebox[0pt][l]{Total magnitude}\phantom{$(0.000, 0.000]$} & 16.36 & 9.08 & 10.93 & 16.29 & 4.63 & 2.53 & 0.81 & 0.63 & 0.23 \\[2pt]
  $\quad$ 1:\phantom{0} \makebox[0pt][l]{$\geq 90.90$}\phantom{$(0.000, 0.000]$} $\lambda_\mathrm{min}$ & 16.04 & 6.88 & 9.46 & 15.36 & 4.48 & 1.87 & 3.75 & 1.35 & 1.16 \\[2pt]
  $\quad$ 2:\phantom{0} \makebox[0pt][l]{$[37.03, 90.90)$}\phantom{$(0.000, 0.000]$} $\lambda_\mathrm{min}$ & 14.38 & 4.20 & 9.95 & 14.26 & 4.26 & 2.38 & 8.24 & 2.13 & 2.51 \\[2pt]
  $\quad$ 3:\phantom{0} \makebox[0pt][l]{$[21.27, 37.03)$}\phantom{$(0.000, 0.000]$} $\lambda_\mathrm{min}$ & 13.77 & 4.26 & 12.56 & 14.55 & 4.42 & 4.83 & 9.90 & 2.29 & 2.87 \\[2pt]
  $\quad$ 4:\phantom{0} \makebox[0pt][l]{$[11.36, 21.27)$}\phantom{$(0.000, 0.000]$} $\lambda_\mathrm{min}$ & 13.40 & 3.66 & 13.20 & 15.55 & 4.75 & 6.71 & 12.03 & 2.46 & 3.61 \\[2pt]
  $\quad$ 5:\phantom{0} \makebox[0pt][l]{$[7.57, 11.36)$}\phantom{$(0.000, 0.000]$} $\lambda_\mathrm{min}$ & 14.52 & 3.07 & 13.19 & 16.51 & 5.13 & 8.23 & 13.30 & 2.51 & 3.85 \\[2pt]
  $\quad$ 6:\phantom{0} \makebox[0pt][l]{$[5.34, 7.57)$}\phantom{$(0.000, 0.000]$} $\lambda_\mathrm{min}$ & 17.73 & 3.05 & 13.74 & 17.81 & 5.25 & 9.44 & 14.33 & 2.64 & 4.24 \\[2pt]
  $\quad$ 7:\phantom{0} \makebox[0pt][l]{$[3.93, 5.34)$}\phantom{$(0.000, 0.000]$} $\lambda_\mathrm{min}$ & 22.63 & 3.58 & 14.56 & 20.03 & 5.42 & 9.92 & 15.92 & 2.65 & 4.49 \\[2pt]
  $\quad$ 8:\phantom{0} \makebox[0pt][l]{$[3.02, 3.93)$}\phantom{$(0.000, 0.000]$} $\lambda_\mathrm{min}$ & 27.20 & 3.83 & 14.90 & 23.87 & 5.44 & 10.38 & 16.68 & 2.77 & 4.69 \\[2pt]
  $\quad$ 9:\phantom{0} \makebox[0pt][l]{$[2.38, 3.02)$}\phantom{$(0.000, 0.000]$} $\lambda_\mathrm{min}$ & 32.04 & 3.95 & 15.16 & 26.85 & 5.62 & 10.81 & 17.08 & 2.82 & 4.94 \\[2pt]
  $\quad$ 10: \makebox[0pt][l]{$[1.94, 2.38)$}\phantom{$(0.000, 0.000]$} $\lambda_\mathrm{min}$ & 36.60 & 4.23 & 15.70 & 30.43 & 5.74 & 10.71 & 18.14 & 2.89 & 5.04 \\[2pt]
  $\quad$ 11: \makebox[0pt][l]{$[1.60, 1.94)$}\phantom{$(0.000, 0.000]$} $\lambda_\mathrm{min}$ & 40.84 & 4.59 & 16.06 & 33.53 & 5.84 & 10.80 & 18.57 & 2.96 & 5.20 \\[2pt]
  $\quad$ 12: \makebox[0pt][l]{$[1.35, 1.60)$}\phantom{$(0.000, 0.000]$} $\lambda_\mathrm{min}$ & 45.52 & 5.24 & 16.45 & 36.55 & 5.97 & 10.66 & 19.61 & 3.01 & 5.35 \\[2pt]
  $\quad$ 13: \makebox[0pt][l]{$[1.15, 1.35)$}\phantom{$(0.000, 0.000]$} $\lambda_\mathrm{min}$ & 50.19 & 6.04 & 16.79 & 39.91 & 5.85 & 10.69 & 20.04 & 3.13 & 5.48 \\[2pt]
  $\quad$ 14: \makebox[0pt][l]{$[1.00, 1.15)$}\phantom{$(0.000, 0.000]$} $\lambda_\mathrm{min}$ & 52.85 & 6.79 & 17.27 & 44.22 & 5.84 & 10.93 & 20.73 & 3.20 & 5.60 \\[2pt]
  \hline
  $\quad$ Mean & 27.61 & 4.83 & \num{14.00} & 24.38 & 5.24 & 8.06 & 13.94 & 2.50 & 3.95 \\[2pt]
  \hline
  \hline 
   Colours & & & & & & & & & \\[2pt]
  $\quad$ $(g-i)_\mathrm{SDSS}$ early-types & 12.60 & 1.15 & 3.16 & - & - & - & - & - & - \\[2pt]
  $\quad$ $(g-i)_\mathrm{SDSS}$ late-types & 12.23 & 1.39 & 2.15 & - & - & - & - & - & - \\[2pt]
  $\quad$ Bulge statistic $F(G,M_{20})$ & 16.89 & 1.63 & 11.89 & - & - & - & - & - & - \\[2pt]
  \hline
  $\quad$ Mean & 13.90 & 1.39 & 5.73 & - & - & - & - & - & - \\[2pt]
  \hline
  \hline
  Computer vision-based & &  &  &  &  &  &  &  & \\[2pt]
  $\quad$ FID & 11737 & 776 & 921 & 18425 & 145 & 1465 & 1088 & 55 & 161 \\[2pt]
  $\quad$ KID & 12.78 & 0.60 & 0.52 & 22.28 & 0.08 & 1.19 & 0.81 & 0.04 & 0.08 \\[2pt]
  \hline
 \end{tabular*}
 \caption{Generative model evaluation. The first group shows the $\mathcal{W}_1$-Wasserstein distance between the 1D distributions of the normalized optical morphological measurements of the generated data and the source data set. The power spectra metrics show the $\mathcal{W}_1$-Wasserstein distance of the radially averaged shifted 2D power spectra between the generated data and source data set for different physical scales. The minimum wavelength $\lambda_\mathrm{min}$ is $0.55\,$kpc or $0.056\,$arcsec for SKIRT, $0.042\,$arcsec for COSMOS and $0.056\,$arcsec for S\'ersic profiles. 
In the colours group, we display the $\mathcal{W}_1$-Wasserstein distance for the $(g-i)_\mathrm{SDSS}$ and bulge statistic, see Fig. \ref{fig:colors_plot} for reference.
 The computer vision-based metrics FID and KID are based on feature similarity from activations of the InceptionV3 network and are typically correlated with a perceptual similarity. }
 \label{tab:quantitative_metrics}
\end{table*}

\section{Results} 
\label{sec:evaluation}

Here, we quantify the relative performance of the three generative models presented in Section \ref{sec:gen_modelling} when applied to the three data sets described in Section \ref{sec:datasets}, in terms of the physically-motivated and computer-vision quantities discussed in the previous section. To this end, we set the size of the latent space $Z$ of the generative models to $512$ for the COSMOS and SKIRT data sets, and $32$ for the  S\'ersic one. For each combination of generative model and training data set, we create a generated data set consisting of $50,000$ samples. For details on the training hyper-parameters of the individual models, we refer the reader to Appendix \ref{sec:detailed_model}.

\subsection{Morphological properties} 
\label{subseq:eval_morph}

Table \ref{tab:quantitative_metrics} lists the $\mathcal{W}_1$-Wasserstein distance for each of the morphological properties considered. For the reader less familiar with this distance, in Table \ref{tab:quantitative_metrics_mean_std} in the appendix, we also report the mean and standard deviation for the distribution of each quantity. In general, we find that for the SKIRT and COSMOS data sets StyleGAN either outperforms both VAE and ALAE for most morphological properties or is close to the best model.
In particular, for the SKIRT data set, StyleGAN achieves the lowest average $\mathcal{W}_1$-Wasserstein distance of $9.04$, which is $85$ per cent lower than VAE $(61.14)$ and $49$ per cent lower than ALAE ($17.85$). Similarly, StyleGAN also has the lowest average $\mathcal{W}_1$-Wasserstein distance for the COSMOS data set with $6.72$, which is $89$ per cent lower than VAE $(63.72)$ and $84$ per cent lower than ALAE ($42.32$). 

For the sample of S\'ersic profiles, different morphological properties are best reproduced by different generative models. In particular, ALAE obtains an average score of $11.07$, which is $7$ per cent lower than StyleGAN and $65$ per cent lower than VAE. 

Interestingly, VAE, which is one of the most commonly used methods in astronomy, performs significantly worse than StyleGAN and ALAE for most morphological parameters and data sets. 
Overall, StyleGAN scores best in 20 out of 30 cases, and ALAE in 9 out of 30 cases. This asssessment, based on the $\mathcal{W}_1$-Wasserstein distance, is consistent with the results based on the mean and standard deviations of the distributions (see Table \ref{tab:quantitative_metrics_mean_std} in the appendix).

In Fig. \ref{fig:morph_SKIRT}, we show histograms of the morphological measurements for the generative models and the SKIRT data set. For example, the second bottom panel from the left shows the smoothness $S$. Generated data from VAE has a lower smoothness value $S$ than the other three data sets, i.e. it is generally less clumpy and smoother. This can also be observed from the SKIRT data set visualizations in Fig. \ref{fig:comparison_overview_tng_stylegan2} and Fig. \ref{fig:comparison_overview_tng_alae_vae} in the appendix. How well the distributions in Fig. \ref{fig:morph_SKIRT} overlap with the SKIRT distribution for the morphological measurements strongly correlates with the $\mathcal{W}_1$-Wasserstein distance score they obtain. In Fig. \ref{fig:morphological_properties_COSMOS} and Fig. \ref{fig:morphological_properties_Sersic} in the appendix we show histograms for the COSMOS and S\'ersic profiles data sets, respectively. Note that since we normalize the morphological measurements for each data set for better comparability of the individual morphological properties, it can be misleading to compare the $\mathcal{W}_1$-Wasserstein distance across different data sets.

\begin{figure*}
	\centering
	\includegraphics[clip=True, trim=5 5 55 5, width=\textwidth]{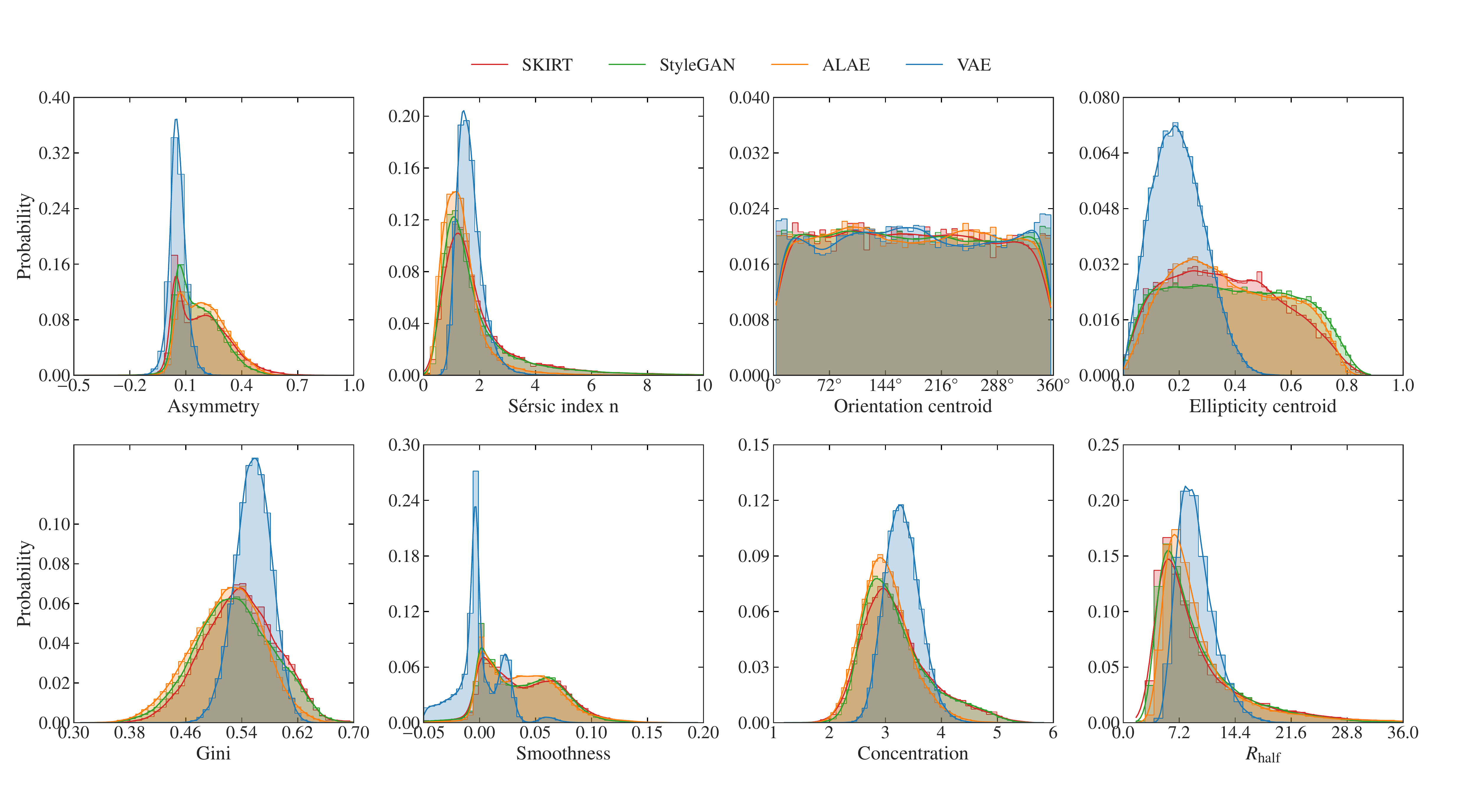}
    \caption{SKIRT data set: Histograms showing selected optical morphological measurement for the SKIRT data set and the generated data sets (StyleGAN, ALAE and VAE). Morphological properties learned very well by the StyleGAN and ALAE models are the orientation, the half-light radius $R_\mathrm{half}$, the smoothness $S$ and the concentration $C$. Slightly harder to learn are the S\'ersic index $n$ and the Gini-coefficient $G$. The models have the most difficulty reproducing the asymmetry $A$ and the ellipticity.}
    \label{fig:morph_SKIRT}
\end{figure*}

\begin{figure*}
	\centering
    \subfloat[VAE]{\includegraphics[clip=True, trim=35 35 70 70, width=.23\textwidth]{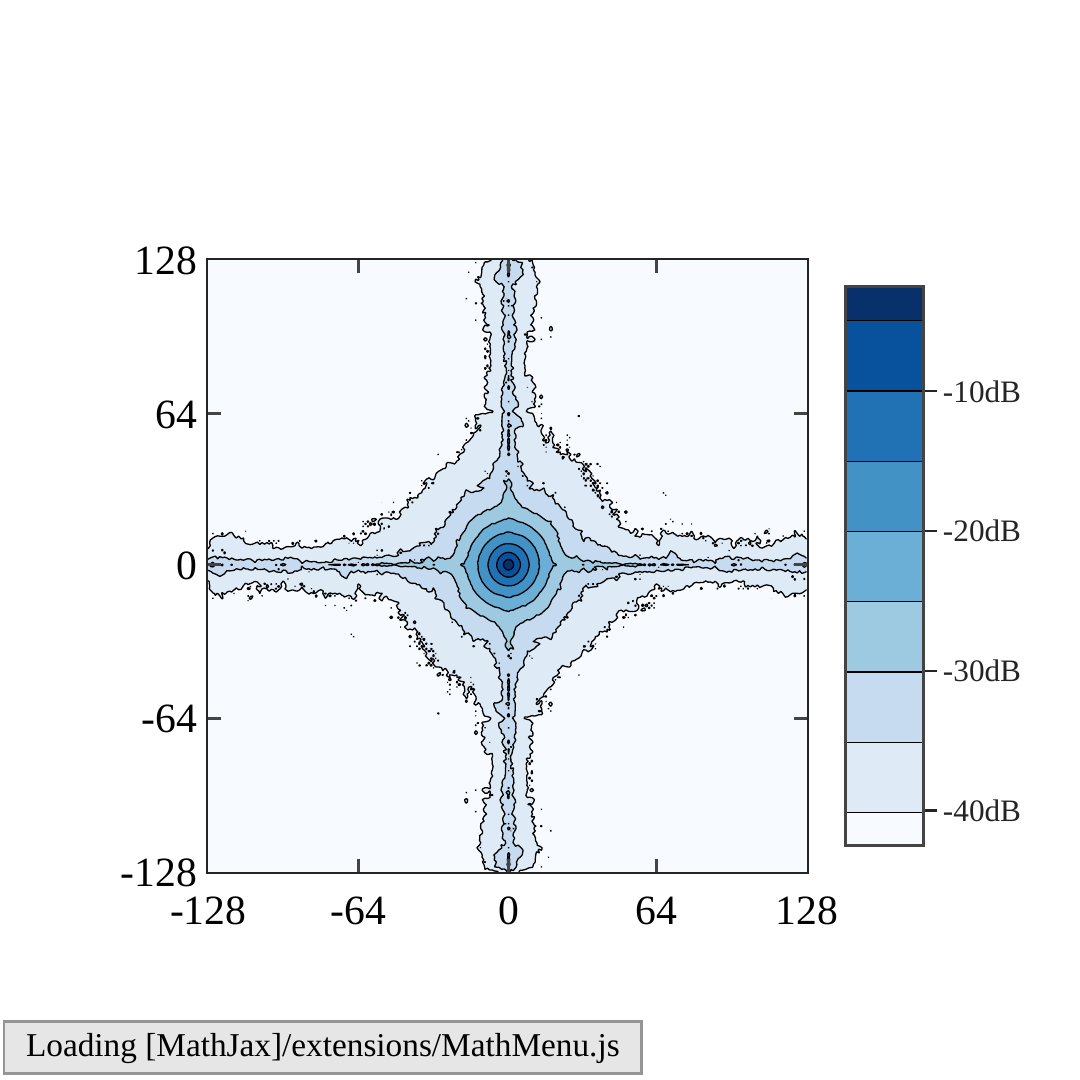}} 
    \subfloat[ALAE]{\includegraphics[clip=True, trim=35 35 70 70, width=.23\textwidth]{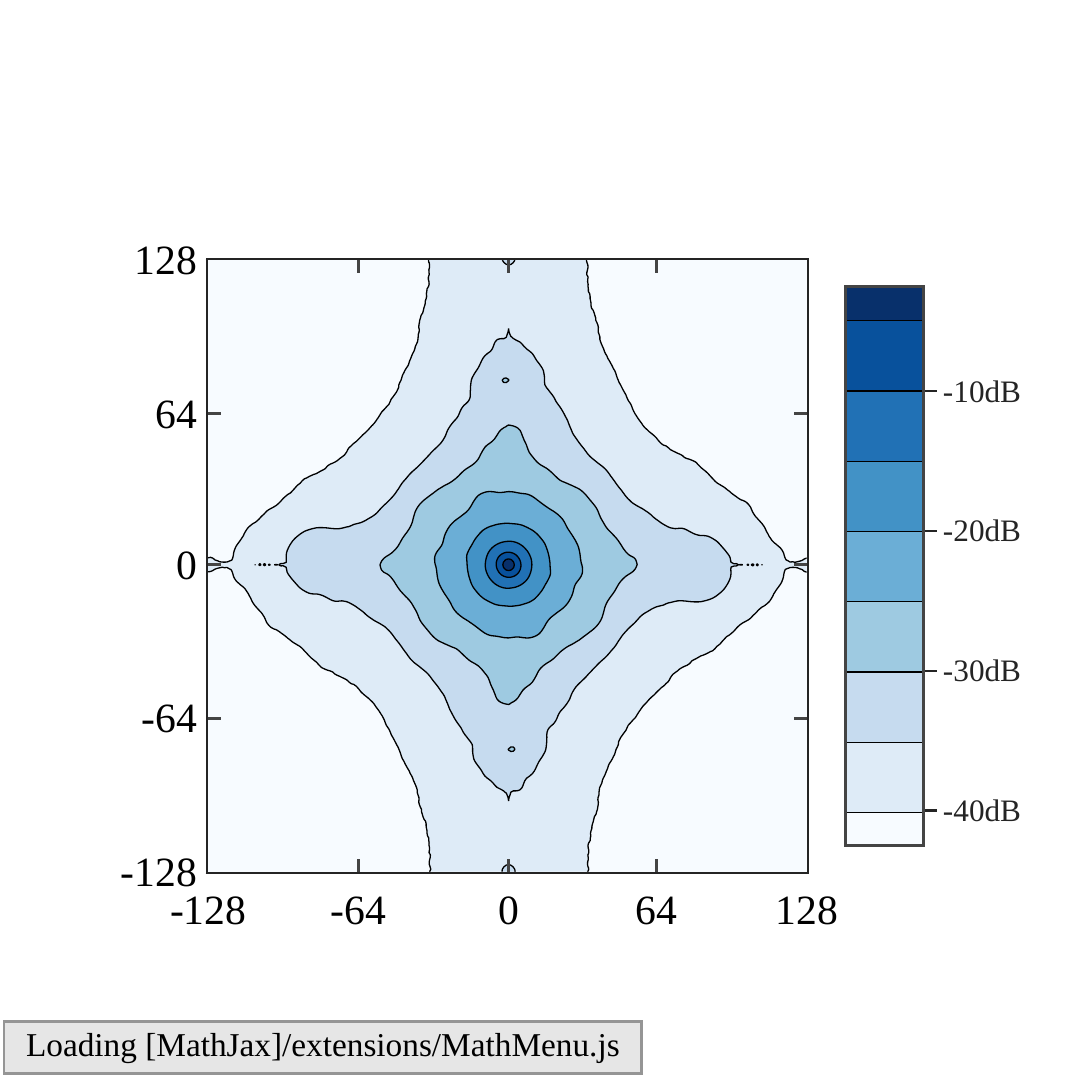}}  
    \subfloat[StyleGAN]{\includegraphics[clip=True, trim=35 35 70 70, width=.23\textwidth]{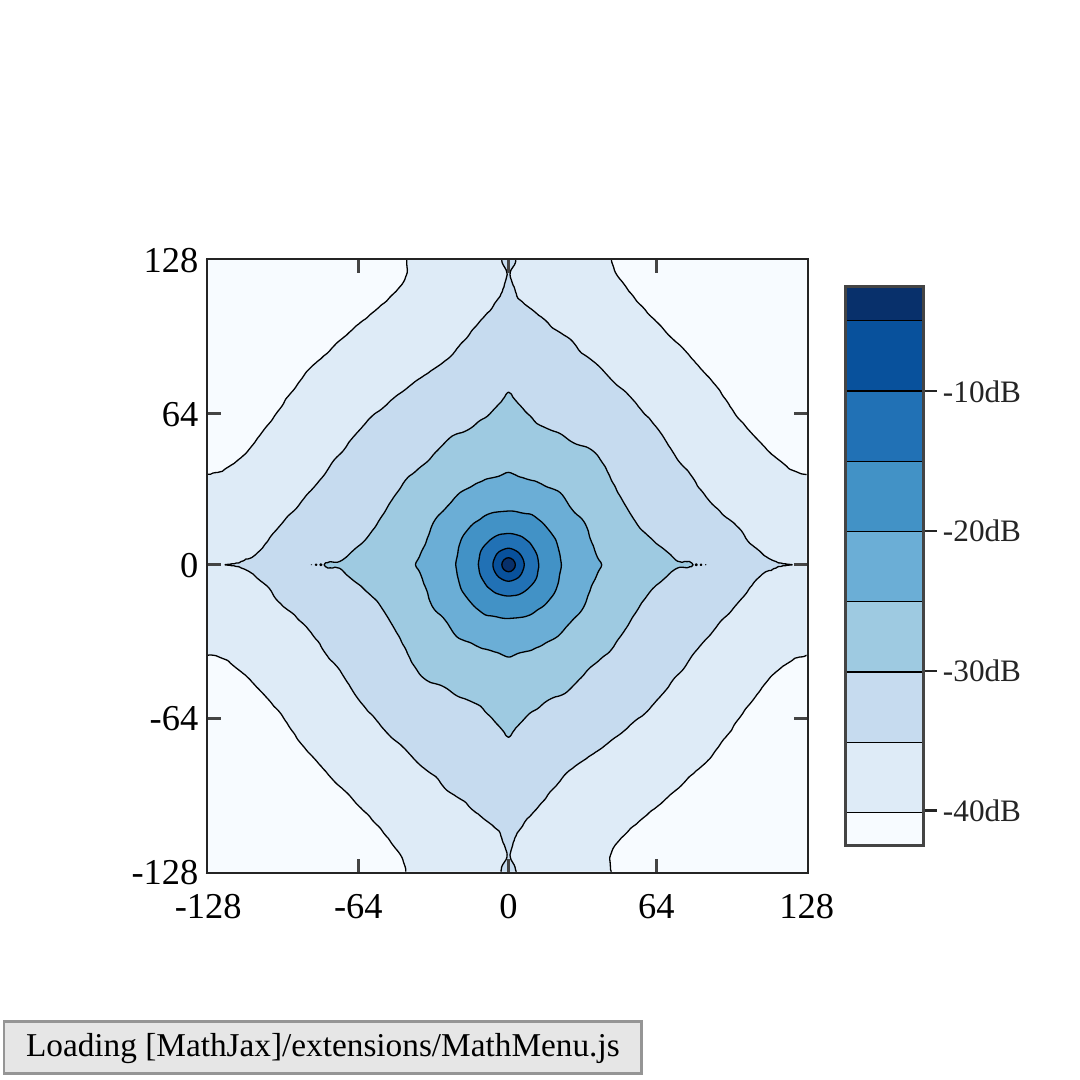}}  
    \subfloat[SKIRT]{\includegraphics[clip=True, trim=35 35 0 70, width=.31\textwidth]{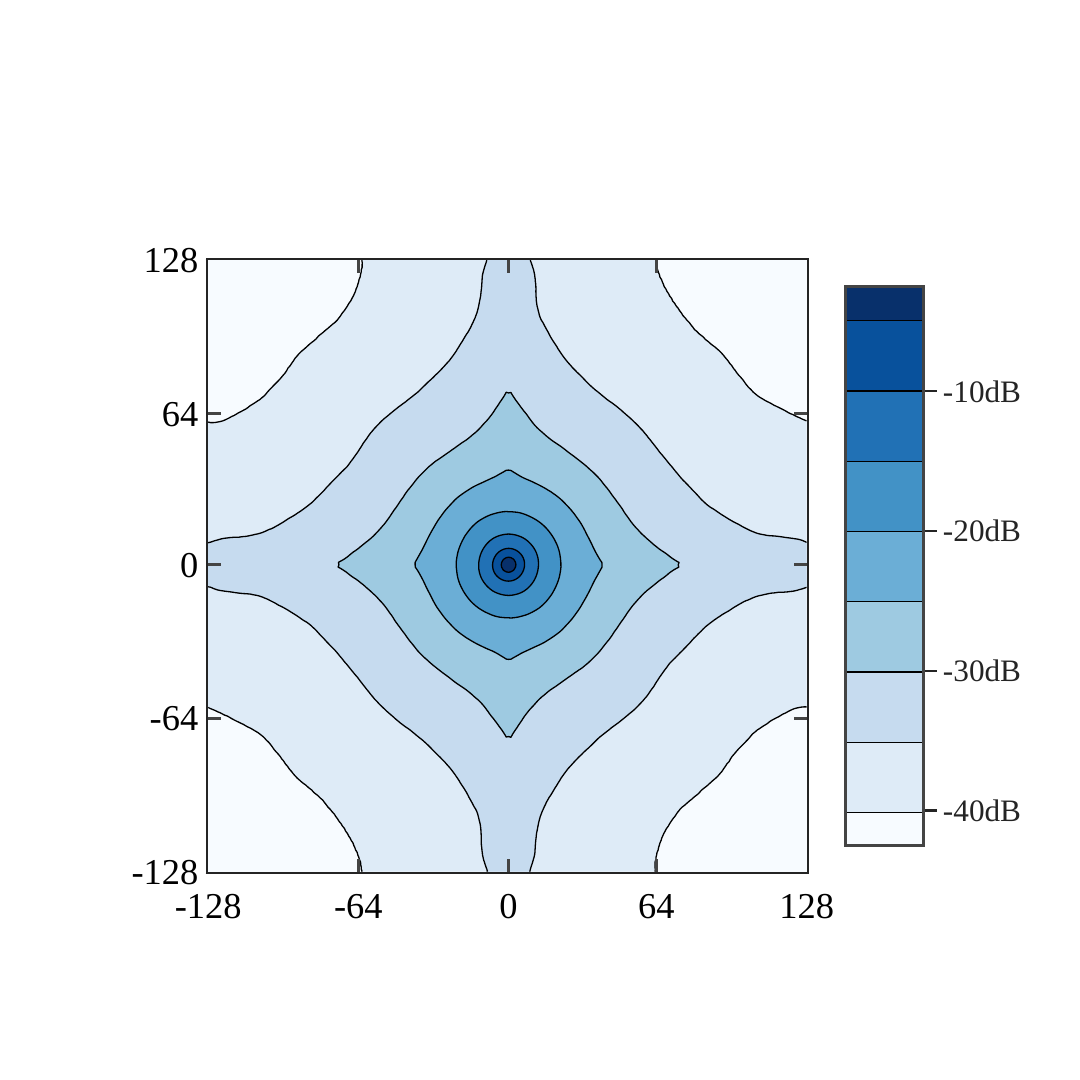}} 
    \caption{Contour plots of the average shifted 2D power spectrum of the $r$-band of the raw network outputs (VAE (a), ALAE (b) and StyleGAN(c)) and of the resized $256 \times 256$ images of the SKIRT data set (d). The shifted 2D power spectrum of individual galaxies is calculated as described in Fig. \ref{fig:powerspectra_example}. Units on the axes are in pixels. We show a contour plot in dB, i.e. values are in $\log_{10}$ scale. While the StyleGAN generated data (c) is visually close to the original SKIRT data set, there are noticeable differences, which affect the higher frequencies in particular. The ALAE generated data (b) deviates even more from the SKIRT data set. Finally, the VAE generated data (a) looks very differently compared to the SKIRT contour plot, since the VAE has an inherent difficulty to generate non-smooth data. The average $\mathcal{W}_1$-Wasserstein distances for VAE, ALAE and StyleGAN are $27.61$, $14.00$ and $4.83$, showing that the $\mathcal{W}_1$-Wasserstein distance captures the different qualities of the contour plots very well.}
    \label{fig:comparison_zoom_avg_powerspectra}
\end{figure*}

\begin{figure*}
	\centering
	\includegraphics[clip=True, trim=0 0 5 5, width=\textwidth]{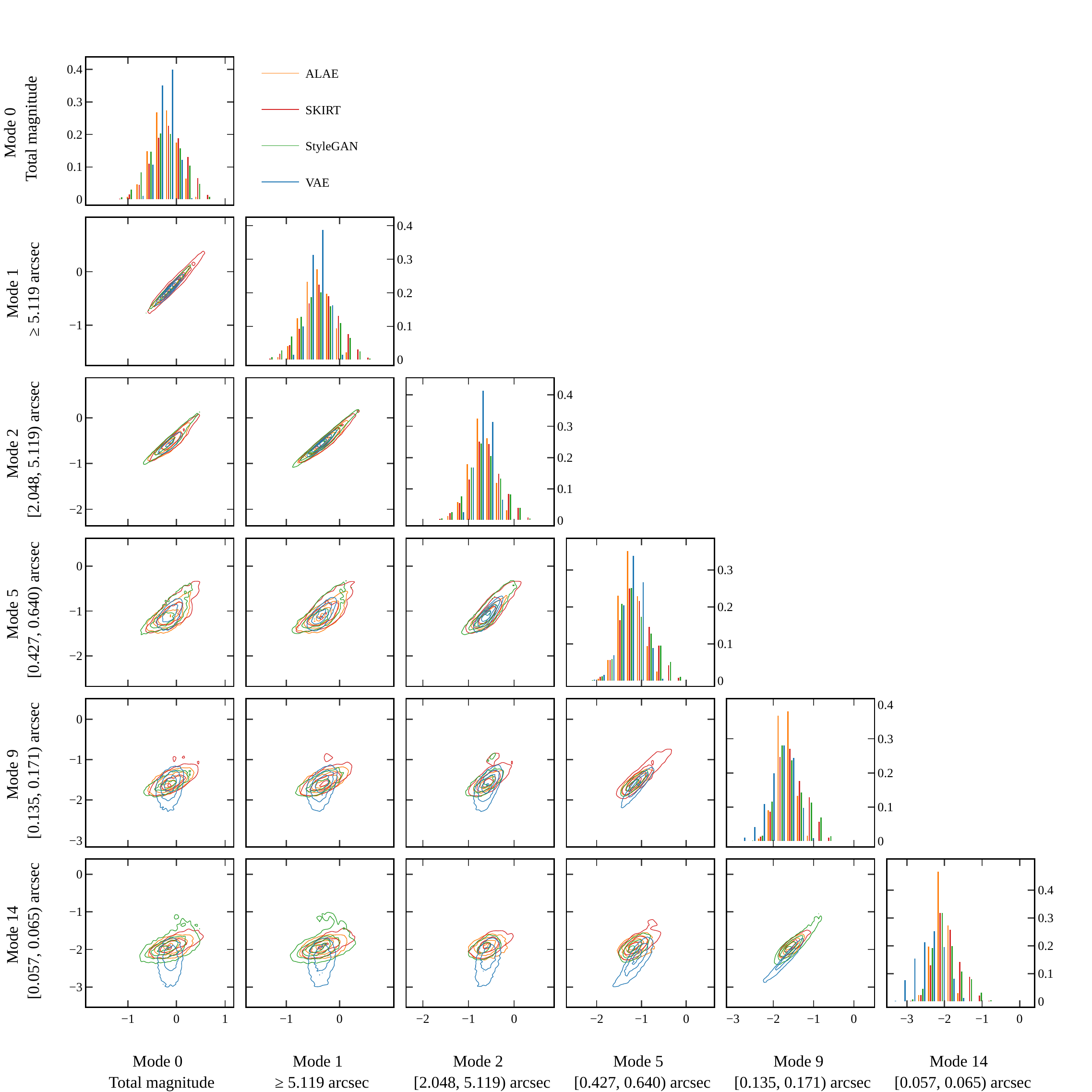}
    \caption{Corner plot for different physical scales of the shifted 2D power spectra of the SKIRT data set. Axes are in $\log_{10}$-scale. The modes correspond to a partition of the wavelength range, see Table \ref{tab:quantitative_metrics}.}
    \label{fig:power_spectra_SKIRT}
\end{figure*}

\begin{figure*}
	\centering
	\includegraphics[clip=True, trim=5 5 5 5, width=\textwidth]{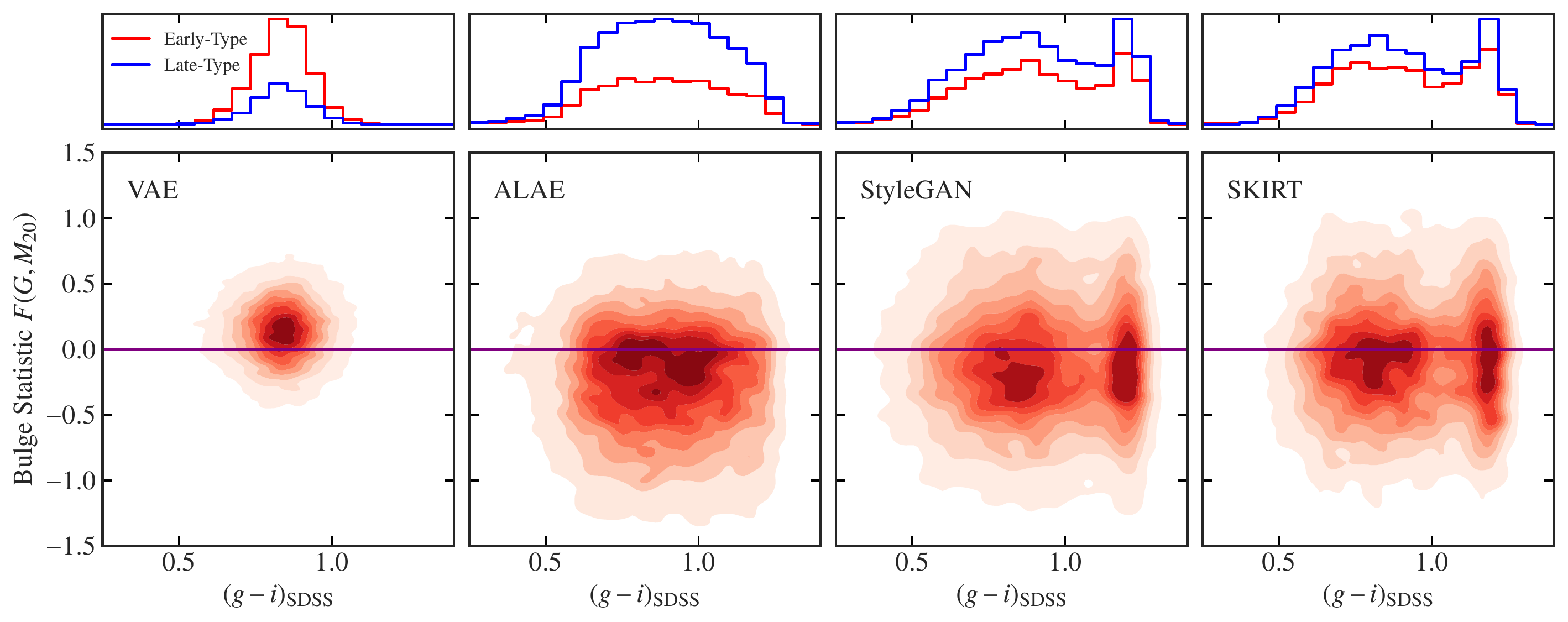}
    \caption{Distribution of the $F(G, M_{20})$ bulge statistic and the $(g-i)_\text{SDSS}$ colour index for the three generated data sets (VAE, ALAE and StyleGAN) and the SKIRT one. See \citetalias[][Fig. 9]{rodriguez2019optical} for reference, but note that we have removed several galaxies as described in Section \ref{subsec:skirt}. The bottom panels show contour plots of the galaxy distribution, while the histograms in the top panels show the marginal distribution of late- and early-type galaxies. A galaxy is classified as early-type if $F(G, M_{20}) \geq 0$ and late-type otherwise.}
    \label{fig:colors_plot}
\end{figure*}

\subsection{Power-spectrum} 
\label{subseq:eval_power_spectra}

In Table \ref{tab:quantitative_metrics} we show the $\mathcal{W}_1$-Wasserstein distance for the distribution of magnitudes of the power spectra for different physical scales between the source and the generated data sets. The different scales are grouped into $15$ intervals, which we call modes.  

For the SKIRT galaxy images, StyleGAN consistently outperforms the other models and attains a mean $\mathcal{W}_1$-Wasserstein distance of $4.83$, which is $82$ per cent lower than VAE ($27.61$) and $65$ per cent lower than ALAE ($14.00$). To analyze our results, we find that it is best to partition the wavelength ranges into three groups. The first group comprises the largest wavelengths, i.e. the total magnitude and wavelengths above $5\,$arcsec (mode 0). In this group, the generative models reproduce the SKIRT data slightly worse than in the second group. The second group corresponds to physical scales at which each model can reproduce the SKIRT data set best and which attain the lowest $\mathcal{W}_1$-Wasserstein distance. For example, for StyleGAN this is the case for scales between 
$0.30$ and $0.66\,$arcsec (mode 4 and 5). Then, for smaller scales in the third group, we notice that it gradually becomes harder for all generative models to reproduce the power spectrum of the SKIRT data set, as reflected by an increasing $\mathcal{W}_1$-Wasserstein distance. We plot and discuss the average power spectrum in Fig. \ref{fig:comparison_zoom_avg_powerspectra}. While we still see imperfections regarding the average power spectrum of the raw network outputs, we believe that these issues will be mitigated with continuing progress in generative modelling research, for example, with approaches that explicitly consider Fourier features \citep[][]{karras2021alias, gal2021swagan}.

In Fig. \ref{fig:power_spectra_SKIRT}, we show contour plots for some selected scale ranges. Our quantitative analysis is confirmed by comparing how well the the generated data sets for each model match the SKIRT data visually. For example, the VAE data set shows significantly less power on smaller physical scales (between mode $9$ and mode $14$) than the SKIRT data set, implying less clumpy and smoother galaxies. On the other side, StyleGAN matches the SKIRT data set best on those scales. Indeed, as seen in Fig. \ref{fig:comparison_overview_tng_stylegan2} and Fig. \ref{fig:comparison_overview_tng_alae_vae} in the appendix, the StyleGAN generated data is in general much sharper, while the VAE produces overly-smooth data lacking any morphological details.  

For the COSMOS observations, StyleGAN attains the lowest average $\mathcal{W}_1$-Wasserstein distance ($5.24$), which is $34$ per cent lower than ALAE ($8.06$) and $78$ per cent lower than VAE ($24.38$). For larger physical scales (total magnitude, mode 1 and mode 2), ALAE outperforms StyleGAN, but it is worse for all smaller physical scales. For all models, we observe that the $\mathcal{W}_1$-Wasserstein distance gradually increases for smaller physical scales.
For the COSMOS data set, we already see a convincing agreement of the average  power spectra between the generated and original data, compare Fig. \ref{fig:comparison_zoom_avg_powerspectra_COSMOS} and Fig. \ref{fig:power_spectra_COSMOS} for contour plots of selected individual scale ranges in the appendix.

Finally, for the S\'ersic profiles, StyleGAN achieves an average $\mathcal{W}_1$-Wasserstein distance of $2.50$, which is $36$ per cent lower than ALAE ($3.95$) and $82$ per cent lower than VAE ($13.94$).
Here, both adversarial-based models can reproduce the power spectrum very well across all scales, but StyleGAN outperforms ALAE for all scales below $5.09\,$arcsec (between mode 2 and 14). Similar to the COSMOS data, the generated data convincingly reproduces the average power spectrum, see Fig. \ref{fig:comparison_zoom_avg_powerspectra_Sersic} and Fig. \ref{fig:power_spectra_Sersic} in the appendix.

Overall, StyleGAN was the best model in $40$ out of $45$ cases and ALAE in $5$.

\subsection{Colour and bulge statistic} 
\label{subseq:eval_colour}

Fig. \ref{fig:colors_plot} shows the Gini-M$_{\rm 20}$ bulge statistic $F(G,M_{\rm 20})$ vs $(g-i)_\mathrm{SDSS}$ colour of the generated data and the SKIRT galaxies. 
The Gini-M$_{\rm 20}$ bulge statistic, $F(G, M_{\rm 20})$, is a linear combination of the Gini-coefficient and the M$_{\rm 20}$ statistic that correlates with optical bulge strength \citep[][]{snyder2015tng, rodriguez2019optical}. We can see from this plot that the VAE model cannot reproduce either quantity, as it mostly produces galaxies with a limited range of colour and $F(G,M_{\rm 20})$ values. 

The ALAE model is better at reproducing the range of colour values. However, it fails to reproduce the bimodality of the colour distribution seen in the SKIRT data and it underestimates the fraction of red (early-type) galaxies. Compared to VAE, it allows for a larger range of values for $F(G,M_{\rm 20})$, but produces results which are biased towards less bulge-dominated galaxies.

The StyleGAN generated data is the one that most closely reproduces the colour distribution and the bulge statistic of the SKIRT galaxies, including the colour bimodality. This result is very encouraging, because it demonstrates that StyleGAN is capable of reproducing statistical properties in the source data set which it was not directly trained to do. 

These qualitative results are corroborated by the more quantitative $\mathcal{W}_1$-Wasserstein distance as can be seen in Table \ref{tab:quantitative_metrics}. Indeed, StyleGAN is reproducing the bulge statistic $F(G,M_{20})$ much better than the other two methods ($1.63$ vs $11.89$ and $16.89$ for ALAE and VAE).
We find similar results for the $(g-i)_\mathrm{SDSS}$ colour distribution of the early-types ($1.15$ vs $3.16$ and $12.60$) and the late-types ($1.39$ vs $2.15$ and $12.23$).

\subsection{Computer vision metrics: FID and KID} 
\label{subseq:eval_fid_kid}

We now turn our attention to the computer vision metrics FID and KID.
Values for these metrics for all considered data sets are listed in Table \ref{tab:quantitative_metrics}. 
We find that the VAE model performs the worst in terms of both FID and KID for all data sets. Conversely, StyleGAN produces the lowest FID for all sample of galaxies. For example, for the SKIRT data set, StyleGAN has a FID which is $93$ per cent lower than that of the VAE and $15$ per cent lower than that of the ALAE. However, for the same data set, the ALAE achieves the best KID which is $13$ per cent lower than StyleGAN and $96$ per cent lower than VAE. 

Interestingly, the low values of FID  achieved here by the StyleGAN and ALAE algorithms for galaxy images are comparable to those reached by recent state-of-the-art networks (e.g. StyleGAN) for natural images \citep[e.g. the FFHQ human faces data set][]{karras2020analyzing}. This result is encouraging as the feature-extraction networks used to obtain the FIDs were trained on natural images and not on physical data. As discussed in Section \ref{sec:metrics}, the FID correlates with a perceptual similarity of two data sets based on basic geometrical shapes and their composition, which are most commonly found in natural images and extracted in the form of activations in the deeper layers of the InceptionV3 network \citep[][]{zhang2018unreasonable}.

\section{Improving robustness using generated data} 
\label{sec:generalization}

In the previous section we demonstrated that generative models, and in particular StyleGAN, can generate galaxy images with physical properties matching those of the input data sets. In this section, we investigate the ability of generated data to complement the training data in machine learning applications. Based on the previous results, we use StyleGAN to generate a data set consisting of $50,000$ galaxies to mix with the existing SKIRT samples. We choose SKIRT, rather than COSMOS, as it is both harder to learn (compare FID $7.76$ vs $1.45$ for StyleGAN) and it has no noise or PSF. This makes it possible to consider a widespread machine learning problem, image denoising, which has applications in astronomy.
From the generated data and SKIRT images we create mock observations by adding a PSF and observational noise. We then train a CNN to denoise the mock observations. Successfully solving this task requires the denoising model to both learn the properties of the PSF and noise as well as the underlying galaxy data distribution. In all cases, the size of the training data set is fixed, although the proportion of generated and original data changes. This ensures that any improvement in robustness we observe is due to the presence of the generated data, and not simply an increase in the number of gradient updates of the model.

\begin{figure*}
	\centering
	\includegraphics[clip=True, trim=5 5 0 5, width=.7\textwidth]{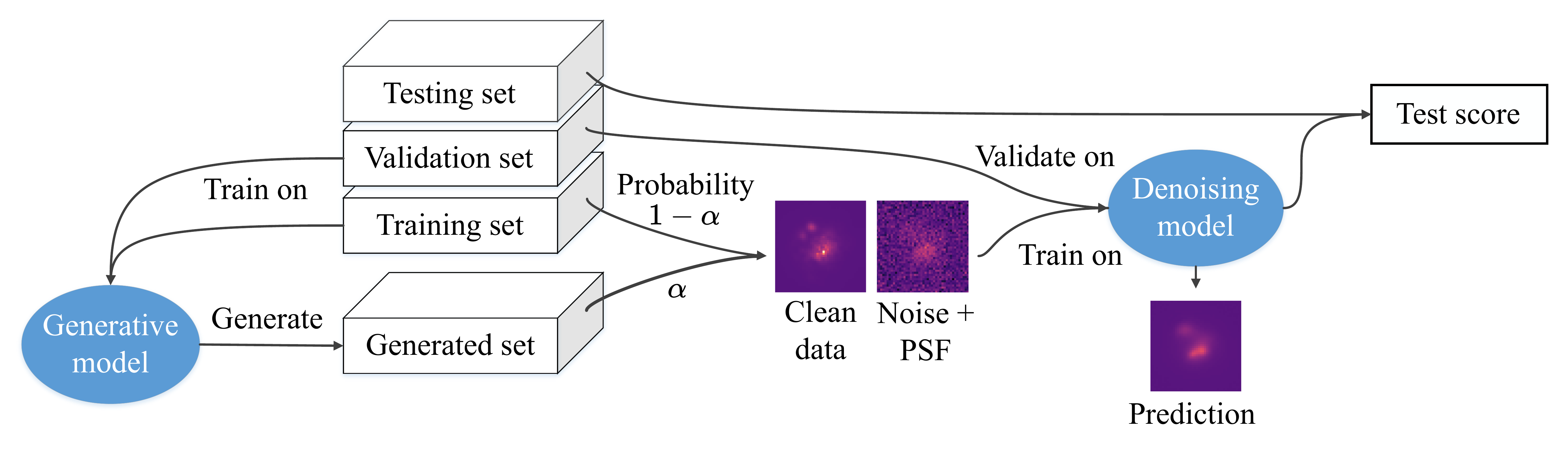}
    \caption{Experiment setup for evaluating the influence of the mixing of source and generated data on the denoising model. The real data is split into a testing set, validation set and generated set. The generative models are trained on the training set and validation set. When training the denoising models, we randomly draw a sample from the training set with probability $1-\alpha$ and from the generated set with probability $\alpha$. The best denoising model is picked based on its performance on the validation set. We evaluate the best denoising model on the testing set along with several augmentations to simulate a domain shift in the data.}
    \label{fig:denoising_setup}
\end{figure*}

\subsection{Methodology} \label{sec:methodology}
 The input to the CNN is an observation with dimensions $N \times M \times 4$, where $N$ and $M$ denote the number of pixels in each direction and $4$ is the number of filters. Since the CNN is fully convolutional, this means that $N$ and $M$ do not need to be fixed to a constant value during training and evaluation, and images do not have to be resized. For the mock observations based on the SKIRT data, we always have $N=M$, i.e. the images are square. It is essential that the physical pixel size remains the same. The objective function that the CNN minimizes is the mean squared error loss between the data without noise $y$ and the network prediction $\hat{y}$, i.e. \begin{equation}
    \text{MSE}(y, \hat{y}) := \frac{1}{n} \sum_{i=1}^n ||y-\hat{y}||_F^2 = \frac{1}{n} \sum_{i=1}^n \sum_{j,k=1}^{N,M} (y_{j,k}-\hat{y}_{j,k})^2,
\end{equation}
where $n$ is the number of samples and $||\cdot||_F$ is the Frobenius norm. 
We split the SKIRT data set into a learning set and a testing set and train the StyleGAN model on the learning set (as already described in Section \ref{subsec:skirt}). We use the term learning set here to refer to both the validation set and training set. We apply the same PSF and noise to the generated data. To test if our network benefits from combining the original and generated data set, we consider the updated loss
\begin{equation}
    \mathcal{L}(\alpha) := (1-\alpha) \mathbb{E}_{y \sim D_\text{training}}[\text{MSE}(y, \hat{y})] + \alpha \mathbb{E}_{y \sim D_\text{generated}}[\text{MSE}(y, \hat{y})],
\end{equation}
where $\alpha$ is a mixing factor between $0$ and $1$ \citep[][]{gowal2021improving}. For each step during training, we draw the next sample from the training data set $D_\mathrm{training}$ with probability $1-\alpha$ or draw it from the generated images $D_\mathrm{generated}$. We choose the PSF and observational noise in such a way that we obtain a challenging denoising problem, without necessarily matching a specific instrument.
The PSF is modelled as a Gaussian distribution with $\sigma_\mathrm{PSF}=2.0$ in pixels ($0.08$ arcsec) that is convolved with the input image. We draw the noise from a Gaussian distribution with $\sigma_\mathrm{noise} = 4.0$ [$e^-s^{-1}\mathrm{pixel}^{-1}$] and add it to each pixel independently. 

\subsection{CNN architecture and training details}

The denoising network consists of an encoder and decoder with ResNet \citep[][]{he2016deep} architecture. In the encoder, the residual blocks consist of a LeakyReLU activation and a Conv2D layer with kernel size $5$, $32$ filters with `same' padding followed by another LeakyReLU activation and Conv2D layer with the same configuration. Residual blocks are chained by skip connections. This gives the following sequence of layers for the encoder: Conv2D (32 filters), $5 \times $  ResidualBlock, LeakyReLU, Conv2D (2 filters). The residual blocks of the decoder are analogous but have Conv2DTranspose layers instead of Conv2D. For the decoder: Conv2DTranspose (32 filters), $5 \times $  ResidualBlock, LeakyReLU, Conv2DTranspose(4 filters). 

The networks are trained using Adam optimizer \citep{kingma2014adam} with an initial learning rate of \num{e-4}, which is decreased by the factor $0.8$ every $50$ epochs. The networks are trained for $1000$ epochs. The batch size is $1$, since different inputs do not need the same number of pixels for the height and width. 
During training, we use several data augmentations; we use flips in $x$- and $y$-direction as well as random rotations by $0, 90, 180$ and $270$ degrees. A new random noise is sampled and added to the network input before each prediction.  

\begin{figure*}
	\centering
	\includegraphics[width=.8\textwidth]{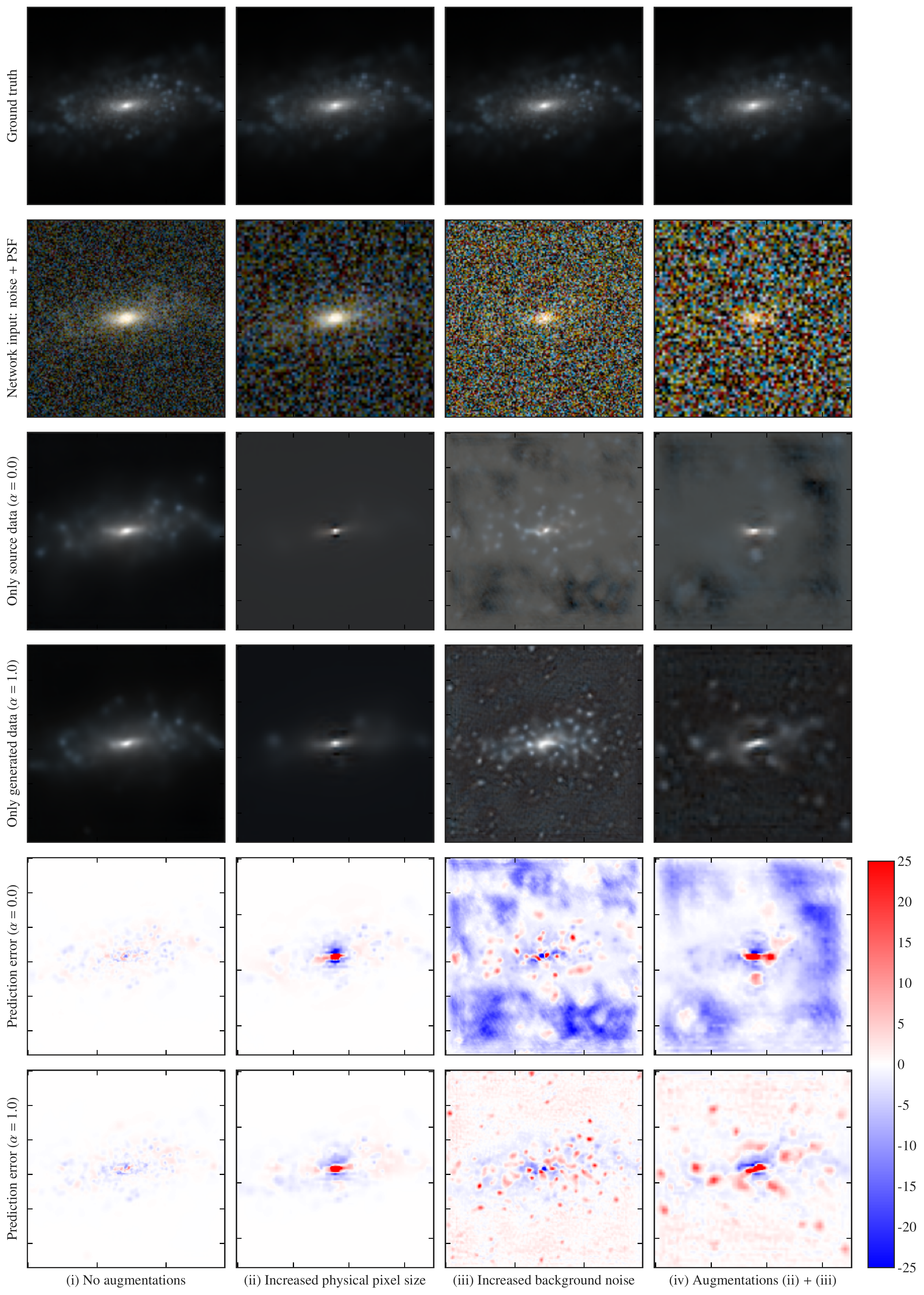}
    \caption{Image denoising of a galaxy from the test data set. The columns show the data with different domain shifts. The first row shows the ground truth data without PSF and noise. The second row shows the data after artificially adding the noise and PSF. This data also serves as the network input. The third row shows the prediction of the denoised galaxy from the model trained on only the source data set without any generated data $(\alpha=0)$, whereas the fourth row shows the prediction by the model trained on only generated data $(\alpha=1)$. The fifth and sixth rows show the prediction error, i.e. ground truth minus prediction, for $\alpha=0$ and $\alpha=1$, respectively. As can be seen both models perform very well for (i). For augmentation (ii), both models show strange artifacts around the peak brightness. We see this behaviour very often for this particular data augmentation and this has a significant impact on the testing loss, predominantly in already very bright galaxies. While the $\alpha=1$ model still manages to predict reasonable outcomes for augmentations (iii) and (iv), given the very substantial domain shift, the $\alpha=0$ model produces large negative pixel outputs, showing that the model cannot deal with this situation very well.}
    \label{fig:denoising_examples}
\end{figure*}

\subsection{Model selection and augmentations} 

To avoid overfitting, we reset the model weights to the point during training at which the model attained the lowest loss on the validation set. Moreover, we apply three different augmentations to the test data set to measure the robustness of the model:
\begin{enumerate}
    \item no domain shift: we evaluate the model on the test data set.
    \item increased physical pixel size: increase the physical pixel size by a factor of $2$ by downsampling the images to half the resolution using linear interpolation, i.e. the physical pixel size of the galaxy images changes from $0.363 \text{kpc}$ to $0.726\text{kpc}$ and the half-light radius of each galaxy decreases by a factor of 2. 
    \item increased background noise: we increase the standard deviation of the Gaussian noise from $\sigma_\text{noise} = 4.0$ to $\sigma_\text{noise} = 16.0$.
    \item increased background noise and physical pixel size: we combine augmentations (i) and (ii).
\end{enumerate}
The augmentations (ii), (ii) and (iv) represent a substantial domain shift making it very hard for the model to denoise faint galaxies. Because of that, they simulate a generic stress test of the model, but are also domain shifts, which are common in astronomy at the same time.

\begin{figure*}
	\centering
	\subfloat[Testing set]{\includegraphics[clip=True, trim=0 0 0 0, width=.249\textwidth]{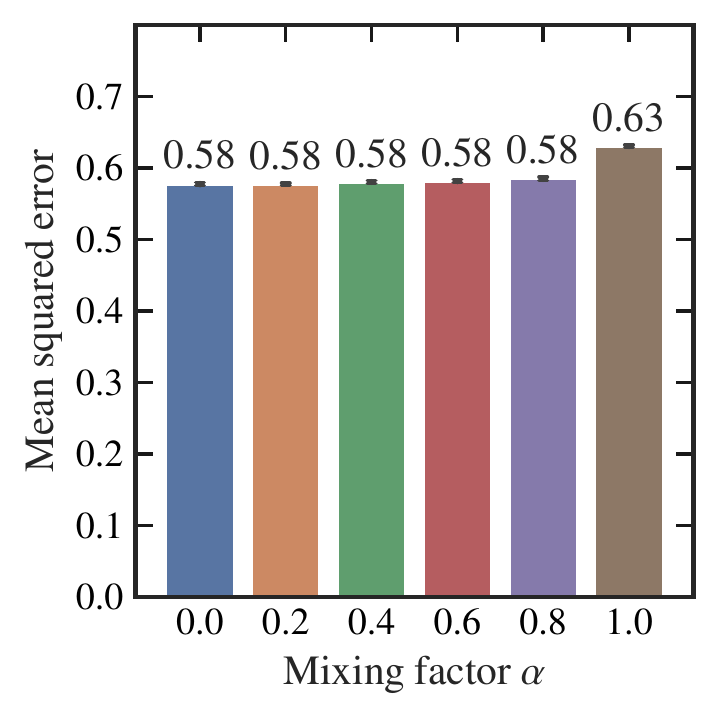}\label{fig:mixing_factor_a}}  \hfill
	\subfloat[Increased physical pixel size]{\includegraphics[clip=True, trim=0 0 0 0, width=.249\textwidth]{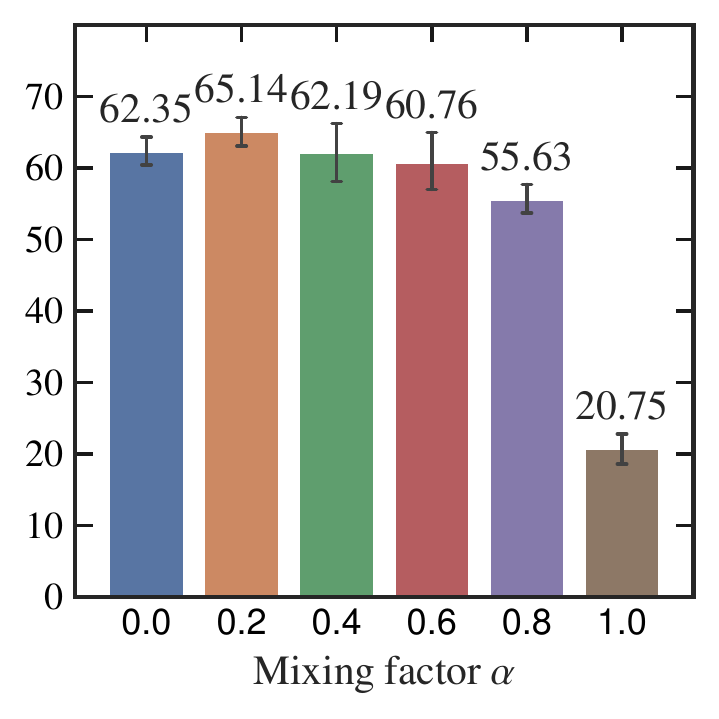}\label{fig:mixing_factor_b}}  \hfill 
	\subfloat[Increased background noise]{\includegraphics[clip=True, trim=0 0 0 0, width=.249\textwidth]{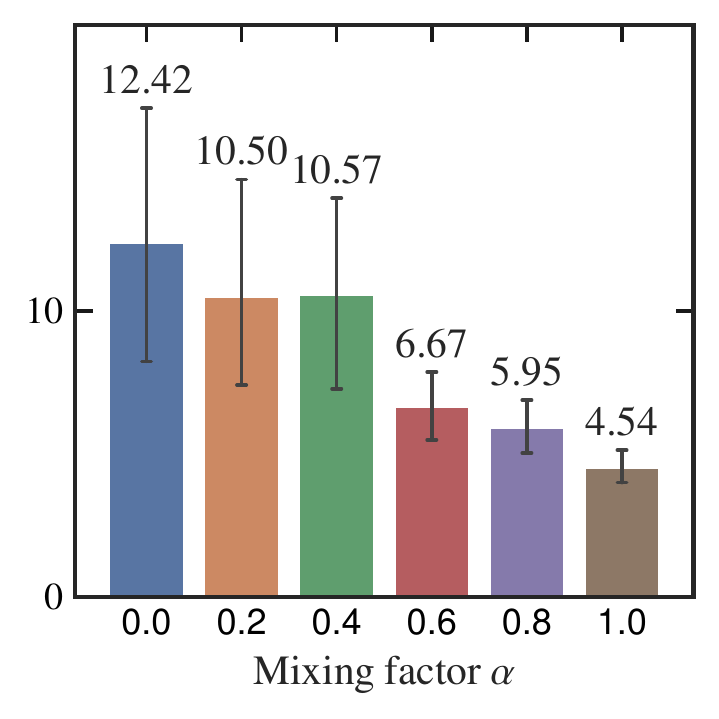}\label{fig:mixing_factor_c} } \hfill 
	\subfloat[Augmentations (ii) and (iii) combined]{\includegraphics[clip=True, trim=0 0 0 0, width=.249\textwidth]{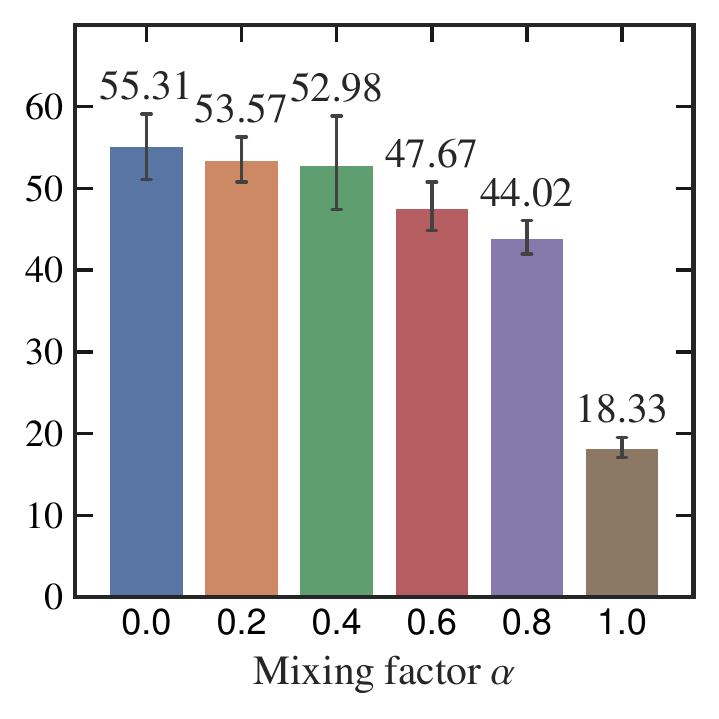}\label{fig:mixing_factor_d}} 
    
    \caption{Mean squared error of the final models trained with different mixing factor $\alpha$ on the testing set and on the testing set with augmentations (i) change in physical pixel size, (ii) change in background noise $\sigma_\text{noise}$ and (iii) with (i) and (ii) combined. We average results over five runs. }
    \label{fig:mixing_factor}
\end{figure*}

\subsection{Robustness results} 
\label{sec:results_denoising}

In Fig. \ref{fig:denoising_examples} we show an example of the predictions obtained by the denoising models trained on only the SKIRT data $(\alpha=0)$ and models trained on only the generated data $(\alpha=1)$ for the different data augmentations. The evaluation of the denoising model regarding the mean squared error is presented in Fig. \ref{fig:mixing_factor}. When there are no augmentations (Fig. \ref{fig:mixing_factor_a}), the model converges to a plateau at $0.58$ if $\alpha \leq 0.8$, i.e. if the original data is seen during training. If the model is trained only on the generated data, the final test loss is $0.63$, which is $8$ per cent higher than the model trained on only the original data set. 

Since all images are weighted equally, very bright galaxies contribute more to the loss. We found that when increasing the physical pixel size (Fig. \ref{fig:mixing_factor_b}) the denoising models generate artifacts close to the peak brightness of the image, which leads to a large error. In this situation, all models with $\alpha \leq 0.6$ plateau with a mean squared error between $60.0$ and $66.0$. There are improvements for the $\alpha = 0.8$ model (decrease in the test loss of $11$ per cent compared to $\alpha = 0$) and for the $\alpha=1$ model (decrease in the test loss of $68$ per cent compared to $\alpha = 0$), suggesting a respective increase in model robustness.

When the background noise is increased (Fig. \ref{fig:mixing_factor_c}), the mean squared error decreases almost linearly from $12.42$ at $\alpha = 0$ to $4.51$ at $\alpha = 1$. For the $\alpha = 0.8$ and $\alpha = 1$ this translates to a decrease in the test loss by $45$ per cent and $63$ per cent. Lastly, when combining the two augmentations above (Fig. \ref{fig:mixing_factor_d}), the mean squared error decreases monotonically from $55.31$ at $\alpha=0$ to $17.81$ at $\alpha=1$. The biggest jump occurs at $\alpha=0.8$ and $\alpha=1$, suggesting an exponential decline. In all cases, the mean squared error is smaller than for augmentation (ii), which only increases the pixel size. The mean squared error decreases for the $\alpha=0.8$ model by $23$ per cent and for the $\alpha=1$ model by $67$ per cent compared to the $\alpha=0$ model. 

To summarise, in the cases where distributional shifts are applied to the test data, i.e. Fig. \ref{fig:mixing_factor_b}, \ref{fig:mixing_factor_c} and \ref{fig:mixing_factor_d}, the model with $\alpha=1$ (only generated data) is in all cases more robust than the $\alpha=0$ (no generated data) version. 
For augmentation (ii), the mean squared error decreases by $68$ per cent, for augmentation (iii) by $63$ per cent and for augmentation (iv) by $67$ per cent.  
Moreover, we find that there is an optimal mixing factor $\alpha = 0.8$ whose corresponding trained models attain the lowest score on the testing set ($0.58$) and are significantly more robust than models with less mixing ($\alpha \leq 0.6$).

It is an interesting observation that the test losses for augmentation (ii) are higher than for augmentation (iv) and this showcases that the effects that domain shifts have on machine learning models can be unexpected and hard to predict. In this case, the higher background noise leads to fewer artifacts close to the peak brightness of the image, which causes a smaller test error, even though the data quality has degraded in general.

Overall, our results show that for the task of image denoising, we can mix generated data and data from the original data source up to larger mixing factors ($\alpha \leq 0.8$) without suffering from any negative impacts regarding the model's prediction quality evaluated on an independent test set ($0.58$). Moreover, when relying entirely on the generated data ($\alpha=1$), the final test score is only slightly higher ($0.63$).

\subsection{Origin of robustness improvements}
In the following, we will discuss possible origins of the increased robustness from generated data. If we regard the source data set as a collection of data points that are drawn independently from the true, but unknown data distribution $P(X)$, then we can increase the information content that the source data set has on $P(X)$ by simple image augmentations. For example, for astrophysical images, we know that the likelihood of drawing a specific image does not depend on the rotation of the image. Therefore, we can explicitly model prior information on $P(X)$ by artificially increasing the size of the source data set with random rotations of the source data. Since the augmented source data set now contains more information about $P(X)$, the models trained on the augmented source data are likely to generalize better to $P(X)$. It is unclear if expanding the source data set with data from generative models has a similar effect on the information content on $P(X)$, when physical prior knowledge is not incorporated in the design of the generative models. There is reason to believe that the convolutional architecture of the generative models considered in this work already represents some type of physical prior information \citep{liu2018image}. 

At the same time, increasing the size of the training data set can have strong regularising effects on the machine learning models \citep{kukavcka2017regularization}. Moreover, in general very large data sets are necessary to train robust models \citep{najafi2019robustness}. Our results demonstrate that the same is true if we consider generated data similar to the source data set for denoising astrophysical images problems. The models trained on a larger proportion of generated data are more robust when dealing with domain shifts in the test data set. In \citet{gowal2021improving}, the authors recently showed a similar result for several popular generative models and classification tasks in computer vision regarding robustness towards adversarial attacks. The surprisingly positive effect that adding potentially imperfect data to the training process is not entirely novel to research in machine learning and can, for example, also be observed with pseudo-labels \citep{lee2013pseudo} and entropy regularization \citep[][]{grandvalet2006entropy}.

It is important to note that while the generated data might not improve the information that we have on $P(x)$, it is still possible that the generated data adds new additional information about any transformations that are applied afterwards. For example, in the case of denoising astropysical images, the background noise and PSF are applied to both the original data set and generated data set. Regardless of whether the generated data is representative of $P(X)$, the generated data set is larger and potentially more diverse. This allows the denoising model to see more pairs of clean and noised galaxies during training, which makes it easier to learn the properties of the PSF and the noise, potentially improving the generalization capabilities and robustness of the denoising model. We believe that this effect may be much stronger for even more complicated transformations than in relatively simple denoising problems.
Given the continuing progress in generative modelling and possibly larger, more diverse source data sets, we expect that the quality of generative models regarding astrophysical images as discussed in Section \ref{sec:evaluation} will improve even more in the future. Eventually, we are confident that the test score when training with only generated data will equalise with the test score obtained by training on the source data alone. 

\section{Conclusion} 
\label{sec:conclusion}

In this paper, we investigated whether three standard generative models, namely variational autoencoders, adversarial latent autoencoders and generative adversarial networks can correctly produce realistic galaxy images. We trained these models on three different data sets consisting of both real and synthetic galaxy images of varying complexity: a sample of S\'ersic profiles, a subsample of real galaxies from the COSMOS survey and a sample of simulated galaxies from the IllustrisTNG simulation obtained by running the SKIRT code.
We quantified the performance of each generative model in terms of the Wasserstein distance between the 1D distributions of a set of physically motivated quantities (morphological parameters, power-spectrum, colour and bulge statistic) as well as metrics traditionally used in computer visions (FID, KID). 

Overall, our evaluation convincingly shows that generative models can very well capture the properties of the source data set. Out of the three generative models we considered in this paper, 
the StyleGAN model was the best performing one, more closely reproducing  71 out of the 84 quantitative metrics considered. 

Our analysis also suggests that, contrary to our initial belief, the StyleGAN model does not suffer much from mode collapse, and it is able to reproduce the distribution of morphological measurements better than ALAE and VAE in most cases without having an explicit reconstruction loss (mean $9.04$ StyleGAN vs $61.14$ VAE vs $17.85$ ALAE for the morphological measurements of the SKIRT data set). 

Interestingly, we find that the VAE model, which is at present commonly used in astrophysical applications, compares poorly to the two adversarial-based models StyleGAN and ALAE in most instances, and especially for the more complex COSMOS and SKIRT data sets.
A possible reason is that the VAE model uses the mean squared error as a loss function. This choice can be problematic for noisy data, such as COSMOS, or data, like SKIRT, for which galaxies are clumpy and show a complex morphology. On the other hand, we found that the hyper-parameter $\lambda$ that weights the reconstruction loss and the regularisation of the latent space for the VAE model significantly improves the results (see Section \ref{subsec:vae_detailed}). However, this hyper-parameter needs to be fined tuned to the data set and size of the latent representation, rendering the performance of the method dependent on subjective user’s choices.

As expected, we find that the performance of each generative model strongly depends on the input data sets, where samples with more complex and diverse properties and fewer objects (e.g. the SKIRT data) are harder to learn than simpler and larger samples (e.g. the S\'ersic and the COSMOS data sets).

In terms of the different metrics used to evaluate generative models, we find them to be mostly consistent. However, we see some inconsistencies as well. For example, while the FID and $\mathcal{W}_1$-Wasserstein distance for the morphological measurements (mean) of StyleGAN is much better than for ALAE ($1.45$ vs $14.65$ and $6,72$ vs $42.32$) for the COSMOS data set, the $\mathcal{W}_1$-Wasserstein distance for the power spectra is much better for ALAE on larger physical scales than for StyleGAN (e.g. the total magnitude $2.53$ vs $4.63$). These results indicate the importance of considering more than one metric. Of course, the choice of metric should also take into consideration the astrophysical problem at hand.

Having demonstrated the capabilities of generative models for generating galaxy images with the correct physical properties, in the second part of this paper, we investigated, whether galaxy images generated with StyleGAN can be used to improve robustness in downstream machine learning problems and in particular for the task of denoising galaxy images. For this purpose, we first created mock observations from the SKIRT galaxies by adding noise and applying a PSF to the images. We then trained a convolutional neural network to learn to denoise the images by using a training data set of mock observations with a varying fraction of SKIRT images and objects generated by StyleGAN. 

Our analysis shows that models trained on larger data sets are generally more robust regarding domain shifts in the data, even when a significant fraction of the data is made of generated rather than real images. Moreover, we show that models trained with additional generated data are more robust against domain shifts like changing the physical pixel size (a) or increasing the background noise level (b). In particular, we find that by mixing generated and original data, it is possible to obtain a $45$ per cent improvement regarding the robustness for (a) and a $23$ per cent improvement for (a) + (b), while keeping the test loss on data without any domain shift the same. Moreover, we show improvements in the robustness by $68$ per cent for (a) and $63$ per cent for (b) when training solely on generated data. In that case, the model evaluated on the test data without domain shifts shows only a slight in the loss by $8$ per cent compared to models trained solely on the original data. These results are particularly interesting for applications of machine learning techniques to the field of astrophysics, where one is often limited by relative small amounts of data, which are possibly subject to domain shifts and selection effects. Our results show that this issue can be mitigated significantly by expanding the data sets with generated data. 

\section*{Acknowledgements}
SV thanks the Max Planck Society for support through a Max Planck Lise Meitner Group. SV acknowledges funding from the European Research Council (ERC) under the European Union’s Horizon 2020 research and innovation programme (LEDA: grant agreement No 758853). This research was partly carried out on the High Performance Computing resources of the FREYA cluster at the Max Planck Computing and Data Facility (MPCDF) in Garching operated by the Max Planck Society (MPG).
This work was supported by the ERC Consolidator Grant \textit{SpaTe} (CoG-2019-863850).

\section*{Public Data and Software}

The IllustrisTNG synthetic galaxy image data \citep[][SKIRT]{nelson2019illustristng} and the COSMOS observations \citep[][]{Mandelbaum.2012} are publicly available. 
This research made use of the deep learning libraries PyTorch \citep[][]{paszke2019pytorch} and Tensorflow \citep[][]{tensorflow2015-whitepaper}, the astronomy packages Astropy \citep[][]{astropy:2013, astropy:2018}, statmorph \citepalias[][]{rodriguez2019optical} and Photutils \citep[][]{larry_bradley_2020_4044744}, and packages for scientific computing NumPy \citep[][]{harris2020array} and SciPy \citep[][]{2020SciPy-NMeth}.
We thank \citet[][]{karras2020analyzing} and \citet[][]{pidhorskyi2020adversarial} for providing baseline implementations for StyleGAN and adversarial latent autoencoders. 

\section*{Data Availability}

The data sets and generated data considered in this work are available upon request to the corresponding author.  

%%%%%%%%%%%%%%%%%%%%%%%%%%%%%%%%%%%%%%%%%%%%%%%%%%

%%%%%%%%%%%%%%%%%%%% REFERENCES %%%%%%%%%%%%%%%%%%

% The best way to enter references is to use BibTeX:

\bibliographystyle{mnras}
\bibliography{references} 

%%%%%%%%%%%%%%%%%%%%%%%%%%%%%%%%%%%%%%%%%%%%%%%%%%

%%%%%%%%%%%%%%%%% APPENDICES %%%%%%%%%%%%%%%%%%%%%

\clearpage

\appendix

\section{Detailed Model Description} \label{sec:detailed_model}

The network architecture and training details are described in this section.  

\subsection{General Training Details} \label{subsec:general_preprocessing}

For all data sets and models, we apply random rotations by $0^{\circ}$, $90^{\circ}$, $180^{\circ}$ and $270^{\circ}$. Moreover, we apply random flips in $x$- and $y$-direction as data augmentations. We estimate the mean and standard deviation for each channel and normalize each channel individually for each data set.  

\subsection{Additional size channel for SKIRT}

\begin{figure*}
	\centering
	\includegraphics[width=\textwidth, clip=True, trim=4 4 4 4]{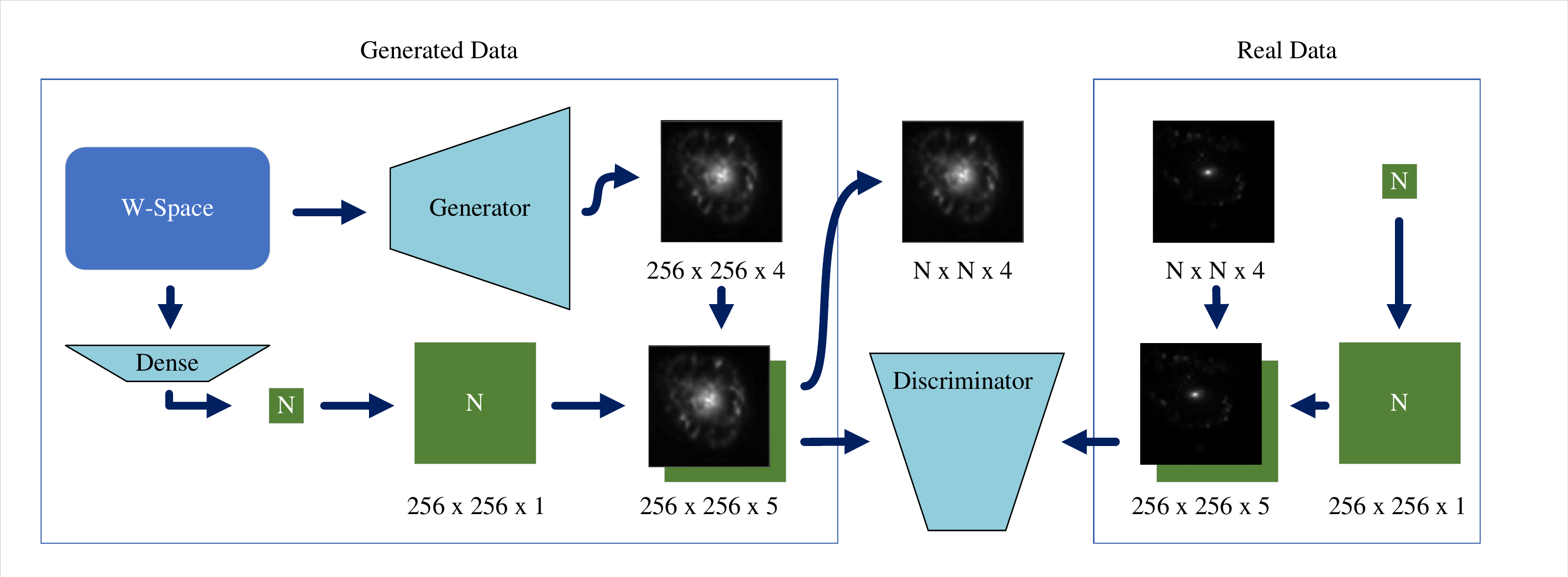} 
    \caption{Adaptation of the training process of SKIRT-like data for the ALAE and StyleGAN models. The SKIRT data has dimension $N \times N \times 4$ based on the physical size of each galaxy. The channels correspond to the four bands $g, r, i$ and $z$. We resize the data to $256 \times 256 \times 256$ using linear interpolation. We append an additional channel of size $256 \times 256 \times 1$ to the resized data, which has constant value $N$. This gives a real data sample of size $256 \times 256 \times 5$ that is fed into the discriminator. Note that the new channel is normalized as described in Appendix \ref{subsec:general_preprocessing}. On the generator side, we use a small dense neural network that predicts the size parameter $N$ from the transformed $z$ latent representation ($W$-space, see Section \ref{subsection:alae}). The output of this network is inflated to $256 \times 256 \times 1$ and concatenated with the output of the generator. This representation can be used as the input to the discriminator or can be back to size $N \times N \times 4$ using linear interpolation.}
    \label{fig:size_channel_construction}
\end{figure*}

The SKIRT data set has a fixed physical pixel size, see Section \ref{sec:datasets}, but the grid size of the $2$D-projections extracted from the IllustrisTNG simulation varies depending on the galaxy's size. One strategy to deal with this data issue is to crop the images to obtain a grid of fixed size. However, we found that there is no clear optimal resolution. For almost all resolutions, there are larger galaxies where cropping would cut out parts of the galaxy. At the same time, there are galaxies which have fewer pixels than the chosen resolution and which would have to be padded. We found that including smaller galaxies, often very faint, incentives mode collapse and confounds the training. 
Therefore, we decided to resize all galaxies to $256 \times 256$ and encode the information about the original size in an additional channel. Fig. \ref{fig:size_channel_construction} gives an overview of how we adapt the original architectures to include the size information channel for the two GAN models. The adaption of the VAE is analogous but the original size is predicted from the mean of the latent $z$ vector. 

\subsection{Adversarial latent autoencoder}

We base the code for the adversarial latent autoencoder on \citet{pidhorskyi2020adversarial}. The code is built around progressively growing the generator and discriminator, i.e. the training begins by training on images of the size $2 \times 2 \times C$, $4 \times 4 \times C$ and so on, where $C$ is the number of channels. The resolution in the progressive growing is increased every $15$ epochs until the final resolution is reached. For the first four resolutions, we train with a learning rate of $\num{1.5e-3}$, which is decreased $\num{1.0e-3}$ for the fifth resolution $32 \times 32 \times C$. For all higher resolutions, we train with a learning rate of $\num{0.5e-3}$. The remaining model parameters are the same as in \citet{pidhorskyi2020adversarial} for the CelebA data set.

\subsection{StyleGAN}

The StyleGAN model is based on \citet[][]{karras2020analyzing}, who provide a public release of their code. We choose not to include improvements of the StyleGAN models regarding adaptive discriminator augmentation \cite{karras2020training}, as some of the augmentations are suited only for natural images and we want to keep the augmentations strategies among the different models the same. We found that using the improved architecture of StyleGAN2 \citep[][]{karras2020analyzing} did not improve the generated data perceptually as well as in terms of the FID compared to the StyleGAN1 architecture \citep[][]{karras2019style}. Based on this, we choose to use the original generator from StyleGAN1 and the updated discriminator from StyleGAN2. Otherwise, the hyperparameters are set by the StyleGAN ´auto' configuration. 

\subsection{VAE} \label{subsec:vae_detailed}

For the variational autoencoder model, we employ the ResNet50 architecture \citep[][]{he2016deep} as the encoder. We do not include the classification head of the ResNet50 architecture and replace it with our implementation, which consists of a simple Conv2D layer with $32$ filters, `same' padding and kernel size $1$ and a final dense layer, which predicts the mean and variance of the standard normal distribution that the latent vector $z$ is drawn from. For the decoder, we use a comparably small and simple architecture consisting of a dense linear layer with the latent vector $z$ as input and output dimension $512$, which is reshaped to $4 \times 4 \times 32$. This is followed by Conv2d with kernel size $1$, `same' padding and $64$ filters. Next, there are 6 blocks, which increase the height and width by a factor of $2$ each. The blocks have the following structure: UpSampling2D, BatchNormalization, LeakyReLU and Conv2D with kernel size $3$, `same padding' and $64$ filters. Finally, there is a last Conv2D layer with kernel size $9$. The number of filters of this last Conv2D layer depends on the number of channels of the data set the VAE is trained on.   
We also tested an implementation of the VAE described by \citet[][]{bretonniere2021}, however we had difficulty training this architecture successfully for our data sets, possibly due to limited training time. When training the VAE, we also found that we have to modify the training objective \eqref{eq:vae_objective}, 
because the models converged to either having good reconstruction capabilities but poor generation capabilities, or the model showed signs of mode collapse. We traced this to a inadequate weighting between the reconstruction loss $p_\phi(x|z)$ and loss on the latent representation $\log p_\phi(z) - \log q_\psi(z|x)$.
To solve this issue, we introduce a weighting factor $\lambda_\mathrm{reconst}$ to the modified VAE objective, which is similar to $\beta$-VAEs \citep[][]{burgess2018understanding} 
\begin{equation}
	\lambda_\text{reconst} \log p_\phi(x|z) + \log p_\phi(z) - \log q_\psi(z|x)\,,
\end{equation}
Lower values of $\lambda_\mathrm{reconst}$ will favour the regularity of the latent space and lead to mode collapse. Higher values will favour reconstruction quality, which leads to weak regularisation of the latent space and poor quality of the generated data. We describe in the next section, how we choose the best $\lambda_\mathrm{reconst}$ and pick the final VAE model. 

\subsection{Model Selection}

Our strategy for deciding when to stop training the models depends on the model type. 
For VAEs, we need to choose an appropriate weighting factor $\lambda_\mathrm{reconst}$. We found that starting with $\lambda_\mathrm{reconst} = 100$ is not too high for all of the three considered data sets. Every $250$ epochs, we increase $\lambda_\mathrm{reconst}$ by a factor of $10$. For each value of $\lambda_\mathrm{reconst}$, we save the best model based on its performance (ELBO) on a validation set that comprises $10\%$ of all training data. Out of all weighting factors, we pick the best one by inspection of the generated data. We obtained the best models for weighting factors corresponding to $500$ epochs for the COSMOS data set and the S\'ersic profiles and $750$ epochs for the SKIRT data set.  
For the adversarial-based models, we pick the models with the best FID, which we compute on the fly every $20$ epochs.

\section{Morphological measurements} \label{sec:appendix_morphological measurements}

In this section, we describe the pre-processing before computing the morphological measurements and give a short overview of the morphological parameters we consider in this work. 

\subsection{Pre-processing} \label{subsec:preprocessing_metrics}

The \verb statmorph $\;$ package requires a segmentation map as an additional input, which classifies each pixel as either part of the background $(0)$ or part of the central galaxy $(1)$. 
We assign each pixel as part of the central galaxy if the pixel value is $1.5\sigma$ above the sky median and smooth the resulting segmentation mask by convolving it with a uniform boxcar filter using the \verb photutil $\,$ package. For the SKIRT data set, we compute morphological measurements on the $i$-band. 
We add background noise to the SKIRT images with $\sigma_\mathrm{noise} = 1/15$. Moreover, we apply a PSF similar to Pan-STARRS, which is $1.18$ arcseconds in the $i$-band, by setting the FWHM of the PSF to $3$ pixels to match the pre-processing by \citetalias[][]{rodriguez2019optical}.
We apply no further pre-processing for the S\'ersic profiles and COSMOS data set. \verb statmorph $\,$ expects another input parameter, which is the \emph{gain} that is used internally to compute a weight map that represents uncertainty in each pixel. For the SKIRT data set, the gain is linked to the exposure time and converts the units of the SKIRT data set $e^{-}s^{-1}\mathrm{pixel}^{-1}$ to $e^{-}\text{pixel}^{-1}$. For the $i$ band, we use the recommended gain of 1200. We set a gain of $1$ for the S\'ersic profiles and the COSMOS data set.  

\subsection{Selected morphological parameters} 

Here, we define several morphological measurements that are computed by \verb statmorph .  
In many cases, calculating the morphological measurements requires knowing the notion of a centre of the galaxy, which is defined by the $x$- and $y$-coordinates that minimize the asymmetry $A$ (see below for a definition). 

\begin{enumerate}
    \item Petrosian radius $r_\mathrm{petro}$: the Petrosian radius $r_\mathrm{petro}$ is defined as the radius inside which the mean surface brightness is equal to some constant $\mu := 0.2$ times the overall mean surface brightness of the image.
    \item Asymmetry $A$: Let $I(i,j)$ denote the pixel values of the image at position $i,j$ and let $I_{180}$ be the image rotated by $180$ degree around the galaxy's centre. Then the asymmetry is defined as 
    \begin{equation} \label{eq:asymmetry} 
         A = \sum_{i,j} \frac{|I(i,j) - I_{180}(i,j)|}{|I(i,j)|} - B_{180}\,, 
    \end{equation}
    where $B_{180}$ is the asymmetry of the background as defined by the segmentation map \citep{lotz2004new}. In \verb statmorph $\;$ the sum only includes those pixels with a radius of $1.5r_\mathrm{petro}$ of the centre. Since the latter is defined as the point that minimizes the asymmetry, the asymmetry $A$ is the minimum asymmetry over all possible centre points of the galaxy.  
    \item Concentration $C$: the concentration $C$ is defined as
    \begin{equation} 
         C = 5\log_{10}\left( \frac{r_{80}}{r_{20}}\right)\,, 
    \end{equation}
    where $r_{20}$ and $r_{80}$ are the circular apertures containing 20 per cent and 80 per cent of the total flux \citep{lotz2004new}. 
    \item Smoothness $S$: the smoothness $S$ is defined by subtracting a smoothed version of the image $I_S$ from the original image $S$, i.e. 
    \begin{equation} \label{eq:smoothness} 
         S = \sum_{i,j} \frac{|I(i,j) - I_{S}(i,j)|}{|I(i,j)|} - B_{S}\,. 
    \end{equation}
    The smoothed image $I_S$ is obtained by a boxcar filter of width of $0.25r_\text{petro}$ \citep{lotz2004new}. $B_S$ denotes the average smoothness of the background. 
    For additional details on the calculation, we refer to \citetalias[][]{rodriguez2019optical}. In general, galaxies that are very smooth (instead of clumpy) have a low value of $S$.  
    
    \item Gini $G$: Let $(X_n)$ denote a list of all pixel values in the image. The Gini-coefficient $G$ is defined as 
    \begin{equation}
        G = \frac{1}{2\overline{X}n(n-1)} \sum_{i=1}{n}\sum_{j=1}{n}|X_i-X_j|\,,
    \end{equation}
    where $\overline{X}$ is the average flux over all pixels. The Gini-coefficient measures how close the actual distribution of pixel flux values is compared to an ideal uniform distribution, where each pixel has the same value. For example, if $G=0$, all pixels have the same value, while $G=1$ implies that all flux is in only $1$ pixel and the other pixels have zero flux \citepalias[][]{rodriguez2019optical}.
    \item Half-light radius $R_\mathrm{half}$: The radius $R_\mathrm{half}$ is the radius that contains half of the total flux of the galaxy. It can be defined for elliptical and circular apertures, but we use the circular version in this paper.
    \item The bulge statistic $F(G,M_{20})$: The bulge statistic $F(G,M_{20})$ is used to classify early-type and late-type galaxies and is defined as 
    \begin{equation}
        F(G, M_{20}) = -0.693 M_{20} + 4.95G - 3.96\,,
    \end{equation}
    where $M_{20}$ is a statistic that relates the second moment of a galaxy's brightest regions (comprising 20 per cent of the total flux) to the total central-order central moment. Galaxies with $F(G, M_{20}) > 0$ are classified as early-types and as late-types otherwise.
\end{enumerate}
Additionally, \verb statmorph $\:$ fits S\'ersic profiles, see Section \ref{subsec:sersic_profiles}, to the galaxy images. This gives more parameters, such as the fitted S\'ersic index $n$, the orientation of the S\'ersic profile and the ellipticity.

\section{Additional quantitative evaluation: Mean and standard deviation}

We provide the mean and standard deviation in addition to Table \ref{tab:quantitative_metrics} in Table \ref{tab:quantitative_metrics_mean_std} for the SKIRT and COSMOS data set. 

\begin{table*}
 \begin{tabular*}{\textwidth}{@{}l@{\hspace*{0pt}}c@{\hspace*{0pt}}c@{\hspace*{0pt}}c@{\hspace*{0pt}}c@{\hspace*{0pt}}|@{\hspace*{0pt}}c@{\hspace*{0pt}}c@{\hspace*{0pt}}c@{\hspace*{0pt}}c@{\hspace*{0pt}}}
  \hline
   & \multicolumn{4}{c}{\hspace*{0pt}SKIRT} & \multicolumn{4}{c}{\hspace*{0pt}COSMOS} \\
  \hline
  & SKIRT & VAE & StyleGAN & ALAE & COSMOS & VAE & StyleGAN & ALAE \\ 
  \hline
  Morphological properties &  &  &  &  &  &  &  & \\[2pt]
  $\;$ \makebox[0pt][l]{Smoothness}\phantom{Orientation xxxxxxxx} $[\num{e-2}]$ & 
  \makebox[0pt][l]{\phantom{$-0$}$3.52\pm3.94$}\phantom{$-00.00\pm00.00$} & 
  \makebox[0pt][l]{\phantom{$0$}$-0.40\pm2.54$}\phantom{$-00.00\pm00.00$} & 
  \makebox[0pt][l]{\phantom{$-0$}$3.25\pm4.78$}\phantom{$-00.00\pm00.00$} & 
  \makebox[0pt][l]{\phantom{$-0$}$3.69 \pm3.35$}\phantom{$-00.00\pm00.00$} & 
  \makebox[0pt][l]{\phantom{$0$}$-4.44\pm10.95$}\phantom{$-00.00\pm00.00$} & 
  \makebox[0pt][l]{\phantom{$-0$}$0.68\pm0.53$}\phantom{$-00.00\pm00.00$} & 
  \makebox[0pt][l]{\phantom{$0$}$-3.70\pm10.43$}\phantom{$-00.00\pm00.00$} & 
  \makebox[0pt][l]{\phantom{$0$}$-4.39 \pm10.24$}\phantom{$-00.00\pm00.00$}  \\[2pt]
  $\;$ \makebox[0pt][l]{Gini coefficient}\phantom{Orientation xxxxxxxx} $[\num{e-2}]$ & 
  \makebox[0pt][l]{\phantom{$-$}$53.85\pm5.22$}\phantom{$-00.00\pm00.00$} & 
  \makebox[0pt][l]{\phantom{$-$}$55.65\pm2.63$}\phantom{$-00.00\pm00.00$} & 
  \makebox[0pt][l]{\phantom{$-$}$53.08\pm5.54$}\phantom{$-00.00\pm00.00$} & 
  \makebox[0pt][l]{\phantom{$-$}$51.89\pm5.11$}\phantom{$-00.00\pm00.00$} & 
  \makebox[0pt][l]{\phantom{$-$}$50.86\pm5.44$}\phantom{$-00.00\pm00.00$} & 
  \makebox[0pt][l]{\phantom{$-$}$51.77\pm2.43$}\phantom{$-00.00\pm00.00$} & 
  \makebox[0pt][l]{\phantom{$-$}$51.13\pm5.46$}\phantom{$-00.00\pm00.00$} & 
  \makebox[0pt][l]{\phantom{$-$}$52.97\pm5.34$}\phantom{$-00.00\pm00.00$} \\[2pt]
  $\;$ \makebox[0pt][l]{S\'ersic index $n$}\phantom{Orientation xxxxxxxx} & 
  \makebox[0pt][l]{\phantom{$-0$}$1.94\pm1.45$}\phantom{$-00.00\pm00.00$} & 
  \makebox[0pt][l]{\phantom{$-0$}$1.68\pm0.50$}\phantom{$-00.00\pm00.00$} & 
  \makebox[0pt][l]{\phantom{$-0$}$1.87\pm1.44$}\phantom{$-00.00\pm00.00$} & 
  \makebox[0pt][l]{\phantom{$-0$}$1.39\pm0.84$}\phantom{$-00.00\pm00.00$} & 
  \makebox[0pt][l]{\phantom{$-0$}$0.81\pm0.51$}\phantom{$-00.00\pm00.00$} & 
  \makebox[0pt][l]{\phantom{$-0$}$1.32\pm0.27$}\phantom{$-00.00\pm00.00$} & 
  \makebox[0pt][l]{\phantom{$-0$}$0.84\pm0.53$}\phantom{$-00.00\pm00.00$} & 
  \makebox[0pt][l]{\phantom{$-0$}$1.05\pm0.50$}\phantom{$-00.00\pm00.00$} \\[2pt]
  $\;$ \makebox[0pt][l]{Orientation}\phantom{Orientation xxxxxxxx} $[\num{e-1}]$ & 
  \makebox[0pt][l]{\phantom{$0$}$-0.20\pm9.04$}\phantom{$-00.00\pm00.00$} & 
  \makebox[0pt][l]{\phantom{$-0$}$0.04\pm9.12$}\phantom{$-00.00\pm00.00$} & 
  \makebox[0pt][l]{\phantom{$0$}$-0.07\pm9.11$}\phantom{$-00.00\pm00.00$} & 
  \makebox[0pt][l]{\phantom{$0$}$-0.05\pm9.04$}\phantom{$-00.00\pm00.00$} & 
  \makebox[0pt][l]{\phantom{$-0$}$0.09\pm9.10$}\phantom{$-00.00\pm00.00$} & 
  \makebox[0pt][l]{\phantom{$-0$}$0.13\pm9.43$}\phantom{$-00.00\pm00.00$} & 
  \makebox[0pt][l]{\phantom{$0$}$-0.12\pm9.09$}\phantom{$-00.00\pm00.00$} & 
  \makebox[0pt][l]{\phantom{$-0$}$0.15\pm9.24$}\phantom{$-00.00\pm00.00$} \\[2pt]
  $\;$ \makebox[0pt][l]{Asymmetry}\phantom{Orientation xxxxxxxx} $[\num{e-1}]$ & 
  \makebox[0pt][l]{\phantom{$-0$}$1.87\pm1.38$}\phantom{$-00.00\pm00.00$} & 
  \makebox[0pt][l]{\phantom{$-0$}$0.56\pm0.44$}\phantom{$-00.00\pm00.00$} & 
  \makebox[0pt][l]{\phantom{$-0$}$1.70\pm1.15$}\phantom{$-00.00\pm00.00$} & 
  \makebox[0pt][l]{\phantom{$-0$}$2.00\pm1.18$}\phantom{$-00.00\pm00.00$} & 
  \makebox[0pt][l]{\phantom{$-0$}$0.56\pm0.89$}\phantom{$-00.00\pm00.00$} & 
  \makebox[0pt][l]{\phantom{$-0$}$1.08\pm0.39$}\phantom{$-00.00\pm00.00$} & 
  \makebox[0pt][l]{\phantom{$-0$}$0.54\pm0.89$}\phantom{$-00.00\pm00.00$} & 
  \makebox[0pt][l]{\phantom{$-0$}$0.34\pm0.59$}\phantom{$-00.00\pm00.00$} \\[2pt]
  $\;$ \makebox[0pt][l]{Concentration}\phantom{Orientation xxxxxxxx} & 
  \makebox[0pt][l]{\phantom{$-0$}$3.20\pm0.61$}\phantom{$-00.00\pm00.00$} & 
  \makebox[0pt][l]{\phantom{$-0$}$3.30\pm0.31$}\phantom{$-00.00\pm00.00$} & 
  \makebox[0pt][l]{\phantom{$-0$}$3.19\pm0.60$}\phantom{$-00.00\pm00.00$} & 
  \makebox[0pt][l]{\phantom{$-0$}$3.01\pm0.44$}\phantom{$-00.00\pm00.00$} & 
  \makebox[0pt][l]{\phantom{$-0$}$2.98\pm0.58$}\phantom{$-00.00\pm00.00$} & 
  \makebox[0pt][l]{\phantom{$-0$}$3.03\pm0.26$}\phantom{$-00.00\pm00.00$} & 
  \makebox[0pt][l]{\phantom{$-0$}$3.04\pm0.59$}\phantom{$-00.00\pm00.00$} & 
  \makebox[0pt][l]{\phantom{$-0$}$3.26\pm0.57$}\phantom{$-00.00\pm00.00$} \\[2pt]
  $\;$ \makebox[0pt][l]{$M_{20}$}\phantom{Orientation xxxxxxxx} & 
  \makebox[0pt][l]{\phantom{$0$}$-1.74\pm0.34$}\phantom{$-00.00\pm00.00$} & 
  \makebox[0pt][l]{\phantom{$0$}$-1.92\pm0.13$}\phantom{$-00.00\pm00.00$} & 
  \makebox[0pt][l]{\phantom{$0$}$-1.75\pm0.32$}\phantom{$-00.00\pm00.00$} & 
  \makebox[0pt][l]{\phantom{$0$}$-1.65\pm0.28$}\phantom{$-00.00\pm00.00$} & 
  \makebox[0pt][l]{\phantom{$0$}$-1.66\pm0.36$}\phantom{$-00.00\pm00.00$} & 
  \makebox[0pt][l]{\phantom{$0$}$-1.88\pm0.10$}\phantom{$-00.00\pm00.00$} & 
  \makebox[0pt][l]{\phantom{$0$}$-1.69\pm0.37$}\phantom{$-00.00\pm00.00$} & 
  \makebox[0pt][l]{\phantom{$0$}$-1.88\pm0.28$}\phantom{$-00.00\pm00.00$} \\[2pt]
  $\;$ \makebox[0pt][l]{Half-light radius}\phantom{Orientation xxxxxxxx} & 
  \makebox[0pt][l]{\phantom{$-0$}$9.79\pm6.54$}\phantom{$-00.00\pm00.00$} & 
  \makebox[0pt][l]{\phantom{$-0$}$9.32\pm2.32$}\phantom{$-00.00\pm00.00$} & 
  \makebox[0pt][l]{\phantom{$-0$}$9.47\pm5.41$}\phantom{$-00.00\pm00.00$} & 
  \makebox[0pt][l]{\phantom{$-0$}$9.89\pm5.76$}\phantom{$-00.00\pm00.00$} & 
  \makebox[0pt][l]{\phantom{$-$}$19.20\pm7.01$}\phantom{$-00.00\pm00.00$} & 
  \makebox[0pt][l]{\phantom{$-$}$22.89\pm4.37$}\phantom{$-00.00\pm00.00$} & 
  \makebox[0pt][l]{\phantom{$-$}$19.38\pm7.42$}\phantom{$-00.00\pm00.00$} & 
  \makebox[0pt][l]{\phantom{$-$}$16.16\pm3.50$}\phantom{$-00.00\pm00.00$} \\[2pt]
  $\;$ \makebox[0pt][l]{Ellipticity}\phantom{Orientation xxxxxxxx} $[\num{e-1}]$ & 
  \makebox[0pt][l]{\phantom{$-0$}$3.65\pm1.96$}\phantom{$-00.00\pm00.00$} & 
  \makebox[0pt][l]{\phantom{$-0$}$1.97\pm0.91$}\phantom{$-00.00\pm00.00$} & 
  \makebox[0pt][l]{\phantom{$-0$}$3.92\pm2.12$}\phantom{$-00.00\pm00.00$} & 
  \makebox[0pt][l]{\phantom{$-0$}$3.78\pm1.93$}\phantom{$-00.00\pm00.00$} & 
  \makebox[0pt][l]{\phantom{$-0$}$3.94\pm2.07$}\phantom{$-00.00\pm00.00$} & 
  \makebox[0pt][l]{\phantom{$-0$}$1.87\pm0.99$}\phantom{$-00.00\pm00.00$} & 
  \makebox[0pt][l]{\phantom{$-0$}$4.25\pm2.08$}\phantom{$-00.00\pm00.00$} & 
  \makebox[0pt][l]{\phantom{$-0$}$3.94\pm1.67$}\phantom{$-00.00\pm00.00$} \\[2pt]
  $\;$ \makebox[0pt][l]{Elongation}\phantom{Orientation xxxxxxxx} & 
  \makebox[0pt][l]{\phantom{$-0$}$1.78\pm0.75$}\phantom{$-00.00\pm00.00$} & 
  \makebox[0pt][l]{\phantom{$-0$}$1.26\pm0.15$}\phantom{$-00.00\pm00.00$} & 
  \makebox[0pt][l]{\phantom{$-0$}$1.92\pm0.89$}\phantom{$-00.00\pm00.00$} & 
  \makebox[0pt][l]{\phantom{$-0$}$1.82\pm0.73$}\phantom{$-00.00\pm00.00$} & 
  \makebox[0pt][l]{\phantom{$-0$}$1.94\pm0.97$}\phantom{$-00.00\pm00.00$} & 
  \makebox[0pt][l]{\phantom{$-0$}$1.25\pm0.17$}\phantom{$-00.00\pm00.00$} & 
  \makebox[0pt][l]{\phantom{$-0$}$2.08\pm1.06$}\phantom{$-00.00\pm00.00$} & 
  \makebox[0pt][l]{\phantom{$-0$}$1.63\pm0.54$}\phantom{$-00.00\pm00.00$} \\[2pt]
  \hline
  \hline
  Wavelength ranges & &  &  &  &  &  &  & \\[2pt]
  $\;$ \makebox[0pt][l]{0:\phantom{0} Total magnitude}\phantom{$10: (0.000, 0.000]\lambda_\mathrm{min}\,$} & 
  \makebox[0pt][l]{\phantom{$0$}$-0.02\pm0.34$}\phantom{$-00.00\pm00.00$} & 
  \makebox[0pt][l]{\phantom{$0$}$-0.13\pm0.17$}\phantom{$-00.00\pm00.00$} &
  \makebox[0pt][l]{\phantom{$0$}$-0.12\pm0.36$}\phantom{$-00.00\pm00.00$} & 
  \makebox[0pt][l]{\phantom{$0$}$-0.12\pm0.26$}\phantom{$-00.00\pm00.00$} & 
  \makebox[0pt][l]{\phantom{$0$}$-2.91\pm0.39$}\phantom{$-00.00\pm00.00$} & 
  \makebox[0pt][l]{\phantom{$0$}$-2.99\pm0.17$}\phantom{$-00.00\pm00.00$} & 
  \makebox[0pt][l]{\phantom{$0$}$-2.86\pm0.43$}\phantom{$-00.00\pm00.00$} & 
  \makebox[0pt][l]{\phantom{$0$}$-2.92\pm0.37$}\phantom{$-00.00\pm00.00$} \\[2pt]
    $\;$ \makebox[0pt][l]{1:\phantom{0} \makebox[0pt][l]{$\geq 90.90$}\phantom{$(0.000, 0.000]$} $\lambda_\mathrm{min}$}\phantom{$10: (0.000, 0.000]\lambda_\mathrm{min}\,$} & 
  \makebox[0pt][l]{\phantom{$0$}$-0.22\pm0.33$}\phantom{$-00.00\pm00.00$} & 
  \makebox[0pt][l]{\phantom{$0$}$-0.34\pm0.17$}\phantom{$-00.00\pm00.00$} &
  \makebox[0pt][l]{\phantom{$0$}$-0.29\pm0.35$}\phantom{$-00.00\pm00.00$} & 
  \makebox[0pt][l]{\phantom{$0$}$-0.31\pm0.25$}\phantom{$-00.00\pm00.00$} & 
  \makebox[0pt][l]{\phantom{$0$}$-3.03\pm0.37$}\phantom{$-00.00\pm00.00$} & 
  \makebox[0pt][l]{\phantom{$0$}$-3.13\pm0.17$}\phantom{$-00.00\pm00.00$} & 
  \makebox[0pt][l]{\phantom{$0$}$-2.99\pm0.41$}\phantom{$-00.00\pm00.00$} & 
  \makebox[0pt][l]{\phantom{$0$}$-3.04\pm0.37$}\phantom{$-00.00\pm00.00$} \\[2pt]
    $\;$ \makebox[0pt][l]{2:\phantom{0} \makebox[0pt][l]{$[37.03, 90.90)$}\phantom{$(0.000, 0.000]$} $\lambda_\mathrm{min}$}\phantom{$10: (0.000, 0.000]\lambda_\mathrm{min}\,$} & 
  \makebox[0pt][l]{\phantom{$0$}$-0.49\pm0.34$}\phantom{$-00.00\pm00.00$} & 
  \makebox[0pt][l]{\phantom{$0$}$-0.59\pm0.18$}\phantom{$-00.00\pm00.00$} &
  \makebox[0pt][l]{\phantom{$0$}$-0.53\pm0.36$}\phantom{$-00.00\pm00.00$} & 
  \makebox[0pt][l]{\phantom{$0$}$-0.58\pm0.26$}\phantom{$-00.00\pm00.00$} & 
  \makebox[0pt][l]{\phantom{$0$}$-3.21\pm0.35$}\phantom{$-00.00\pm00.00$} & 
  \makebox[0pt][l]{\phantom{$0$}$-3.31\pm0.17$}\phantom{$-00.00\pm00.00$} & 
  \makebox[0pt][l]{\phantom{$0$}$-3.17\pm0.40$}\phantom{$-00.00\pm00.00$} & 
  \makebox[0pt][l]{\phantom{$0$}$-3.19\pm0.38$}\phantom{$-00.00\pm00.00$} \\[2pt]
    $\;$ \makebox[0pt][l]{3:\phantom{0} \makebox[0pt][l]{$[21.27, 37.03)$}\phantom{$(0.000, 0.000]$} $\lambda_\mathrm{min}$}\phantom{$10: (0.000, 0.000]\lambda_\mathrm{min}\,$} & 
  \makebox[0pt][l]{\phantom{$0$}$-0.68\pm0.35$}\phantom{$-00.00\pm00.00$} & 
  \makebox[0pt][l]{\phantom{$0$}$-0.79\pm0.20$}\phantom{$-00.00\pm00.00$} &
  \makebox[0pt][l]{\phantom{$0$}$-0.72\pm0.37$}\phantom{$-00.00\pm00.00$} & 
  \makebox[0pt][l]{\phantom{$0$}$-0.80\pm0.26$}\phantom{$-00.00\pm00.00$} & 
  \makebox[0pt][l]{\phantom{$0$}$-3.37\pm0.37$}\phantom{$-00.00\pm00.00$} & 
  \makebox[0pt][l]{\phantom{$0$}$-3.47\pm0.18$}\phantom{$-00.00\pm00.00$} & 
  \makebox[0pt][l]{\phantom{$0$}$-3.33\pm0.41$}\phantom{$-00.00\pm00.00$} & 
  \makebox[0pt][l]{\phantom{$0$}$-3.32\pm0.40$}\phantom{$-00.00\pm00.00$} \\[2pt]
    $\;$ \makebox[0pt][l]{4:\phantom{0} \makebox[0pt][l]{$[11.36, 21.27)$}\phantom{$(0.000, 0.000]$} $\lambda_\mathrm{min}$}\phantom{$10: (0.000, 0.000]\lambda_\mathrm{min}\,$} & 
  \makebox[0pt][l]{\phantom{$0$}$-0.83\pm0.35$}\phantom{$-00.00\pm00.00$} & 
  \makebox[0pt][l]{\phantom{$0$}$-0.96\pm0.22$}\phantom{$-00.00\pm00.00$} &
  \makebox[0pt][l]{\phantom{$0$}$-0.87\pm0.37$}\phantom{$-00.00\pm00.00$} & 
  \makebox[0pt][l]{\phantom{$0$}$-0.96\pm0.25$}\phantom{$-00.00\pm00.00$} & 
  \makebox[0pt][l]{\phantom{$0$}$-3.52\pm0.39$}\phantom{$-00.00\pm00.00$} & 
  \makebox[0pt][l]{\phantom{$0$}$-3.63\pm0.20$}\phantom{$-00.00\pm00.00$} & 
  \makebox[0pt][l]{\phantom{$0$}$-3.47\pm0.43$}\phantom{$-00.00\pm00.00$} & 
  \makebox[0pt][l]{\phantom{$0$}$-3.45\pm0.43$}\phantom{$-00.00\pm00.00$} \\[2pt]
    $\;$ \makebox[0pt][l]{5:\phantom{0} \makebox[0pt][l]{$[7.57, 11.36)$}\phantom{$(0.000, 0.000]$} $\lambda_\mathrm{min}$}\phantom{$10: (0.000, 0.000]\lambda_\mathrm{min}\,$} & 
  \makebox[0pt][l]{\phantom{$0$}$-0.97\pm0.35$}\phantom{$-00.00\pm00.00$} & 
  \makebox[0pt][l]{\phantom{$0$}$-1.12\pm0.24$}\phantom{$-00.00\pm00.00$} & 
  \makebox[0pt][l]{\phantom{$0$}$-1.00\pm0.36$}\phantom{$-00.00\pm00.00$} & 
  \makebox[0pt][l]{\phantom{$0$}$-1.10\pm0.24$}\phantom{$-00.00\pm00.00$} & 
  \makebox[0pt][l]{\phantom{$0$}$-3.66\pm0.41$}\phantom{$-00.00\pm00.00$} & 
  \makebox[0pt][l]{\phantom{$0$}$-3.80\pm0.23$}\phantom{$-00.00\pm00.00$} & 
  \makebox[0pt][l]{\phantom{$0$}$-3.61\pm0.46$}\phantom{$-00.00\pm00.00$} & 
  \makebox[0pt][l]{\phantom{$0$}$-3.58\pm0.47$}\phantom{$-00.00\pm00.00$} \\[2pt]
    $\;$ \makebox[0pt][l]{6:\phantom{0} \makebox[0pt][l]{$[5.34, 7.57)$}\phantom{$(0.000, 0.000]$} $\lambda_\mathrm{min}$}\phantom{$10: (0.000, 0.000]\lambda_\mathrm{min}\,$} & 
  \makebox[0pt][l]{\phantom{$0$}$-1.10\pm0.35$}\phantom{$-00.00\pm00.00$} & 
  \makebox[0pt][l]{\phantom{$0$}$-1.27\pm0.26$}\phantom{$-00.00\pm00.00$} & 
  \makebox[0pt][l]{\phantom{$0$}$-1.12\pm0.36$}\phantom{$-00.00\pm00.00$} & 
  \makebox[0pt][l]{\phantom{$0$}$-1.23\pm0.23$}\phantom{$-00.00\pm00.00$} & 
  \makebox[0pt][l]{\phantom{$0$}$-3.77\pm0.42$}\phantom{$-00.00\pm00.00$} & 
  \makebox[0pt][l]{\phantom{$0$}$-3.94\pm0.25$}\phantom{$-00.00\pm00.00$} & 
  \makebox[0pt][l]{\phantom{$0$}$-3.72\pm0.47$}\phantom{$-00.00\pm00.00$} & 
  \makebox[0pt][l]{\phantom{$0$}$-3.69\pm0.50$}\phantom{$-00.00\pm00.00$} \\[2pt]
    $\;$ \makebox[0pt][l]{7:\phantom{0} \makebox[0pt][l]{$[3.93, 5.34)$}\phantom{$(0.000, 0.000]$} $\lambda_\mathrm{min}$}\phantom{$10: (0.000, 0.000]\lambda_\mathrm{min}\,$} & 
  \makebox[0pt][l]{\phantom{$0$}$-1.20\pm0.35$}\phantom{$-00.00\pm00.00$} & 
  \makebox[0pt][l]{\phantom{$0$}$-1.23\pm0.36$}\phantom{$-00.00\pm00.00$} & 
  \makebox[0pt][l]{\phantom{$0$}$-1.34\pm0.22$}\phantom{$-00.00\pm00.00$} & 
  \makebox[0pt][l]{\phantom{$0$}$-0.05\pm9.04$}\phantom{$-00.00\pm00.00$} & 
  \makebox[0pt][l]{\phantom{$0$}$-3.86\pm0.43$}\phantom{$-00.00\pm00.00$} & 
  \makebox[0pt][l]{\phantom{$0$}$-4.06\pm0.27$}\phantom{$-00.00\pm00.00$} & 
  \makebox[0pt][l]{\phantom{$0$}$-3.81\pm0.48$}\phantom{$-00.00\pm00.00$} & 
  \makebox[0pt][l]{\phantom{$0$}$-3.76\pm0.50$}\phantom{$-00.00\pm00.00$} \\[2pt]
    $\;$ \makebox[0pt][l]{8:\phantom{0} \makebox[0pt][l]{$[3.02, 3.93)$}\phantom{$(0.000, 0.000]$} $\lambda_\mathrm{min}$}\phantom{$10: (0.000, 0.000]\lambda_\mathrm{min}\,$} & 
  \makebox[0pt][l]{\phantom{$0$}$-1.30\pm0.35$}\phantom{$-00.00\pm00.00$} & 
  \makebox[0pt][l]{\phantom{$0$}$-1.58\pm0.31$}\phantom{$-00.00\pm00.00$} &
  \makebox[0pt][l]{\phantom{$0$}$-1.33\pm0.37$}\phantom{$-00.00\pm00.00$} & 
  \makebox[0pt][l]{\phantom{$0$}$-1.45\pm0.22$}\phantom{$-00.00\pm00.00$} & 
  \makebox[0pt][l]{\phantom{$0$}$-4.07\pm0.43$}\phantom{$-00.00\pm00.00$} & 
  \makebox[0pt][l]{\phantom{$0$}$-4.18\pm0.29$}\phantom{$-00.00\pm00.00$} & 
  \makebox[0pt][l]{\phantom{$0$}$-3.89\pm0.48$}\phantom{$-00.00\pm00.00$} & 
  \makebox[0pt][l]{\phantom{$0$}$-3.96\pm0.51$}\phantom{$-00.00\pm00.00$} \\[2pt]
    $\;$ \makebox[0pt][l]{9:\phantom{0} \makebox[0pt][l]{$[2.38, 3.02)$}\phantom{$(0.000, 0.000]$} $\lambda_\mathrm{min}$}\phantom{$10: (0.000, 0.000]\lambda_\mathrm{min}\,$} & 
  \makebox[0pt][l]{\phantom{$0$}$-1.39\pm0.35$}\phantom{$-00.00\pm00.00$} & 
  \makebox[0pt][l]{\phantom{$0$}$-1.71\pm0.33$}\phantom{$-00.00\pm00.00$} &
  \makebox[0pt][l]{\phantom{$0$}$-1.42\pm0.37$}\phantom{$-00.00\pm00.00$} & 
  \makebox[0pt][l]{\phantom{$0$}$-1.54\pm0.21$}\phantom{$-00.00\pm00.00$} & 
  \makebox[0pt][l]{\phantom{$0$}$-4.01\pm0.43$}\phantom{$-00.00\pm00.00$} & 
  \makebox[0pt][l]{\phantom{$0$}$-4.28\pm0.30$}\phantom{$-00.00\pm00.00$} & 
  \makebox[0pt][l]{\phantom{$0$}$-3.96\pm0.48$}\phantom{$-00.00\pm00.00$} & 
  \makebox[0pt][l]{\phantom{$0$}$-3.91\pm0.51$}\phantom{$-00.00\pm00.00$} \\[2pt]
    $\;$ \makebox[0pt][l]{10: \makebox[0pt][l]{$[1.94, 2.38)$}\phantom{$(0.000, 0.000]$} $\lambda_\mathrm{min}$}\phantom{$10: (0.000, 0.000]\lambda_\mathrm{min}\,$} & 
  \makebox[0pt][l]{\phantom{$0$}$-1.47\pm0.35$}\phantom{$-00.00\pm00.00$} & 
  \makebox[0pt][l]{\phantom{$0$}$-1.84\pm0.34$}\phantom{$-00.00\pm00.00$} &
  \makebox[0pt][l]{\phantom{$0$}$-1.51\pm0.37$}\phantom{$-00.00\pm00.00$} & 
  \makebox[0pt][l]{\phantom{$0$}$-1.63\pm0.20$}\phantom{$-00.00\pm00.00$} & 
  \makebox[0pt][l]{\phantom{$0$}$-4.07\pm0.43$}\phantom{$-00.00\pm00.00$} & 
  \makebox[0pt][l]{\phantom{$0$}$-4.37\pm0.32$}\phantom{$-00.00\pm00.00$} & 
  \makebox[0pt][l]{\phantom{$0$}$-4.01\pm0.48$}\phantom{$-00.00\pm00.00$} & 
  \makebox[0pt][l]{\phantom{$0$}$-3.96\pm0.51$}\phantom{$-00.00\pm00.00$} \\[2pt]
      $\;$ \makebox[0pt][l]{11: \makebox[0pt][l]{$[1.60, 1.94)$}\phantom{$(0.000, 0.000]$} $\lambda_\mathrm{min}$}\phantom{$10: (0.000, 0.000]\lambda_\mathrm{min}\,$} & 
  \makebox[0pt][l]{\phantom{$0$}$-1.55\pm0.34$}\phantom{$-00.00\pm00.00$} & 
  \makebox[0pt][l]{\phantom{$0$}$-1.96\pm0.35$}\phantom{$-00.00\pm00.00$} &
  \makebox[0pt][l]{\phantom{$0$}$-1.59\pm0.37$}\phantom{$-00.00\pm00.00$} & 
  \makebox[0pt][l]{\phantom{$0$}$-1.70\pm0.20$}\phantom{$-00.00\pm00.00$} & 
  \makebox[0pt][l]{\phantom{$0$}$-4.12\pm0.42$}\phantom{$-00.00\pm00.00$} & 
  \makebox[0pt][l]{\phantom{$0$}$-4.45\pm0.33$}\phantom{$-00.00\pm00.00$} & 
  \makebox[0pt][l]{\phantom{$0$}$-4.06\pm0.47$}\phantom{$-00.00\pm00.00$} & 
  \makebox[0pt][l]{\phantom{$0$}$-4.01\pm0.50$}\phantom{$-00.00\pm00.00$} \\[2pt]
    $\;$ \makebox[0pt][l]{12: \makebox[0pt][l]{$[1.35, 1.60)$}\phantom{$(0.000, 0.000]$} $\lambda_\mathrm{min}$}\phantom{$10: (0.000, 0.000]\lambda_\mathrm{min}\,$} & 
  \makebox[0pt][l]{\phantom{$0$}$-1.62\pm0.34$}\phantom{$-00.00\pm00.00$} & 
  \makebox[0pt][l]{\phantom{$0$}$-2.08\pm0.36$}\phantom{$-00.00\pm00.00$} & 
  \makebox[0pt][l]{\phantom{$0$}$-1.67\pm0.37$}\phantom{$-00.00\pm00.00$} & 
  \makebox[0pt][l]{\phantom{$0$}$-1.78\pm0.20$}\phantom{$-00.00\pm00.00$} & 
  \makebox[0pt][l]{\phantom{$0$}$-4.16\pm0.42$}\phantom{$-00.00\pm00.00$} & 
  \makebox[0pt][l]{\phantom{$0$}$-4.53\pm0.34$}\phantom{$-00.00\pm00.00$} & 
  \makebox[0pt][l]{\phantom{$0$}$-4.10\pm0.47$}\phantom{$-00.00\pm00.00$} & 
  \makebox[0pt][l]{\phantom{$0$}$-4.06\pm0.50$}\phantom{$-00.00\pm00.00$} \\[2pt]
    $\;$ \makebox[0pt][l]{13: \makebox[0pt][l]{$[1.15, 1.35)$}\phantom{$(0.000, 0.000]$} $\lambda_\mathrm{min}$}\phantom{$10: (0.000, 0.000]\lambda_\mathrm{min}\,$} & 
  \makebox[0pt][l]{\phantom{$0$}$-1.69\pm0.34$}\phantom{$-00.00\pm00.00$} & 
  \makebox[0pt][l]{\phantom{$0$}$-2.21\pm0.37$}\phantom{$-00.00\pm00.00$} & 
  \makebox[0pt][l]{\phantom{$0$}$-1.75\pm0.38$}\phantom{$-00.00\pm00.00$} & 
  \makebox[0pt][l]{\phantom{$0$}$-1.86\pm0.20$}\phantom{$-00.00\pm00.00$} & 
  \makebox[0pt][l]{\phantom{$0$}$-4.21\pm0.41$}\phantom{$-00.00\pm00.00$} & 
  \makebox[0pt][l]{\phantom{$0$}$-4.61\pm0.35$}\phantom{$-00.00\pm00.00$} & 
  \makebox[0pt][l]{\phantom{$0$}$-4.15\pm0.46$}\phantom{$-00.00\pm00.00$} & 
  \makebox[0pt][l]{\phantom{$0$}$-4.11\pm0.50$}\phantom{$-00.00\pm00.00$} \\[2pt]
    $\;$ \makebox[0pt][l]{14: \makebox[0pt][l]{$[1.00, 1.15)$}\phantom{$(0.000, 0.000]$} $\lambda_\mathrm{min}$}\phantom{$10: (0.000, 0.000]\lambda_\mathrm{min}\,$} & 
  \makebox[0pt][l]{\phantom{$0$}$-1.76\pm0.35$}\phantom{$-00.00\pm00.00$} & 
  \makebox[0pt][l]{\phantom{$0$}$-2.31\pm0.37$}\phantom{$-00.00\pm00.00$} &
  \makebox[0pt][l]{\phantom{$0$}$-1.82\pm0.38$}\phantom{$-00.00\pm00.00$} & 
  \makebox[0pt][l]{\phantom{$0$}$-1.93\pm0.20$}\phantom{$-00.00\pm00.00$} & 
  \makebox[0pt][l]{\phantom{$0$}$-4.25\pm0.40$}\phantom{$-00.00\pm00.00$} & 
  \makebox[0pt][l]{\phantom{$0$}$-4.69\pm0.36$}\phantom{$-00.00\pm00.00$} & 
  \makebox[0pt][l]{\phantom{$0$}$-4.19\pm0.46$}\phantom{$-00.00\pm00.00$} & 
  \makebox[0pt][l]{\phantom{$0$}$-4.15\pm0.50$}\phantom{$-00.00\pm00.00$} \\[2pt]
  \hline
  \hline
  Colours & & & & & & & &  \\[2pt]
  $\;$ $(g-i)_\mathrm{SDSS}$ early-types & 
  \makebox[0pt][l]{\phantom{$-0$}$0.98\pm0.20$}\phantom{$-00.00\pm00.00$} & 
  \makebox[0pt][l]{\phantom{$-0$}$0.92\pm0.06$}\phantom{$-00.00\pm00.00$} & 
  \makebox[0pt][l]{\phantom{$-0$}$0.98\pm0.25$}\phantom{$-00.00\pm00.00$} & 
  \makebox[0pt][l]{\phantom{$-0$}$0.95\pm0.24$}\phantom{$-00.00\pm00.00$} & 
  \makebox[0pt][l]{\phantom{$-00.00$}-}\phantom{$-00.00\pm00.00$} & 
  \makebox[0pt][l]{\phantom{$-00.00$}-}\phantom{$-00.00\pm00.00$} & 
  \makebox[0pt][l]{\phantom{$-00.00$}-}\phantom{$-00.00\pm00.00$} & 
  \makebox[0pt][l]{\phantom{$-00.00$}-}\phantom{$-00.00\pm00.00$} \\[2pt]
  $\;$ $(g-i)_\mathrm{SDSS}$ late-types & 
  \makebox[0pt][l]{\phantom{$-0$}$0.97\pm0.19$}\phantom{$-00.00\pm00.00$} & 
  \makebox[0pt][l]{\phantom{$-0$}$0.92\pm0.06$}\phantom{$-00.00\pm00.00$} & 
  \makebox[0pt][l]{\phantom{$-0$}$0.98\pm0.26$}\phantom{$-00.00\pm00.00$} & 
  \makebox[0pt][l]{\phantom{$-0$}$0.96\pm0.26$}\phantom{$-00.00\pm00.00$} & 
  \makebox[0pt][l]{\phantom{$-00.00$}-}\phantom{$-00.00\pm00.00$} & 
  \makebox[0pt][l]{\phantom{$-00.00$}-}\phantom{$-00.00\pm00.00$} & 
  \makebox[0pt][l]{\phantom{$-00.00$}-}\phantom{$-00.00\pm00.00$} & 
  \makebox[0pt][l]{\phantom{$-00.00$}-}\phantom{$-00.00\pm00.00$} \\[2pt]
  $\;$ Bulge statistic $F(G,M_{20})$ & 
  \makebox[0pt][l]{\phantom{$0$}$-0.07\pm0.45$}\phantom{$-00.00\pm00.00$} & 
  \makebox[0pt][l]{\phantom{$-0$}$0.13\pm0.20$}\phantom{$-00.00\pm00.00$} & 
  \makebox[0pt][l]{\phantom{$0$}$-0.10\pm0.45$}\phantom{$-00.00\pm00.00$} & 
  \makebox[0pt][l]{\phantom{$0$}$-0.24\pm0.41$}\phantom{$-00.00\pm00.00$} & 
  \makebox[0pt][l]{\phantom{$-00.00$}-}\phantom{$-00.00\pm00.00$} & 
  \makebox[0pt][l]{\phantom{$-00.00$}-}\phantom{$-00.00\pm00.00$} & 
  \makebox[0pt][l]{\phantom{$-00.00$}-}\phantom{$-00.00\pm00.00$} & 
  \makebox[0pt][l]{\phantom{$-00.00$}-}\phantom{$-00.00\pm00.00$} \\[2pt]
  \hline
 \end{tabular*}
 \caption{Mean and standard deviation of selected morphological measurements, the power spectrum on different physical scales and colour properties of the generated data and the source data set for SKIRT and COSMOS.}
 \label{tab:quantitative_metrics_mean_std}
\end{table*}

\clearpage

\section{Data set visualisations}
\subsection{SKIRT}

\begin{figure*}
	\centering
	\subfloat[SKIRT]{\includegraphics[width=\textwidth]{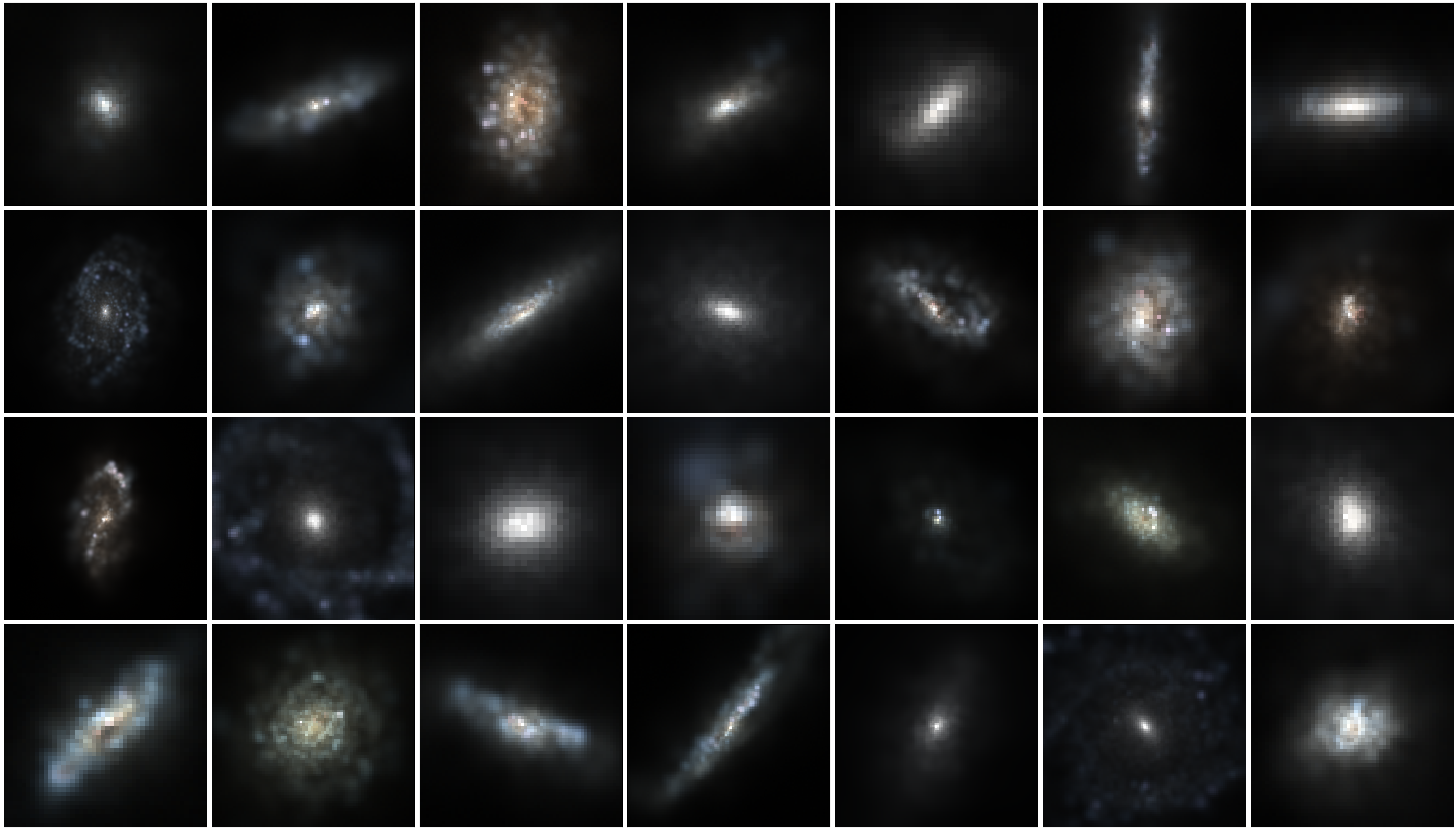}} \\
    \subfloat[StyleGAN]{\includegraphics[width=\textwidth]{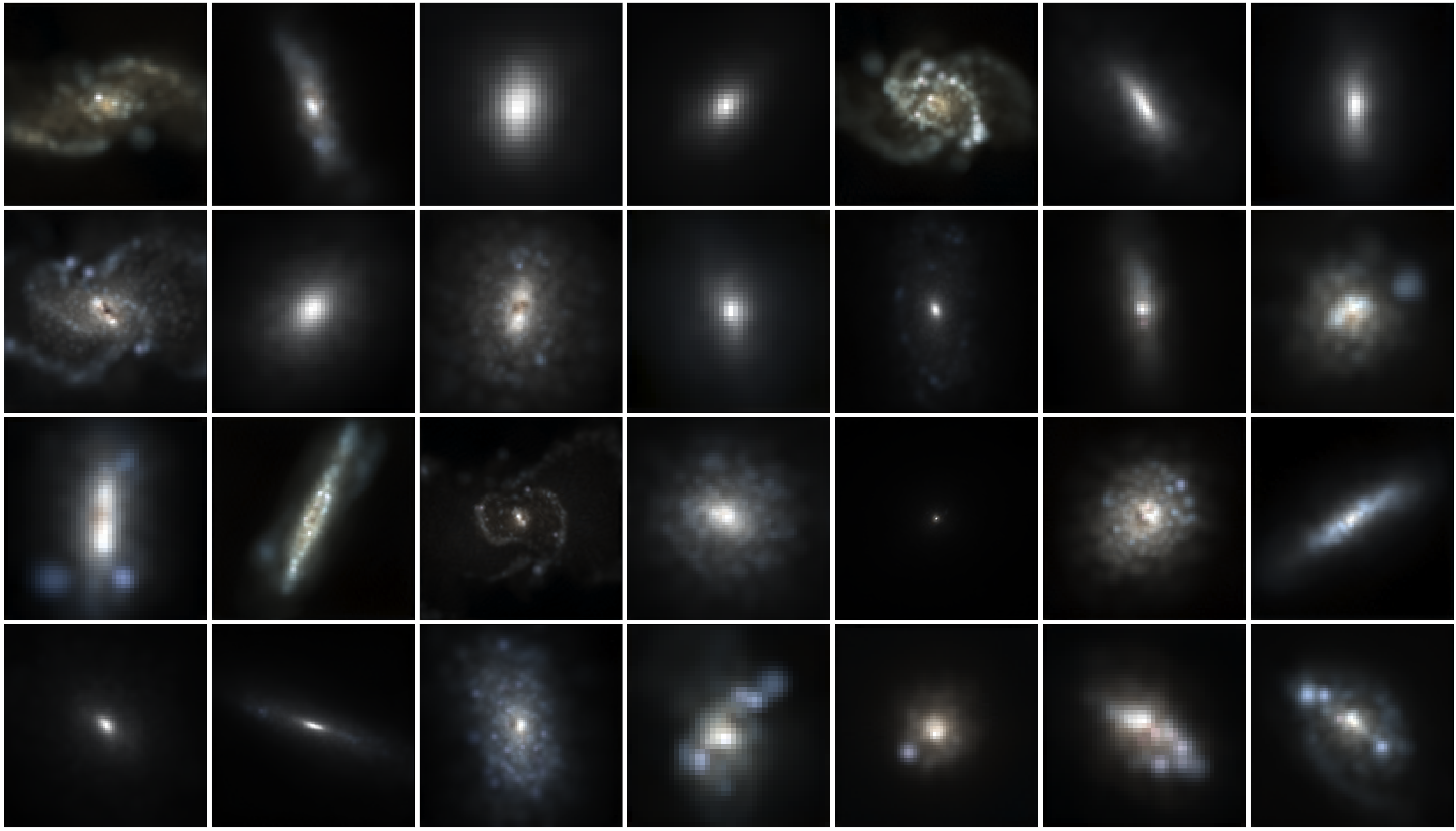}}\hfill%
    \caption{SKIRT: source data set (a) and StyleGAN generated images (b)}
    \label{fig:comparison_overview_tng_stylegan2}
\end{figure*}

\begin{figure*}
	\centering
	\subfloat[ALAE]{\includegraphics[width=\textwidth]{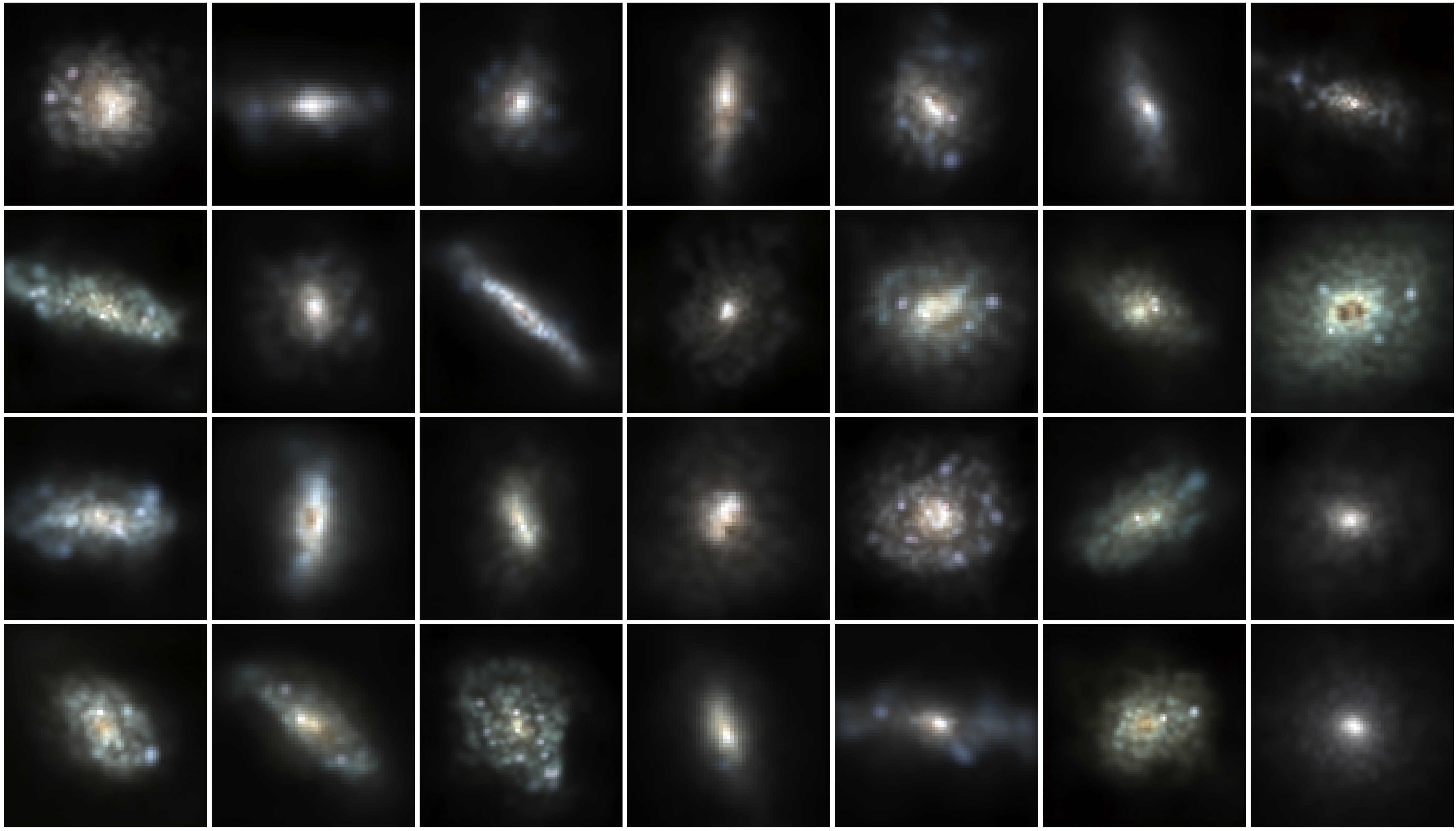}} \\
    \subfloat[VAE]{\includegraphics[width=\textwidth]{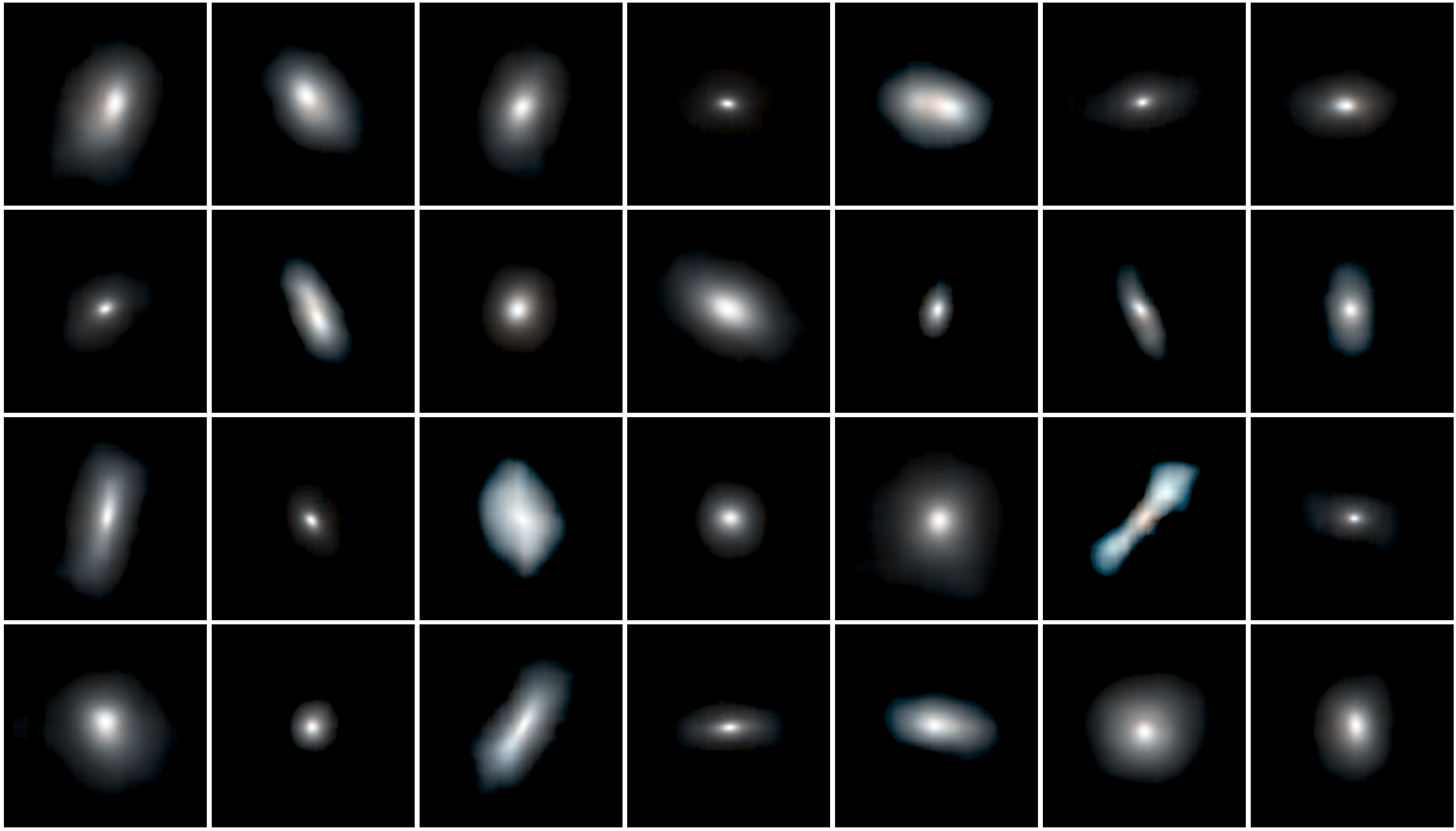}}  
    \caption{SKIRT: ALAE generated images (a) and VAE generated images (b)}
    \label{fig:comparison_overview_tng_alae_vae}
\end{figure*}

\clearpage

\subsection{COSMOS}

\begin{figure*}
	\centering
	\subfloat[COSMOS]{\includegraphics[width=\textwidth]{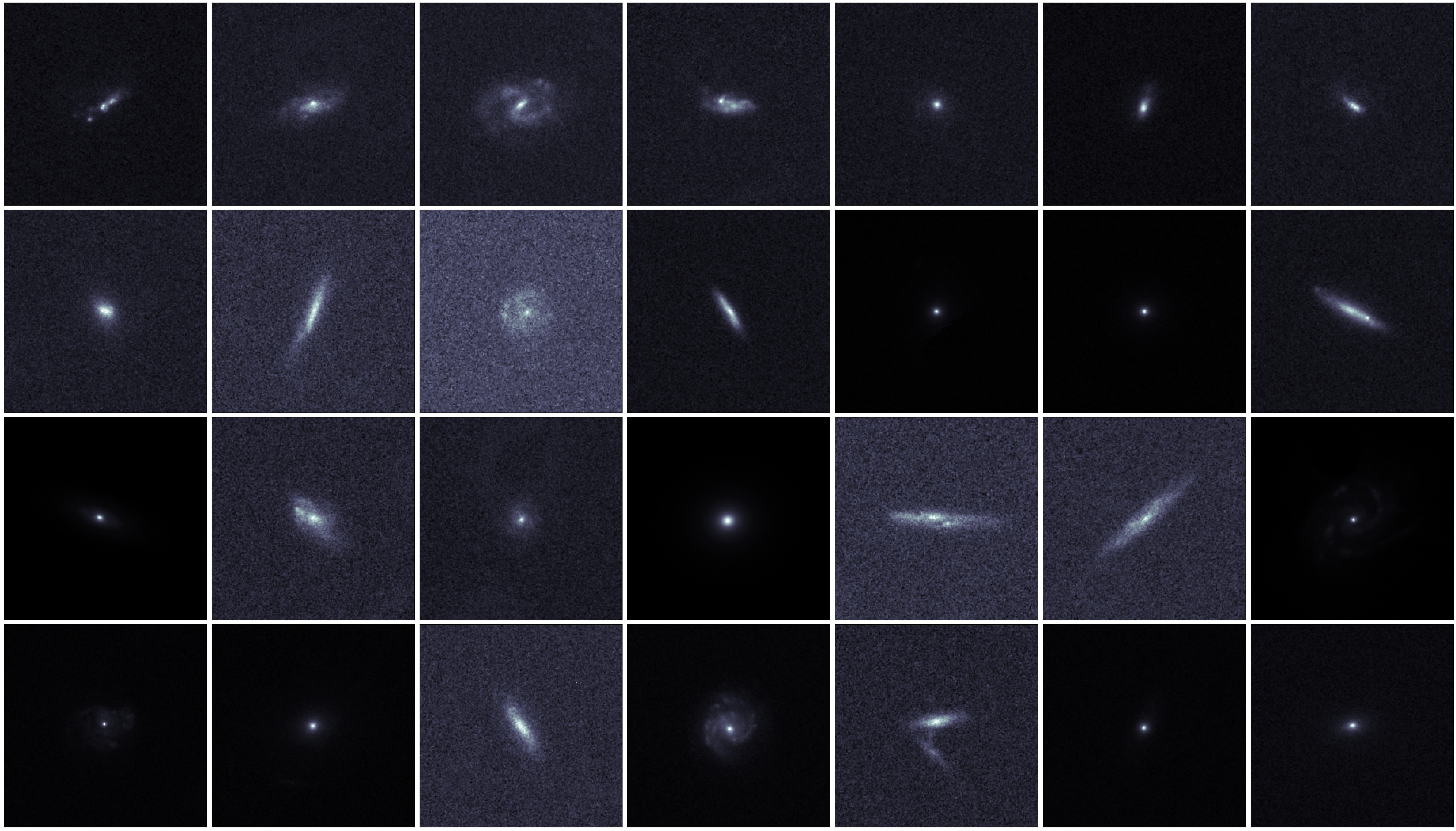}} \\
	\subfloat[StyleGAN]{\includegraphics[width=\textwidth]{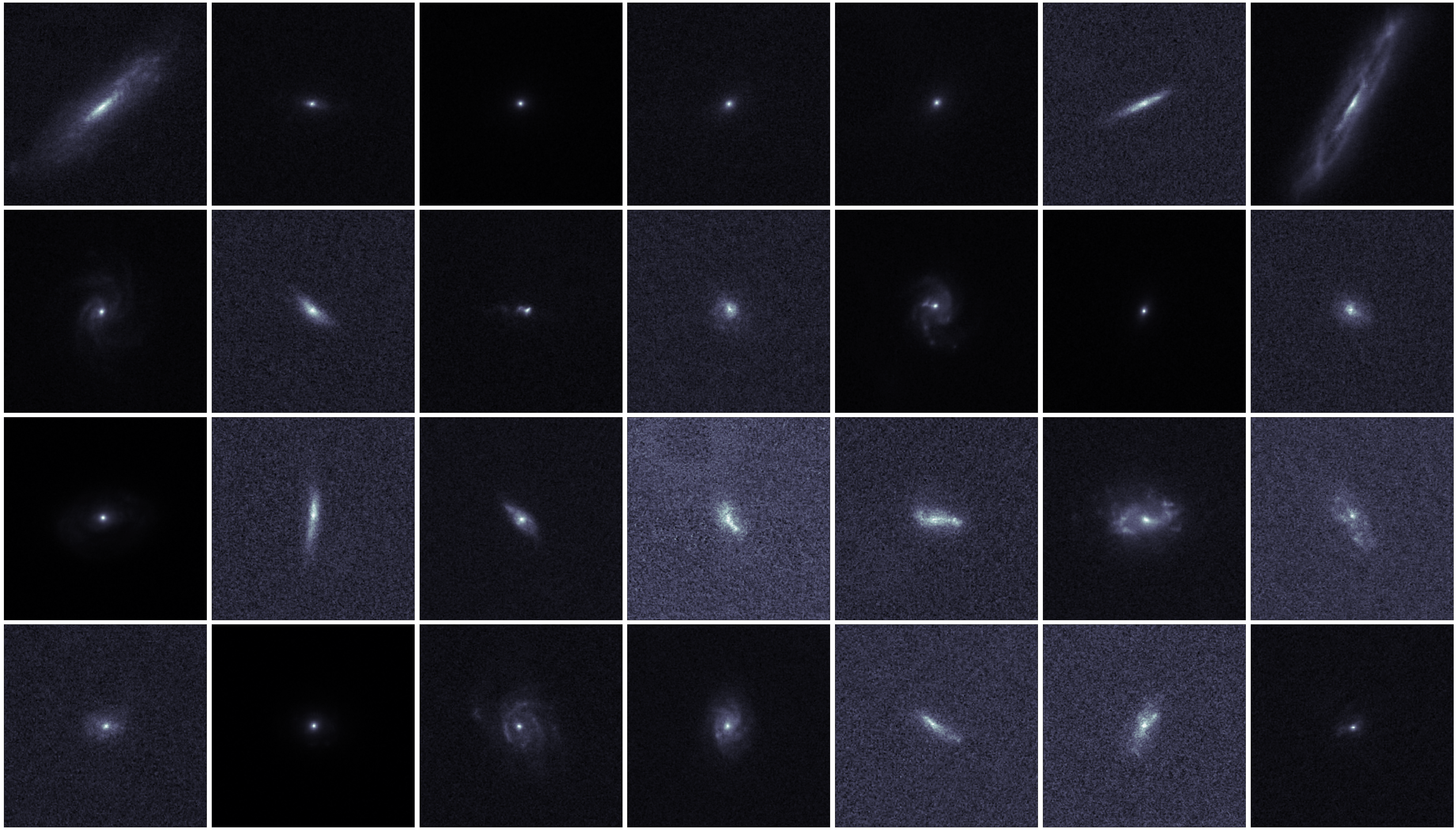}} 
    \caption{COSMOS: source data set (a) and StyleGAN generated images (b)}
    \label{fig:comparison_overview_cosmos_stylegan2}
\end{figure*}

\begin{figure*}
	\centering
	\subfloat[ALAE]{\includegraphics[width=\textwidth]{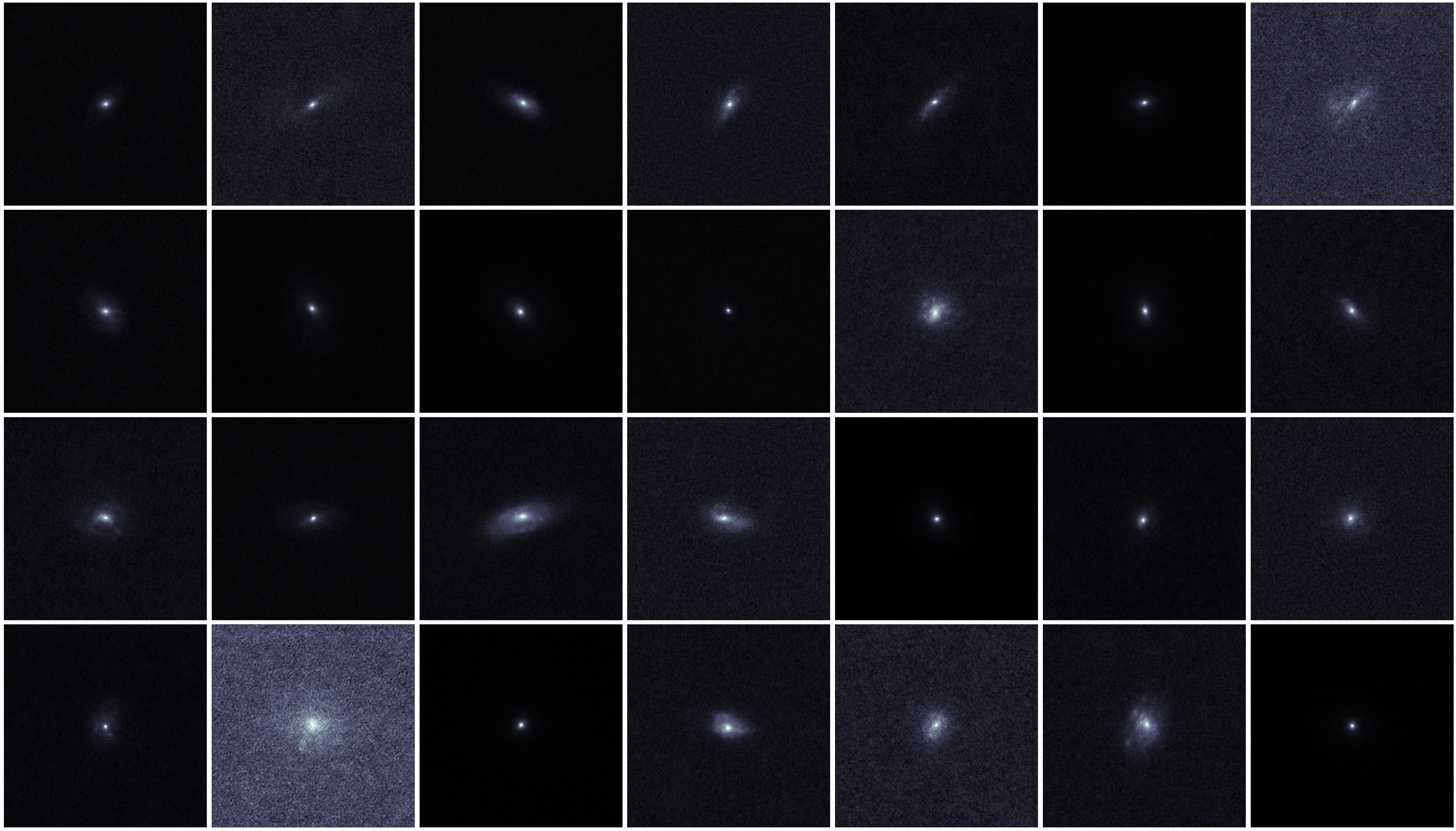}} \\
    \subfloat[VAE]{\includegraphics[width=\textwidth]{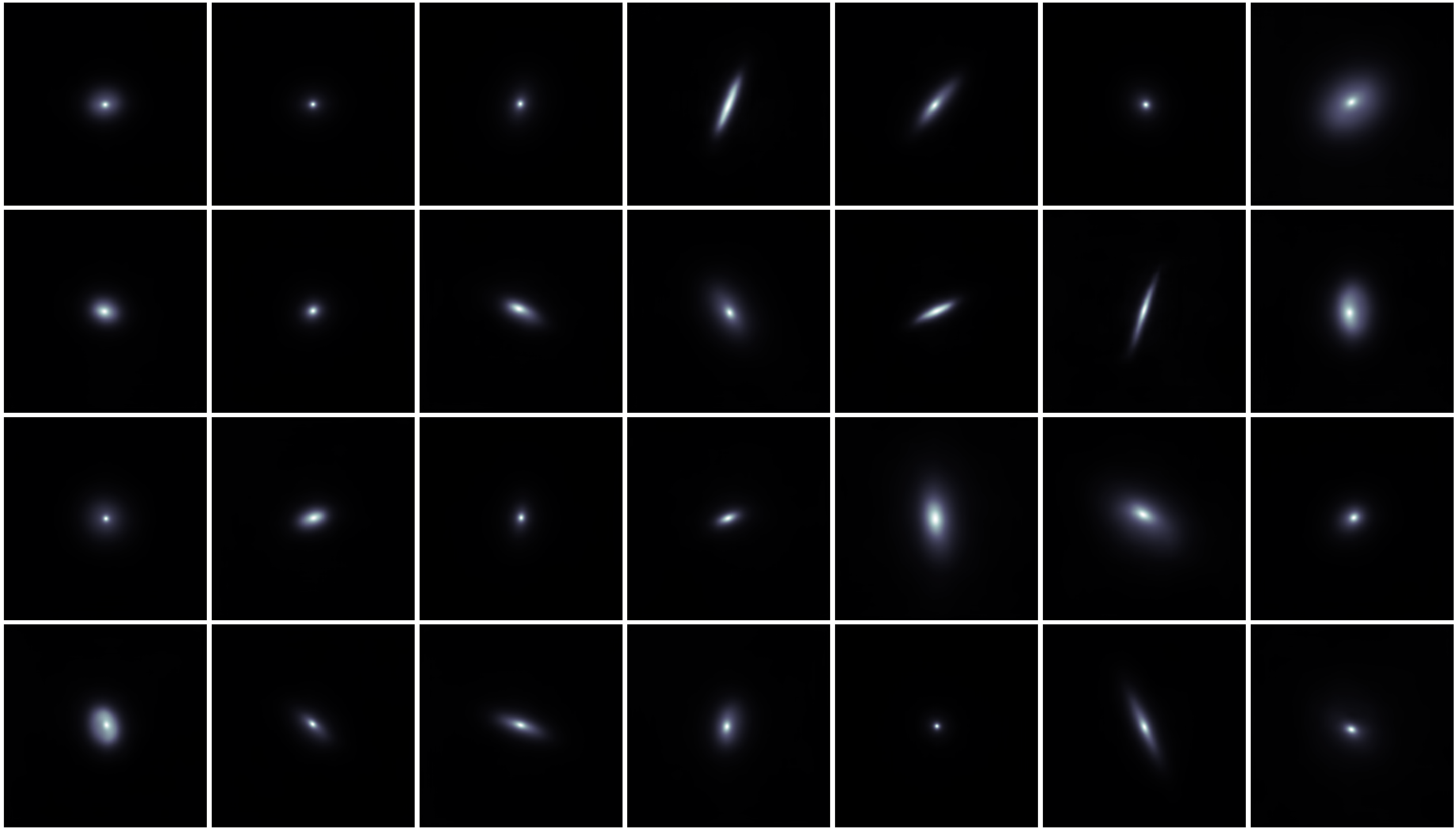}}  
    \caption{COSMOS: ALAE generated images (a) and VAE generated images (b)}
    \label{fig:comparison_overview_cosmos_alae_vae}
\end{figure*}

\clearpage

\section{Optical morphological measurements: COSMOS and S\'ersic profiles}

\begin{figure*}
    \includegraphics[width=\textwidth]{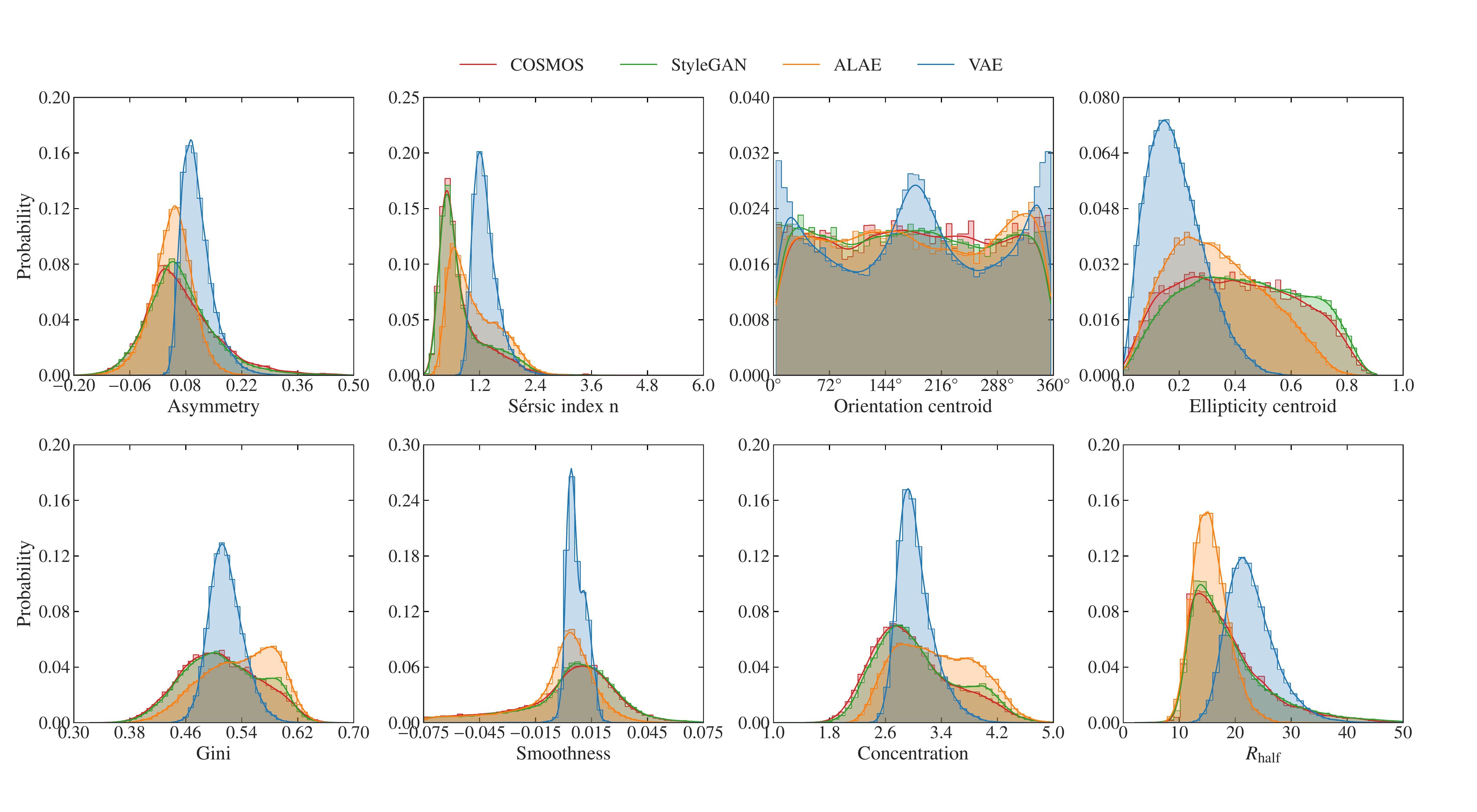}
    \caption{COSMOS: Histograms showing selected optical morphological measurement for the source data set (COSMOS) and the generated data sets (StyleGAN, ALAE and VAE).}
    \label{fig:morphological_properties_COSMOS}
\end{figure*}

\begin{figure*}
    \includegraphics[width=\textwidth]{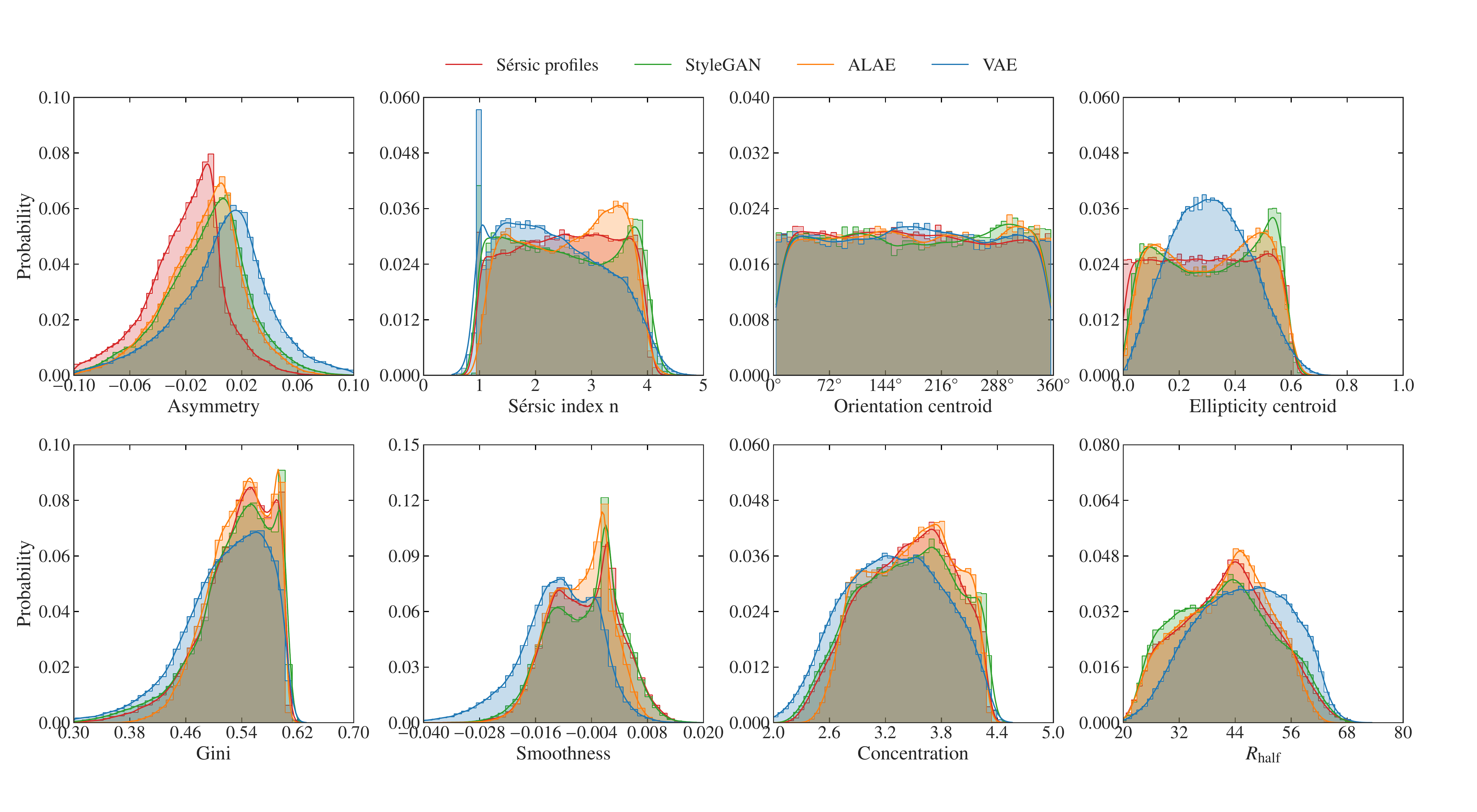}
    \caption{S\'ersic profiles: Histograms showing selected optical morphological measurement for the source data set (S\'ersic profiles) and the generated data sets (StyleGAN, ALAE and VAE).}
    \label{fig:morphological_properties_Sersic}
\end{figure*}

We show 1D plots of selected morphological measurements as described in Section \ref{sec:metrics} for the COSMOS data set in Fig. \ref{fig:morphological_properties_COSMOS} and the S\'ersic profiles in Fig. \ref{fig:morphological_properties_Sersic}.

\section{Power spectra contour and corner plots: S\'ersic and COSMOS}

\begin{figure*}
	\centering
    \subfloat[VAE]{\includegraphics[clip=True, trim=35 35 70 70, width=.23\textwidth]{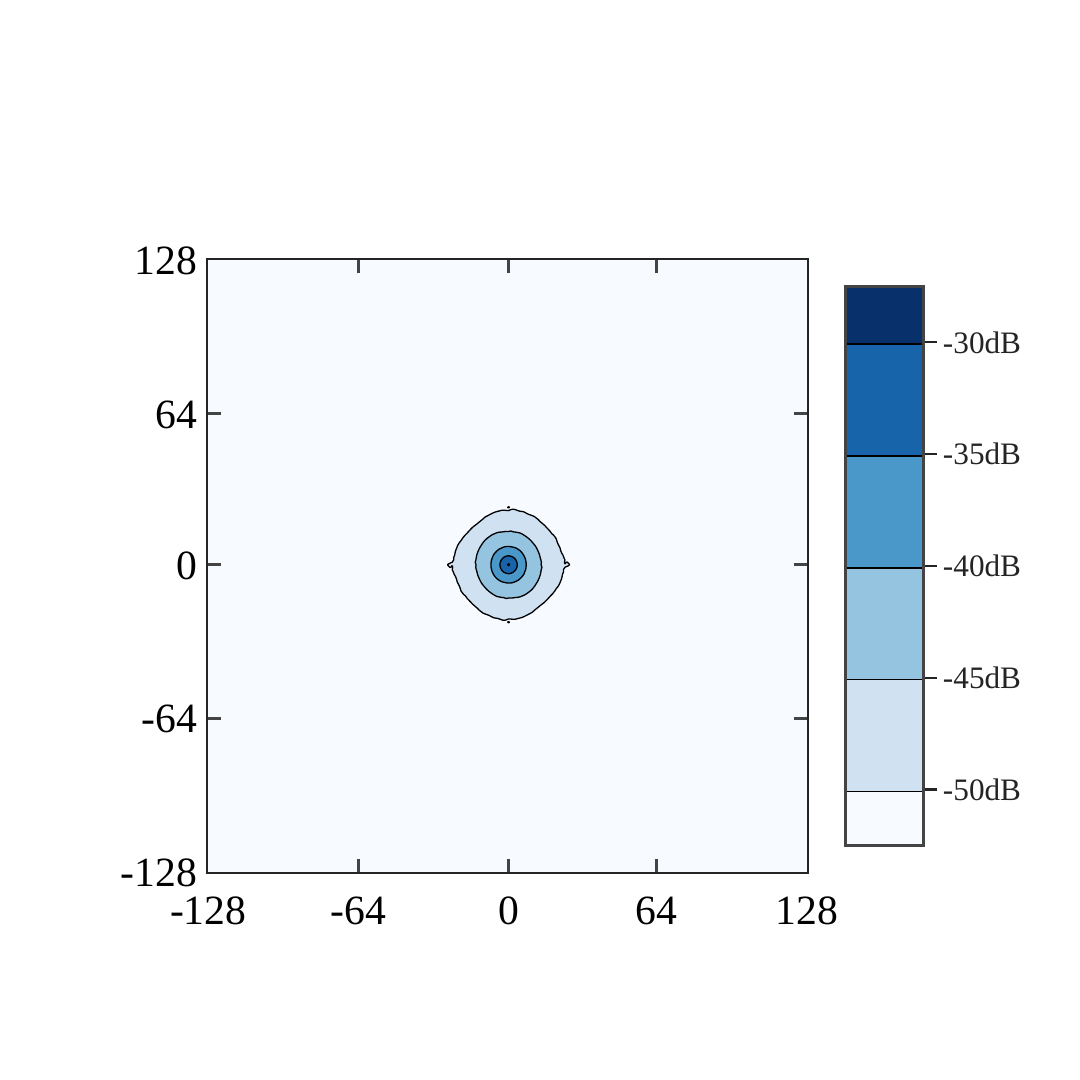}} 
    \subfloat[ALAE]{\includegraphics[clip=True, trim=35 35 70 70, width=.23\textwidth]{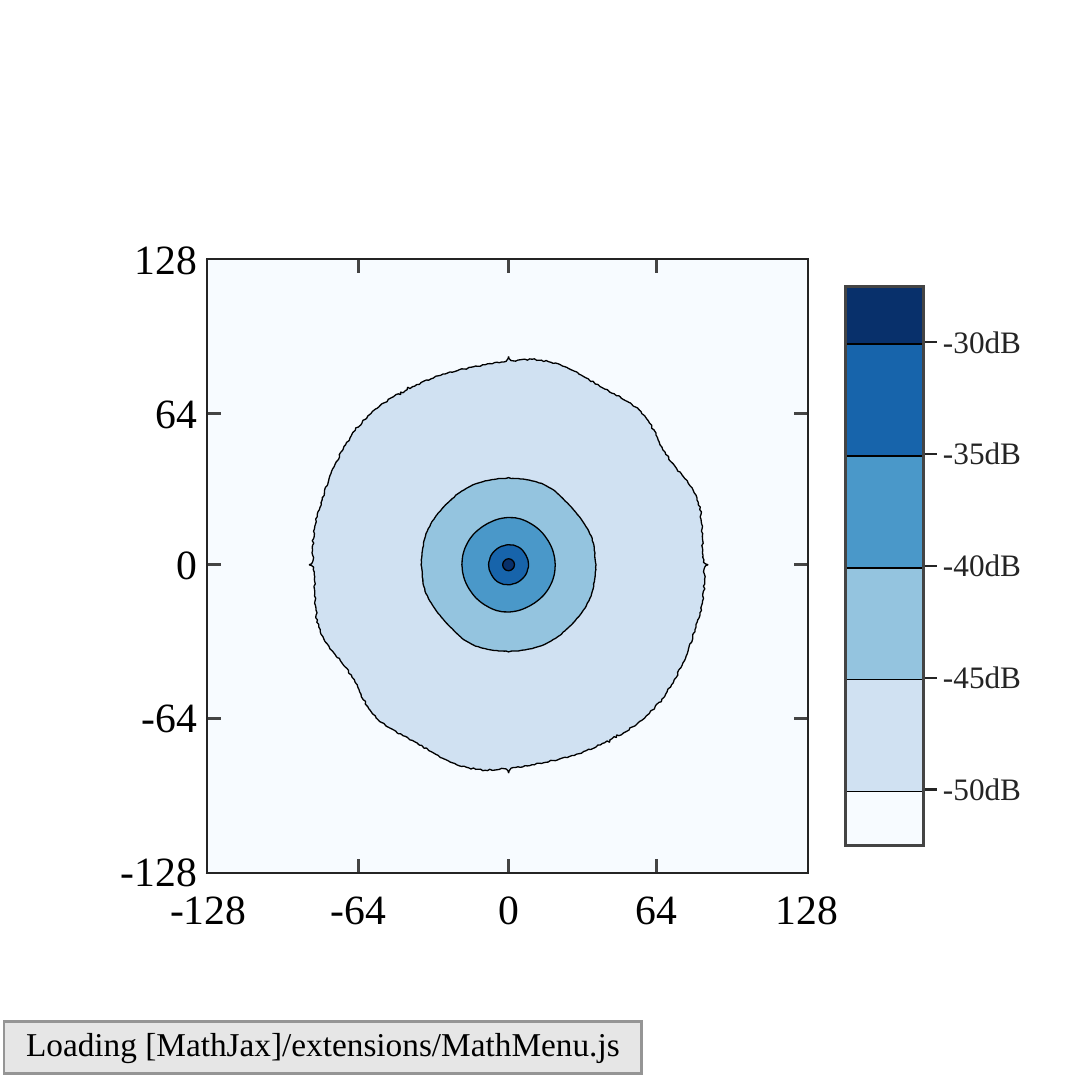}}  
    \subfloat[StyleGAN]{\includegraphics[clip=True, trim=35 35 70 70, width=.23\textwidth]{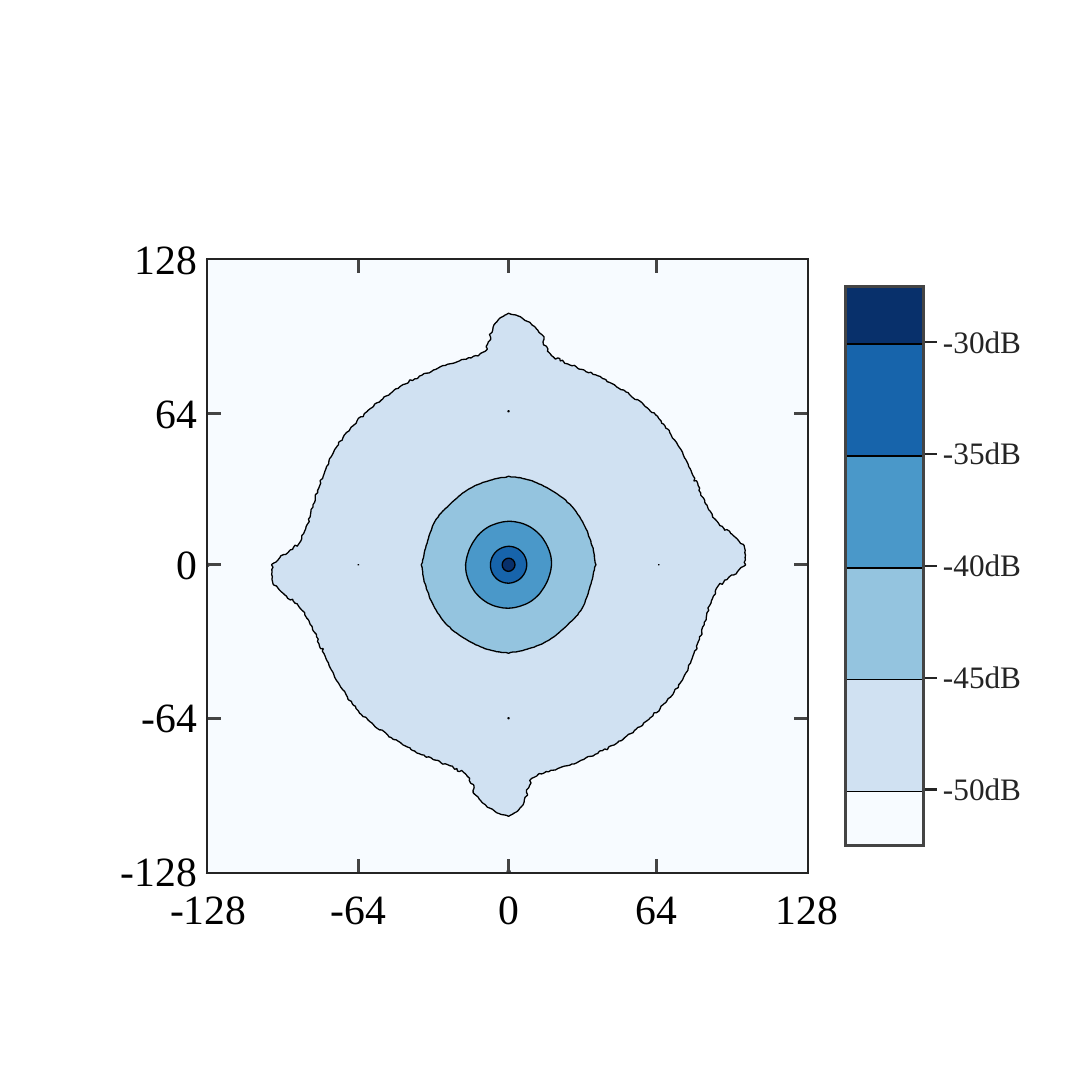}}  
    \subfloat[COSMOS]{\includegraphics[clip=True, trim=35 35 0 70, width=.31\textwidth]{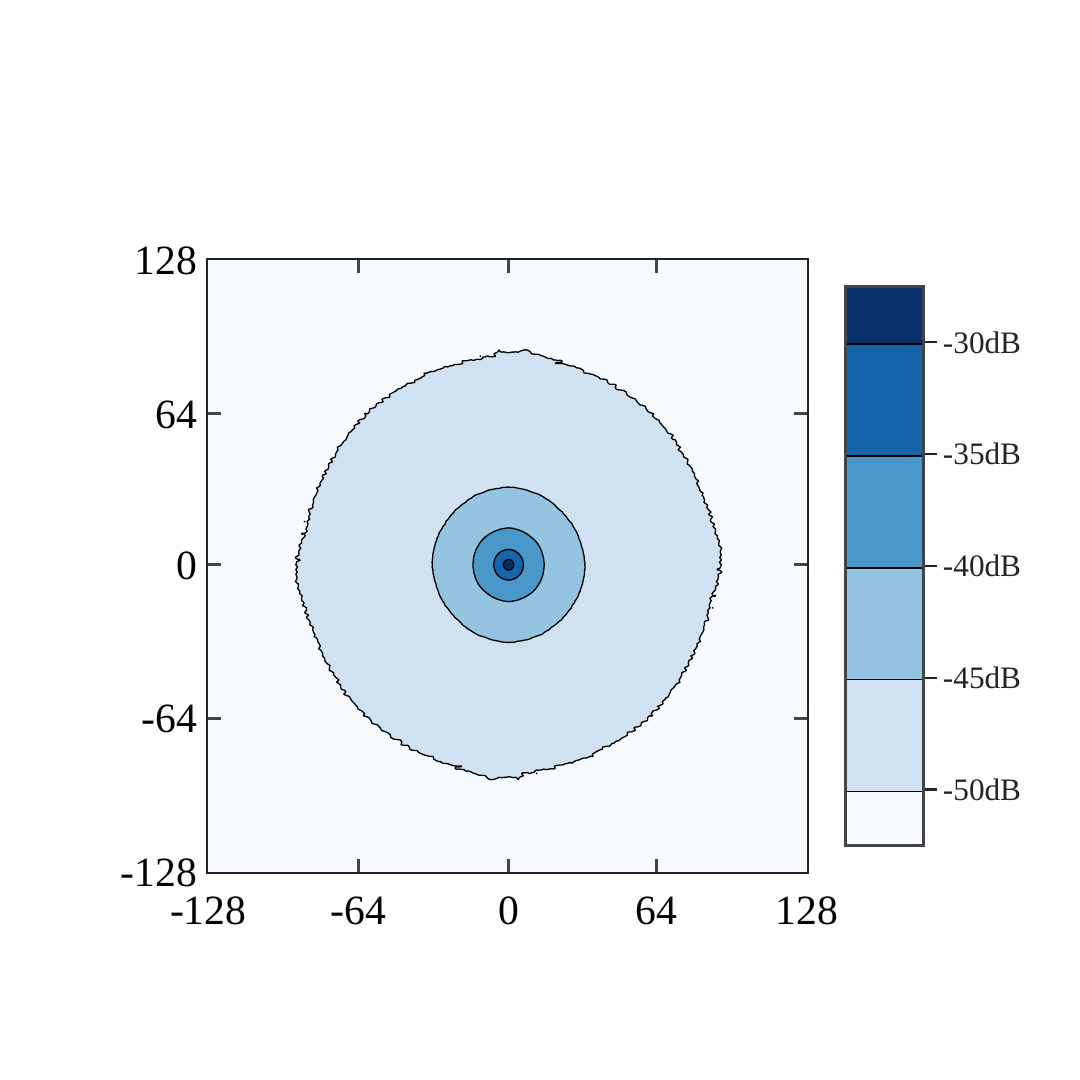}} 
    \caption{Average shifted 2D power spectrum of the COSMOS data set and raw network outputs. The shifted 2D power spectrum of individual galaxies is calculated as described in Fig. \ref{fig:powerspectra_example}. Units on the axes are in pixels. We show a contour plot in dB, i.e. values are in $\log_{10}$ scale.}
    \label{fig:comparison_zoom_avg_powerspectra_COSMOS}
\end{figure*}

\begin{figure*}
    \includegraphics[width=\textwidth]{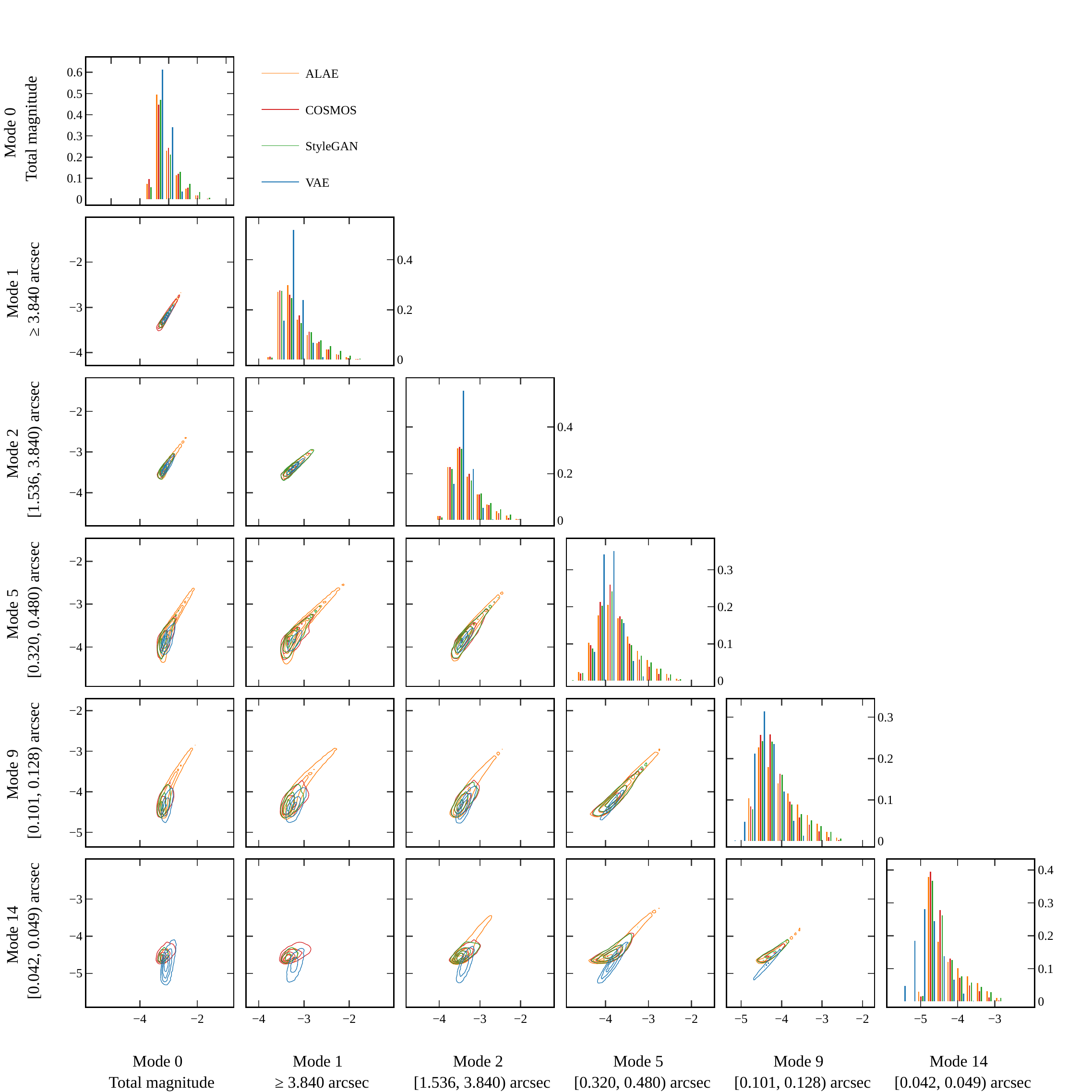}
    \caption{Corner plot for different wavelength ranges of the shifted 2D power spectra of the COSMOS data set. Axes are in $\log_{10}$-scale. The modes correspond to a partition of the wavelength range, see Table \ref{tab:quantitative_metrics}.}
    \label{fig:power_spectra_COSMOS}
\end{figure*}

\begin{figure*}
	\centering
    \subfloat[VAE]{\includegraphics[clip=True, trim=35 35 67 70, width=.23\textwidth]{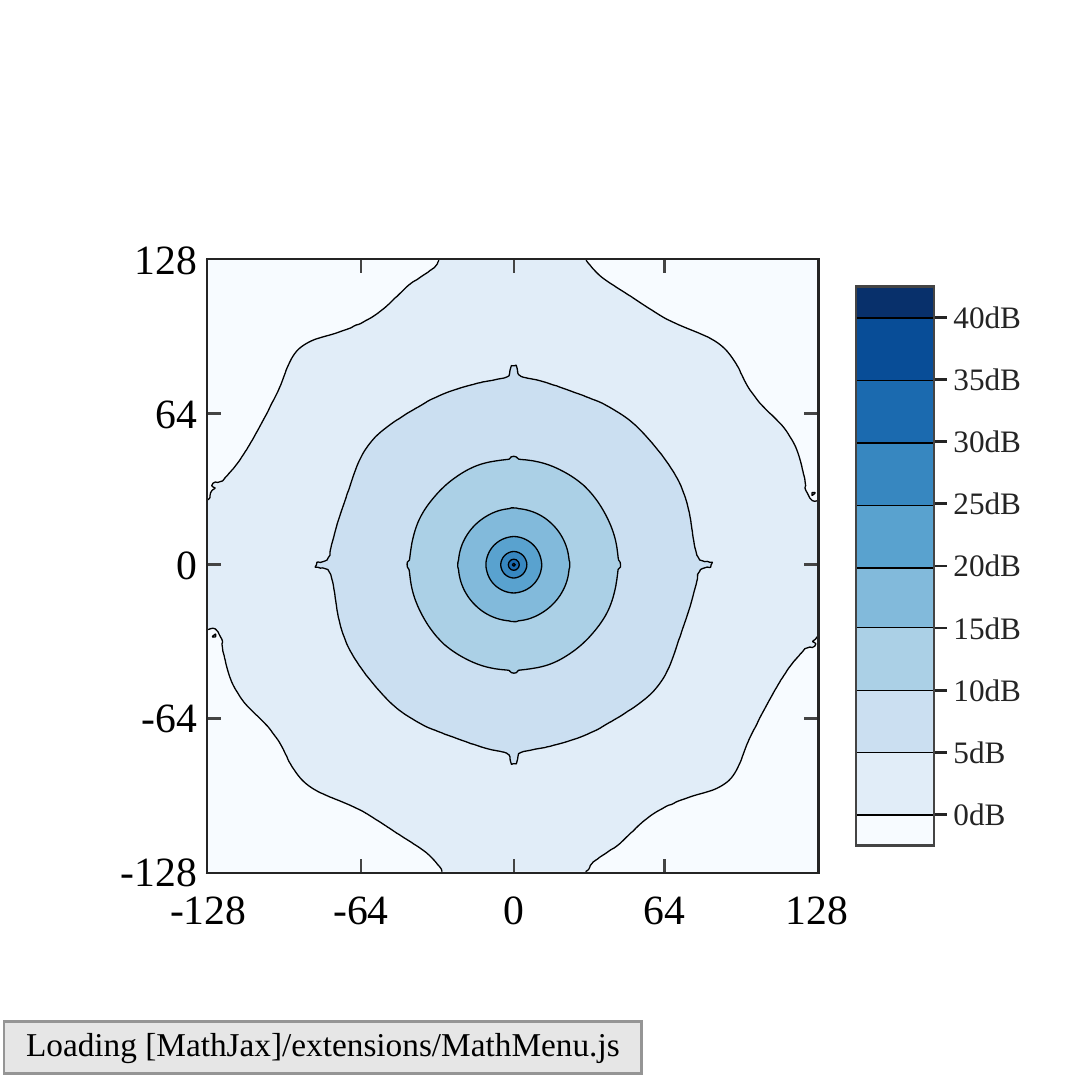}} 
    \subfloat[ALAE]{\includegraphics[clip=True, trim=35 35 67 70, width=.23\textwidth]{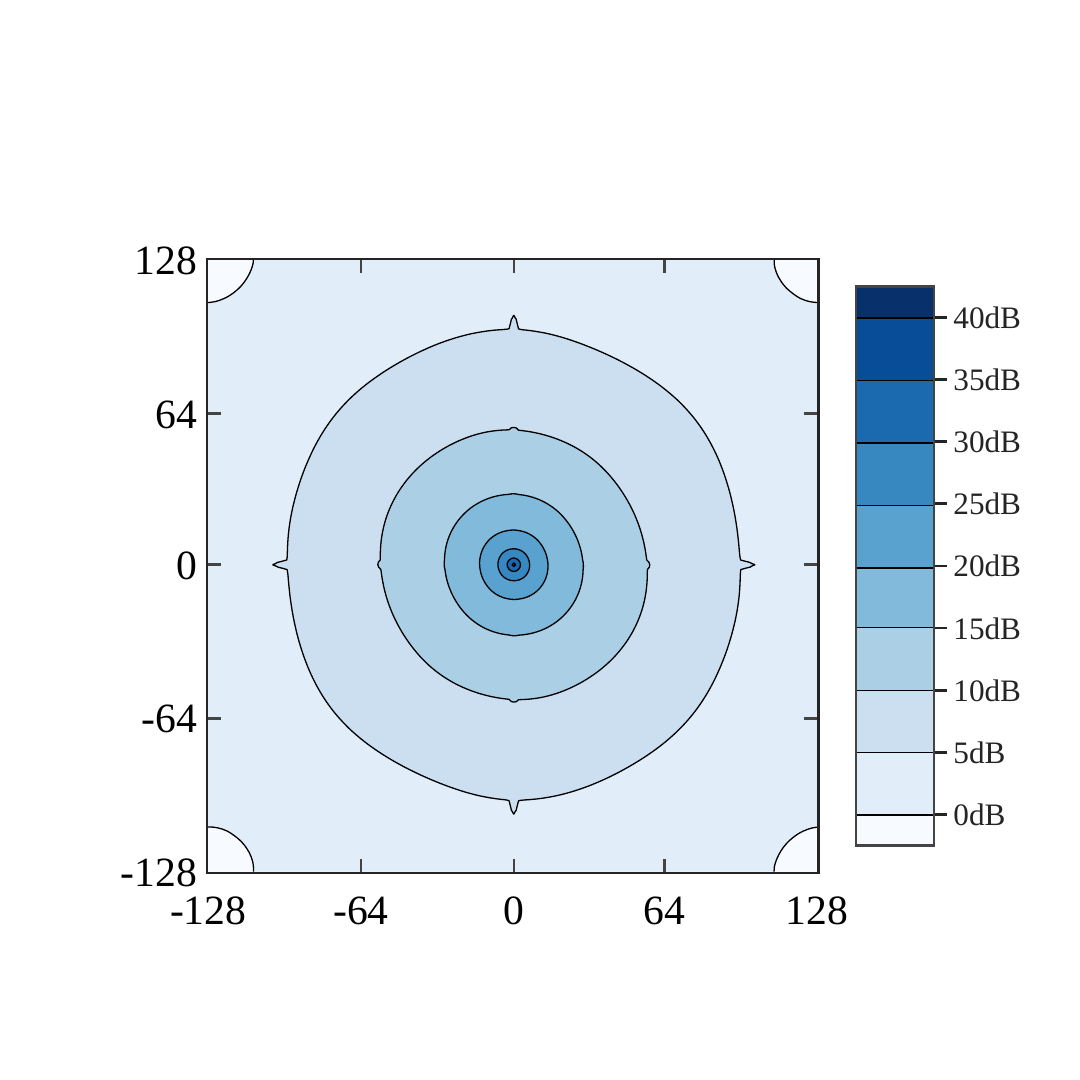}}  
    \subfloat[StyleGAN]{\includegraphics[clip=True, trim=35 35 67 70, width=.23\textwidth]{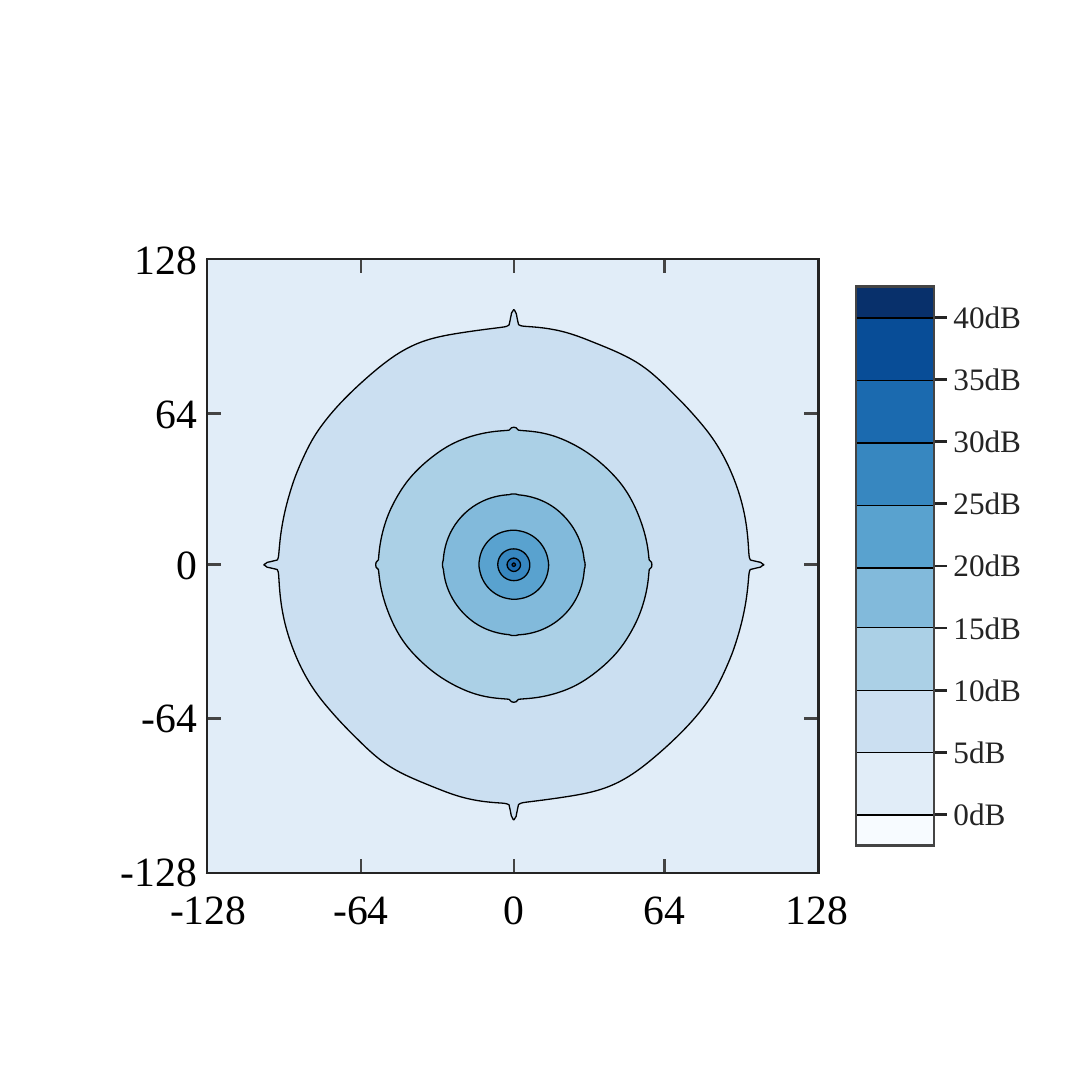}}  
    \subfloat[S\'ersic profiles]{\includegraphics[clip=True, trim=35 35 0 70, width=.31\textwidth]{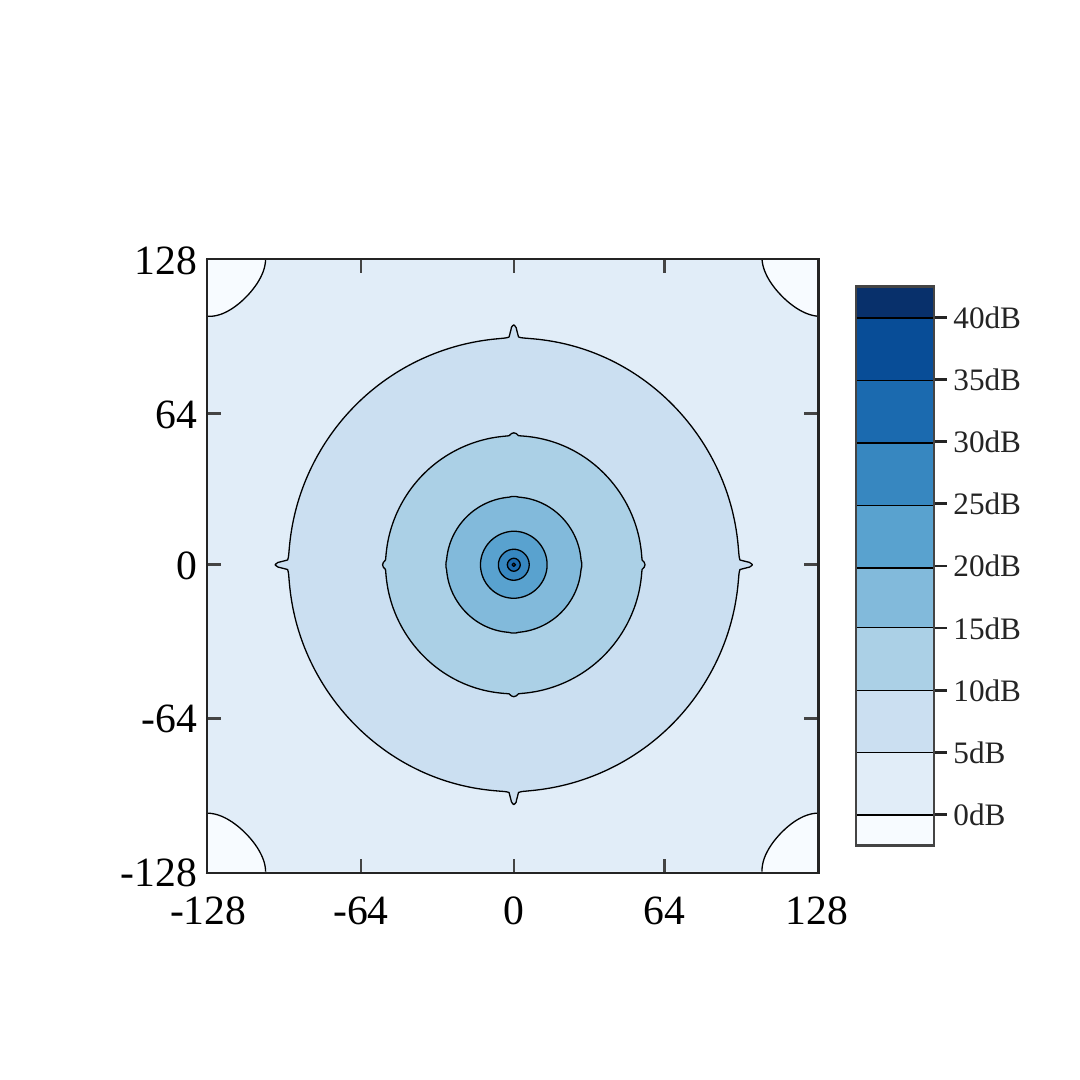}} 
    \caption{Average shifted 2D power spectrum of the S\'ersic profiles data set and raw network outputs. The shifted 2D power spectrum of individual galaxies is calculated as described in Fig. \ref{fig:powerspectra_example}. Units on the axes are in pixels. We show a contour plot in dB, i.e. values are in $\log_{10}$ scale.}
    \label{fig:comparison_zoom_avg_powerspectra_Sersic}
\end{figure*}

\begin{figure*}
    \includegraphics[width=\textwidth]{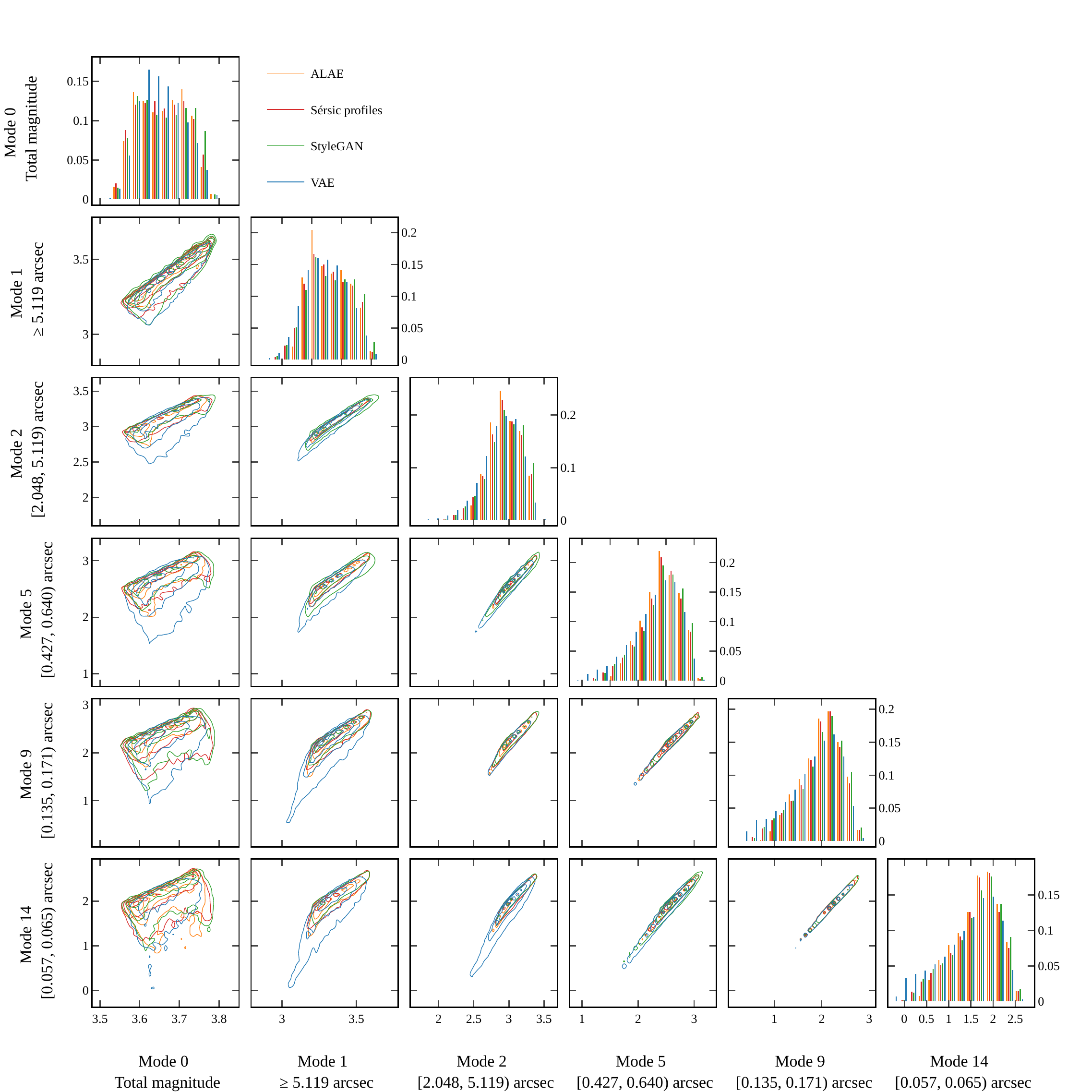}
    \caption{Corner plot for different wavelength ranges of the shifted 2D power spectra of the S\'ersic profiles. Axes are in $\log_{10}$-scale. The modes correspond to a partition of the wavelength range, see Table \ref{tab:quantitative_metrics}.}
    \label{fig:power_spectra_Sersic}
\end{figure*}

We show contour and corner plots for the COSMOS data set in Fig. \ref{fig:comparison_zoom_avg_powerspectra_COSMOS} and Fig. \ref{fig:power_spectra_COSMOS} as well as for the S\'ersic profiles in Fig. \ref{fig:comparison_zoom_avg_powerspectra_Sersic} and in Fig. \ref{fig:power_spectra_Sersic}.

\clearpage

%%%%%%%%%%%%%%%%%%%%%%%%%%%%%%%%%%%%%%%%%%%%%%%%%%

% Don't change these lines
\bsp	% typesetting comment
\label{lastpage}
\end{document}